%% file: main.tex
\documentclass[
 twocolumn,
 aps,
 prx,
 10pt,
 superscriptaddress,
 reprint,
 bibnotes,
 amsmath,amssymb,
 floatfix
]{revtex4-2}

\usepackage{times}
\usepackage{stmaryrd}
\usepackage[normalem]{ulem}
\usepackage{comment}
\usepackage[usenames,dvipsnames]{color}
\definecolor{dgreen}{rgb}{0.0, 0.5, 0.0}

\usepackage{color}
\usepackage[pdftex]{graphicx}
\usepackage{dcolumn}
\usepackage{bm}
\usepackage{here}
\usepackage{comment}
\usepackage{siunitx}
\usepackage[colorlinks,linkcolor=blue,citecolor=blue,urlcolor=black]{hyperref}

\include{commands}

\begin{document}
\title{Finite-time thermodynamic bounds and tradeoff relations for information processing}
\author{Takuya Kamijima}
\email{kamijima@noneq.t.u-tokyo.ac.jp}
\affiliation{
 Department of Applied Physics, The University of Tokyo, 7-3-1 Hongo, Bunkyo-ku, Tokyo 113-8656, Japan}

\author{Ken Funo}
\affiliation{
 Department of Applied Physics, The University of Tokyo, 7-3-1 Hongo, Bunkyo-ku, Tokyo 113-8656, Japan}

\author{Takahiro Sagawa}
\affiliation{
 Department of Applied Physics, The University of Tokyo, 7-3-1 Hongo, Bunkyo-ku, Tokyo 113-8656, Japan}
\affiliation{
 Quantum-Phase Electronics Center (QPEC), The University of Tokyo, 7-3-1 Hongo, Bunkyo-ku, Tokyo 113-8656, Japan}

\begin{abstract}
In thermal environments, information processing requires thermodynamic costs determined by the second law of thermodynamics.
Information processing within finite time is particularly important, since fast information processing has practical significance but is inevitably accompanied by additional dissipation.
In this paper, we reveal the fundamental thermodynamic costs and the tradeoff relations between incompatible information processing such as measurement and feedback in the finite-time regime.
To this end, we introduce a general framework based on the concept of the Pareto front for thermodynamic costs, revealing the existence of fundamental tradeoff relations between them.
Focusing on discrete Markov jump processes, we consider the tradeoff relation between thermodynamic activities, which in turn determines the tradeoff relation between entropy productions.
To identify the Pareto fronts, we introduce a new Wasserstein distance that captures the thermodynamic costs of subsystems, providing a geometrical perspective on their structure.
Our framework enables us to find the optimal entropy production of subsystems and the optimal time evolution to realize it.
In an illustrative example, we find that even in situations where naive optimization of total dissipation cannot realize the function of Maxwell's demon, reduction of the dissipation in the feedback system according to the tradeoff relation enables the realization of the demon.
We also show that an optimal Maxwell's demon can be implemented by using double quantum dots.
Furthermore, our framework is applicable to larger scale systems with multiple states, as demonstrated by a model of chemotaxis.
Our results would serve as a designing principle of efficient thermodynamic machines performing information processing, from single electron devices to biochemical signal transduction. 
\end{abstract}


\maketitle
\section{Introduction}
\subsection{Background}
With advancements in measurement and control techniques in microscopic systems, stochastic thermodynamics has seen progress in both theoretical and experimental aspects over the past few decades \cite{sekimoto2010stochastic,Seifert2012,peliti2021stochastic,Ciliberto2017experiment-history}.
Although the second law of thermodynamics has been well established, the maximum efficiency can only be achieved in infinite time (i.e., the quasi-static limit).
Therefore, exploring thermodynamic bounds in the finite-time regime has become a topic of active research \cite{Curzon-Ahlborn1975,Broeck2005EMP,Schmiedl-Seifert2007optimal,Crooks2007length,Esposito-2010EMP-low,Sivak-Crooks2012metrics,Blaber-Sivak2023optimal,Ma-2020exp-scaling}, especially in terms of thermodynamic speed limit \cite{Shiraishi-Funo-Saito2018speed,Ito2018geometry,Ito-Dechant2020info,Falasco-Esposito2020dissipation-time,Hamazaki2022speed} and thermodynamic uncertainty relations \cite{Barato-Seifert2015,Gingrich-Horowitz2016,pietzonka-Barato-Seifert2016,Shiraishi-Saito-Tasaki2016heatengine,proesmans2017discrete,Maes2017Frenetic,dechant2018multidimensional,Brandner-Hanazato-Saito2018,Pietzonka-Seifert2018,Hasegawa-Van2019,koyuk2020thermodynamic,Liu-Gong-Ueda2020,Otsubo-Ito-Dechant-Sagawa2020}.
The central problem is estimating the additional entropy production (EP) that accompanies finite-time thermodynamic processes, compared to the infinite-time limit.
In particular, it has been revealed that optimal transport theory \cite{villani2009,peyre2019computational} provides the minimum amount of EP in finite-time processes for both overdamped Langevin systems \cite{Aurell2011Wasserstein,Aurell2012refined,Dechant-Sakurai2019,Nakazato-Ito2021} and discrete systems \cite{Muratore2013heat,dechant2022minimum,Van-Saito-unification2023,Van-Saito2023topological}, where the optimal thermodynamic cost can always be achieved by designing optimal protocols for any finite time interval.
The Landauer principle \cite{Landauer1991information,Esposito2011Landauer,berut2012Landauer_exp,Jun-Gavrilov-Bechhoefer2014highprecision,Jeongmin2016Landauernanomagnetic,Dago-2021exp-Landauer-underdamped} for finite-time information erasure has been addressed by this approach \cite{Proesmans2020finiteLandauer,Zhen2021boundbitreset,Lee-Park_highly2022,Van-Saito2022quantumLandauer,Sacdi-2022erasurelength}, but more general information processing in finite time remains elusive.

\begin{figure*}
    \centering
    \includegraphics[width=0.95\linewidth]{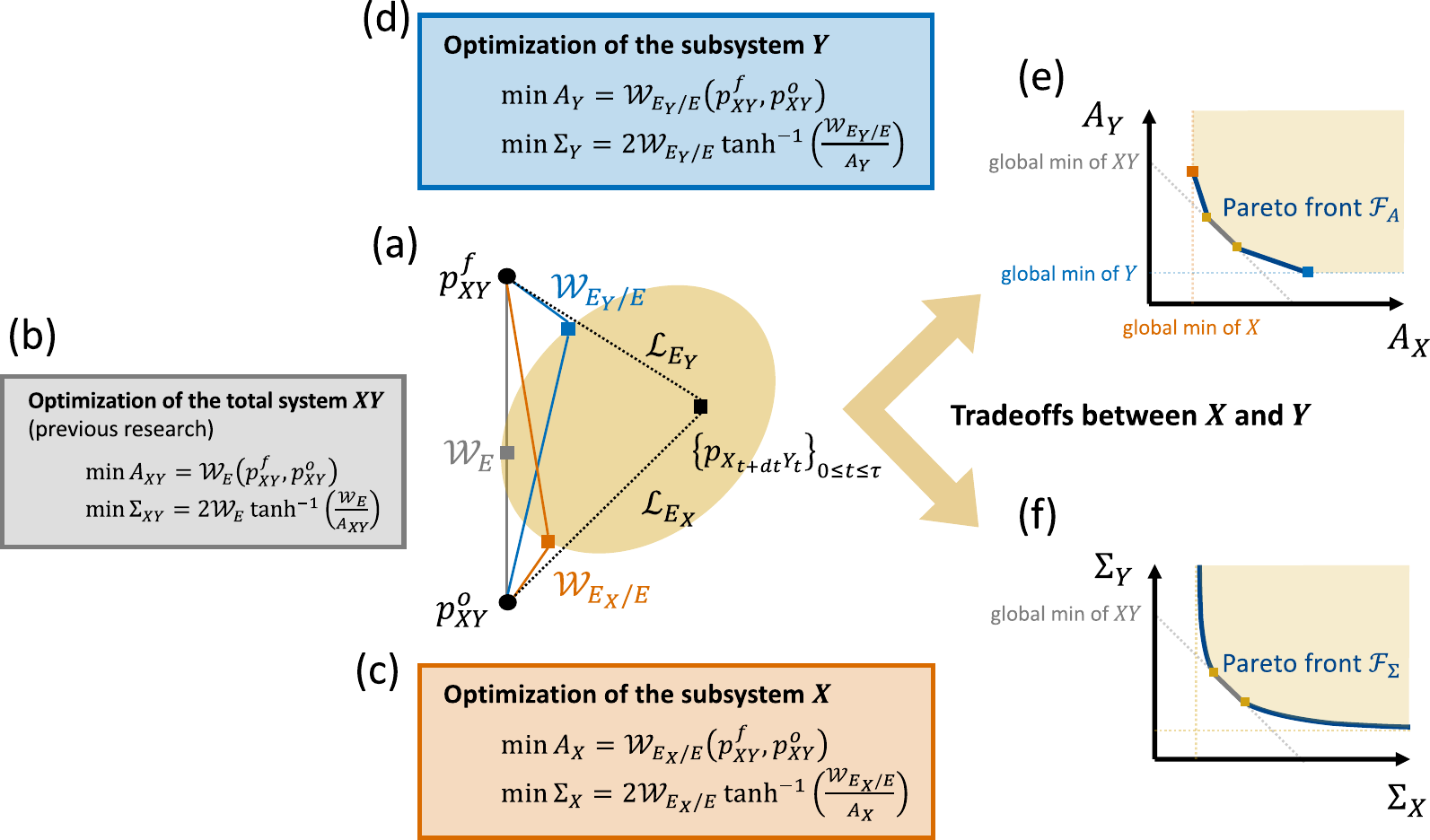}
    \caption{\label{fig: summary}
    The summary of main results and visualization of the origin of the thermodynamic tradeoff relations.
    (a) The central schematic diagram illustrates the difference between local and global optimization using partial time evolution (black squares).
    Partial time evolution specifies the extent to which the transition rates of subsystems (such as the memory and the engine) contribute to the overall time evolution (see \fref{fig: partial_evolution} and \esref{partial evolution: X}\eqref{partial evolution: Y}). 
    In local optimization, the black square is fixed, and the transition rates within each subsystem are optimized.
    The length $\calL_{E_X}$ ($\calL_{E_Y}$) from the initial distribution $\poo$ (final distribution $\pff$) to the black square represents the local minimum of the partial activity for $X$ ($Y$).
    In global optimization, this partial time evolution is optimized as well.
    The yellow area in the figure schematically represents the range within which the black square can be taken.
    By choosing the time evolution represented by the gray square, the total activity can reach the minimum value \cite{dechant2022minimum} (gray box (b)).
    To globally minimize the partial activity of $X$, one should choose the time evolution represented by the orange square.
    Conversely, selecting the blue square will globally minimize the partial activity of $Y$.
    (c)(d) The orange and blue boxes represent the global minimum thermodynamic costs of subsystems $X$ and $Y$, respectively, during the time evolution $\poo \rightarrow \pff$.
    Since the partial time evolutions (orange and blue squares) that realize these two minima generally do not coincide, it is not possible to simultaneously achieve the minimum activity or dissipation for subsystems $X$ and $Y$.
    Therefore, there exist tradeoff relations between these thermodynamic costs.
    (e)(f) The schematic diagrams on the right represent the typical tradeoff relation between the activities of the subsystems, as well as between the dissipations of the subsystems.
    The yellow-shaded areas in the diagrams represent the feasible regions of thermodynamic costs, and the lower left polyline and curved line indicate the optimal bounds (Pareto fronts $\calF_A$ and $\calF_\Sigma$).
}
\end{figure*}

\begin{figure}[H]
    \centering
    \includegraphics[width=0.90\linewidth]{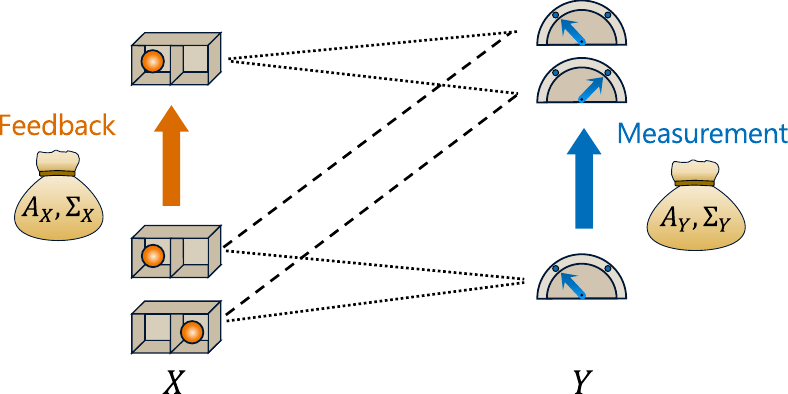}
    \caption{\label{fig: partial_evolution}
    A schematic of the simplest setup for measurement and feedback.
    Subsystem $Y$ serves as the memory and measures subsystem $X$, which functions as the engine.
    Based on the measurement results, $X$ receives feedback. 
    Both the measurement and feedback processes, performed in finite time, require the thermodynamic costs such as activities $A_X,A_Y$ and entropy productions $\Sigma_X,\Sigma_Y$.
    The overall time evolution is divided into separate steps of measurement and feedback, allowing $X$ and $Y$ to be considered as evolving partially in time.
    We however emphasize that our theoretical framework developed in this paper is applicable to a broader class of information processing, where the measurement and feedback steps are not necessarily temporally separated.
    }
\end{figure}

In general, information plays a central role in thermodynamics \cite{Parrondo-Horowitz-Sagawa2015}, as illustrated by a thought experiment of ``Maxwell's demon" \cite{leff2002maxwell}.
A typical setup consists of measurement and feedback as shown in \fref{fig: partial_evolution}, where subsystem $Y$ (the memory of a demon) performs measurement and feedback on subsystem $X$ (the engine).
As in this example, cooperative behavior of subsystems enables information processing and work extraction.
In recent decades, the fundamental energy costs (or the extractable work) in such information processing processes have been revealed from the perspective of the generalized second law of thermodynamics \cite{Sagawa-Ueda2008quantumfeedback,Sagawa-Ueda2009erasure,Sagawa-Ueda2010feedback,ItoSagawa2013causal,infoflow-Holowitz-2014,toyabe-2010exp-Jarzynski,Koski-Sagawa2014experiment,Ribezzi-Marco2019exp-Maxwell}.

However, the thermodynamic costs for Maxwell's setups, including processes such as measurement and feedback, in finite time have been rarely addressed \cite{abreu-Seifert2011feedback,Nakazato-Ito2021,Taghvaei-2022-demon,FujimotoIto2023,Nagase-Sagawa2023infogain} and yet to be understood.
In fact, optimal transport theory cannot be directly applied to subsystems such as the memory and engine, while as mentioned before, the thermodynamic costs such as thermodynamic activity and EP for the total system have recently been studied in terms of optimal transport theory.

\subsection{Summary of the Results}
In this paper, we explore the fundamental bounds of thermodynamic costs for information processing in finite time.
To this end, we introduce a general framework based on the concept of the Pareto front for thermodynamic costs to characterize the tradeoff relations between them.
Based on this framework, we reveal that there exist tradeoff relations for both entropy productions and thermodynamic activities.
To determine the Pareto fronts, we generalize optimal transport theory by newly introducing a Wasserstein pseudo-distance associated with subsystems.
Applying our framework to a measurement and feedback setting, we demonstrate that the thermodynamic costs required for measurement and feedback are incompatible.
We then apply our framework to a simple model of information processing, elucidating a design principle of Maxwell's demon.
Furthermore, we demonstrate that our framework can determine the optimal thermodynamic costs even in more complex systems, exemplified by a model of chemotaxis.

Below, we briefly summarize our results of this paper, which are schematically shown in \fref{fig: summary}.
We focus on Markov jump processes as a theoretical description of thermodynamic processes. 
The transition rates are manipulated over a finite operation time $\tau$ to evolve the system from the initial distribution $\poo$ to the final distribution $\pff$.
The transition rates of the total system are defined only on the edge set $E$, which consists of the edge sets $E_X$ and $E_Y$ of subsystems $X$ and $Y$.
Figure \ref{fig: summary}(a) illustrates the differences in local and global optimizations and the origin of the tradeoff relations in thermodynamic costs.
For these optimizations, we introduce the concept of partial time evolution, as illustrated in \fref{fig: partial_evolution}.
In this setting, the time evolution of the total system can be decomposed into changes in subsystems $X$ and $Y$, corresponding to measurement and feedback, respectively.
Such a decomposition can also be specified at each time point if we consider an infinitesimal time step, referred to as partial time evolution $\ptt \rightarrow \pdt, \ptd (0 \leq t \leq \tau)$.
In other words, $\pdt$ describes how subsystem $X$ contributes to the overall time evolution, given $\ptt \rightarrow \pdd$ (see \esref{partial evolution: X}\eqref{partial evolution: Y} for definition).
The black square in \fref{fig:  summary}(a) represents one such partial time evolution.
The yellow-shaded area schematically represents the set of possible partial time evolutions.


\vskip\baselineskip
In \sref{subsec: Local optimization}, we perform local optimization of the thermodynamic costs of the subsystems.
To achieve this, we fix a specific partial time evolution $\ptt \rightarrow \pdt, \ptd (0 \leq t \leq \tau)$.
From the perspective of information processing, this specifies the information exchange between subsystems, determining whether $X$ or $Y$ performs measurement or feedback at each moment.
In this scenario, we can optimize the transition rates $R_X$ ($R_Y$) during the partial time evolution $\ptt \rightarrow \pdt$ ($\ptt \rightarrow \ptd$) to minimize the thermodynamic costs of subsystem $X$ ($Y$).
The minimal cost obtained through this optimization is feasible for each subsystem, and we refer to this as \textit{local optimization} in this paper.
In \fref{fig: summary}(a), the length $\calL_{E_X}$ ($\calL_{E_Y}$) from the initial distribution $\poo$ (final distribution $\pff$) to the black square represents the partial activity required for $X$ ($Y$) during this time evolution (see \eref{min AX rate local}).
Using these, the minimal partial EPs can also be determined (see \eref{min EP_X rate local}).
However, since there is still room to optimize the partial time evolutions $\pdt$ and $\ptd$, these minimum values can potentially be further reduced.

\vskip\baselineskip
In \sref{subsec: Global optimization}, we perform global optimization of the thermodynamic costs for the subsystems.
This optimization involves not only the procedures of local optimization but also optimizing the partial time evolutions.
This corresponds to moving the black square (partial time evolution) within the yellow-shaded area in \fref{fig: summary}(a).
Therefore, to minimize the partial activity of $X$, one should choose the time evolution (orange square) that minimizes $\calL_{E_X}$.
This optimization involves adjusting the $Y$ component of the transition rates, $R_Y$, thereby optimizing the overall transition rates, $R$.
Physically, this means selecting the interactions between $X$ and $Y$ such that the changes in $X$ are minimized.
We refer to this as \textit{global optimization} in this paper.
We generalize the Wasserstein distance to solve this optimization.
The global minimum of the partial activity for $X$ is given by the first equation in \fref{fig: summary}(c) (see \eref{min AX global}).
Conversely, the global minimum of the partial EP for $X$ is given by the second equation in \fref{fig: summary}(c) (see \eref{min EP_X global}).
To ensure that the timescale of $X$'s dynamics remains finite, we set an upper bound $A_X$ on $X$'s partial activity.
Similarly, global optimization can be performed for $Y$ (see \fref{fig: summary}(d)).

The global minimum costs for $X$ and $Y$ in general cannot be achieved simultaneously.
This is because the protocols that realize these minimum costs are distinct between $X$ and $Y$.
This discrepancy arises from the different preferences in partial time evolution (represented by the orange and blue squares in \fref{fig: summary}(a)).
Consequently, there exist tradeoff relations between the thermodynamic costs for $X$ and $Y$.

\vskip\baselineskip
In \sref{sec: Tradeoff relations in the thermodynamic costs of subsystems}, we determine the bounds of the tradeoff relations between the thermodynamic costs of subsystems $X$ and $Y$.
The bounds of such tradeoff relations can be explicitly represented by the Pareto front, which is the set of optimal cost combinations (see \sref{subsec: Pareto front}).
We derive the Pareto front of partial activities in \sref{subsec: partial activity Pareto}, based on which we derive the Pareto front of partial EPs in \sref{subsec: partial EP Pareto}.
Similar to global optimization, obtaining the Pareto front $\calF_A$ of the partial activities requires optimizing the partial time evolutions, which reduces to solving an optimal transport problem.
In general, $\calF_A$ forms a convex polyline with a finite number of vertices and edges (see \fref{fig: summary}(e)).
A vertex appears when the solution to the optimal transport problem is unique, whereas an edge appears when it is not.
Along the edges, the operational speed required for time evolution can be adjusted between subsystems.
The shape of $\calF_A$ depends on the initial and final distributions but is mainly determined by the structure of the graph on which the dynamics takes place.

The Pareto front $\calF_\Sigma$ of the partial EPs can be obtained using the Pareto front $\calF_A$ of the partial activities in \fref{fig: summary}(e).
To maintain finite timescales for the dynamics, we set an upper bound on the activity of the total system instead of the partial activities.
In general, $\calF_\Sigma$ forms a convex curve (see \fref{fig: summary}(f)), reflecting the shape of $\calF_A$. 
When the dissipation of the total system is minimized, the dissipations of the subsystems are not necessarily minimized and can potentially be further reduced.
However, the tradeoff relation is not linear, and reducing the dissipation in one subsystem requires significantly increasing the dissipation in the other subsystem.
If multiple protocols minimize the activity of the total system, this tradeoff relation qualitatively changes.
As a result, $\calF_\Sigma$ exhibits a linear tradeoff relation with an edge of slope $-1$.
Thus, it is possible to exchange dissipation among the subsystems while maintaining the optimality of the total dissipation.

\vskip\baselineskip
In \sref{sec: Optimal information processing}, we apply our general results to bipartite systems to determine the optimal information processing in finite time.
We consider the situation where measurement is performed before feedback, as well as the situation where they are carried out simultaneously.
In such information processing, there exists a tradeoff relation between the dissipation arising from measurement and that from feedback.
Moreover, increasing the activity corresponds to increasing the number of steps in information processing and enables more diverse information processing.
Specifically, as the activity decreases, it becomes impossible to realize Maxwell's demon while optimizing the total dissipation.
Further reducing the activity makes it impossible to realize Maxwell's demon.
Finally, we present a specific protocol to implement this optimal information processing using double quantum dots.

\vskip\baselineskip
In \sref{sec: multistate}, our results are applied to a multi-state system.
We examine a model of \textit{E. coli} chemotaxis and optimize the time evolution of sensory adaptation, where one steady state relaxes to another when the ligand concentration is suddenly increased.
The Wasserstein distance between these states is calculated numerically, and the optimal protocol is derived.
We demonstrate that the partial EP associated with methylation and demethylation can be reduced by more than $200$ times compared to the autonomous situation.
This highlights the cost of achieving a specific time evolution using autonomous transition rates that violate the detailed balance condition.

\section{Setup}
\label{sec: Setup}
In this section, we provide an overview of stochastic thermodynamics in discrete systems and minimization of thermodynamic costs using optimal transport theory.
Throughout this paper, the operation time $\tau$, initial distribution $\poo$, and final distribution $\pff$ are fixed.
For simplicity, we will consider bipartite systems throughout this paper.
However, results that do not involve information flow are applicable to general subsystems.

\subsection{Stochastic Thermodynamics in Discrete Systems}
We introduce the dynamics of discrete systems.
State transitions can be represented as a graph $G(N, E)$, where the node set is $N = \{r\}$ and the edge set is $E = \{[r, r']\}$.
Here, node $r$ represents a state, and edge $[r, r']$ represents the transition from state $r'$ to $r$.
It is assumed that if $[r, r'] \in E$, then $[r', r] \in E$.
The time evolution is Markovian and can be described by the master equation using transition rates \cite{vanKampen, gardiner2009stochastic}.
The transition rates have nonzero values only on $E$.
Furthermore, to make the dynamics consistent with thermodynamics, the transition rates are assumed to satisfy the local detailed balance condition \cite{Seifert2012}.

\begin{figure}
    \centering
    \includegraphics[width=0.95\linewidth]{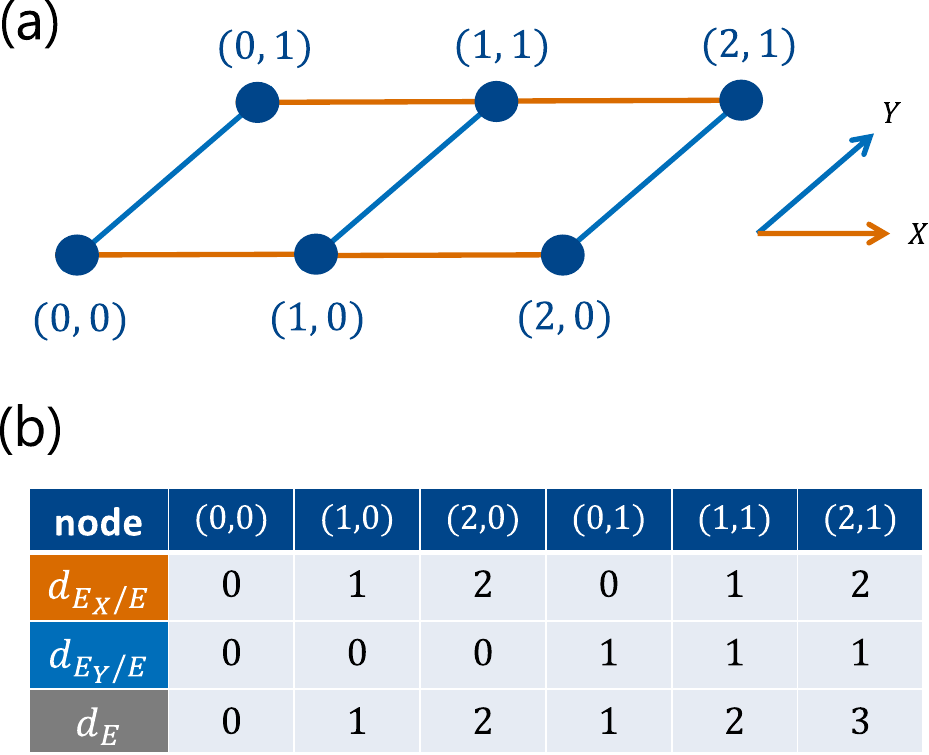}
    \caption{\label{fig: cost_func}
    (a) An example of a bipartite graph $G(N, E)$.
    There are no transitions where both $X$ and $Y$ change simultaneously.
    (b) The differences in the cost of transporting probability from the node $(0,0)$ to each node in $G(N, E)$ (see \esref{def of d_E}\eqref{def of d_EX/E}).
    }
\end{figure}

We consider a graph $G(N, E)$ with a bipartite structure.
The state $r$ is specified by two indices of subsystems $X$ and $Y$, that is, $r = (x, y)$.
The transition rate from node $r'$ to $r$ is denoted as $R(r, r')$ and satisfies
\cite{hartich2014transfer,infoflow-Holowitz-2014}
\begin{align}
    \label{bipartite}
    R(r,r')=
    \begin{cases}
    R_{xx'}^{y}\geq0 \ &(x\neq x',y=y') \\
    R_{x}^{yy'}\geq0 \ &(x=x',y\neq y') \\
    0 \ &(\text{otherwise})
    \end{cases}.
\end{align}
In other words, there are no transitions where both $X$ and $Y$ change simultaneously (see \fref{fig: cost_func}(a)).
The set $E$ is expressed as the disjoint union of the sets of edges representing transitions related to $X$ and $Y$, i.e., $E = E_X \sqcup E_Y$, which is defined as $E = E_X \cup E_Y$ and $E_X\cap E_Y=\emptyset$.

The probability distribution starts from the initial distribution $\poo$ at time $t = 0$ and evolves in time according to the master equation, reaching the final distribution $\pff$ at time $t = \tau (>0)$.
The master equation can be expressed as 
\begin{align}
    \label{master eq}
    d_t\ptt(r)&=\sum_{r'(\neq r)}\big(R(r,r')\ptt(r')-R(r',r)\ptt(r)\big)\nonumber\\
    &=\sum_{x'(\neq x)}J_{xx'}^{y}+\sum_{y'(\neq y)}J_{x}^{yy'}.
\end{align}
Here, $\ptt(r)$ is the probability that the total system is in state $r = (x, y)$ at time $t$.
The probability current from state $(x', y)$ to $(x, y)$ is given by $J_{xx'}^{y} = R_{xx'}^{y} \ptt(x', y) - R_{x'x}^{y} \ptt(x, y)$, and the probability current from state $(x, y')$ to $(x, y)$ is given by $J_{x}^{yy'} = R_{x}^{yy'} \ptt(x, y') - R_{x}^{y'y} \ptt(x, y)$.
These quantities may depend on time, which is not explicitly denoted for simplicity of notations.

First, we consider partial activity as a thermodynamic cost.
The partial activity rates $\dot{A}_X$ and $\dot{A}_Y$ for subsystems $X$ and $Y$ are defined as \cite{Maes2020}
\begin{align}
    \label{partial activity rate}
    \dot{A}_X&:=\sum_{x\neq x',y}R_{xx'}^{y}\ptt(x',y),\nonumber\\ 
    \dot{A}_Y&:=\sum_{y\neq y',x}R_{x}^{yy'}\ptt(x,y'),
\end{align}
which correspond to the average number of transitions per unit time within each subsystem.
These rates characterize the inverse of the relaxation times of the subsystems.
The partial activities $A_X$ and $A_Y$ are obtained by integrating \eref{partial activity rate} over time from $t = 0$ to $t = \tau$.
Increasing $A_X$ and $A_Y$ for a fixed $\tau$ implies operating the subsystems more quickly, thereby allowing partial activity to be considered as a cost.

Second, we consider partial EP as the other thermodynamic cost.
The partial EP rates $\dot{\Sigma}_X$ and $\dot{\Sigma}_Y$ for subsystems $X$ and $Y$ are defined as \cite{infoflow-Holowitz-2014}
\begin{align}
    \label{partial EP rate}
    \dot{\Sigma}_X&:=\sum_{x>x',y}J_{xx'}^{y}\ln\frac{R_{xx'}^{y}\ptt(x',y)}{R_{x'x}^{y}\ptt(x,y)}\nonumber,\\ 
    \dot{\Sigma}_Y&:=\sum_{y>y',x}J_{x}^{yy'}\ln\frac{R_{x}^{yy'}\ptt(x,y')}{R_{x}^{y'y}\ptt(x,y)}.
\end{align}
These rates are nonnegative due to the inequality $(a - b) \ln (a/b) \geq 0$ for $a, b > 0$, representing the second law of thermodynamics for individual subsystems.
Their magnitudes represent the irreversibility of the dynamics and the degree of energy dissipation within the subsystems.
The partial EPs $\Sigma_X$ and $\Sigma_Y$ are obtained by integrating \eref{partial EP rate} over time.

Since $E = E_X \sqcup E_Y$, the sums of the partial activities and partial EPs equal the activity and EP of the total system, respectively: $A_{XY} = A_X + A_Y$ and $\Sigma_{XY} = \Sigma_X + \Sigma_Y$.
We note that partial activities and partial EPs can be defined for any subsystem $E' \subset E$, not limited to bipartite systems \cite{Shiraishi-Sagawa2015masked}.

\subsection{Optimal Transport Theory and the Wasserstein Distance}
Here we briefly review optimal transport theory \cite{villani2009, peyre2019computational}, in which optimal transport cost for the time evolution of probability distributions is discussed.
Specifically, probability distribution $q$ on the node set $N$ is transformed into $p$ by transporting probabilities through the edge set $E$.
However, it is not necessary to transport the probability on each node injectively; it can be split and transported to multiple nodes.
These transports can be represented by a transport matrix $\Pi$, where $\Pi(r, r') (\geq 0)$ denotes the amount transported from node $r'$ to $r$.
To ensure that $q$ is completely transformed into $p$, the transport matrix must satisfy $\sum_{r} \Pi(r, r') = q(r')$ and $\sum_{r'} \Pi(r, r') = p(r)$.

To consider the optimality of transport, we define the cost associated with transport between nodes.
Here, we adopt the length $d_E$ of the shortest path on the graph as the cost function.
The cost $d_E(r, r')$ is explicitly defined as
\begin{align}
\label{def of d_E}
    d_{E}(r,r'):=\min_{\substack{P:r'\rightarrow r\\ \textrm{on }E}} l_{E}(P),
\end{align}
where $P$ represents a path on $E$ connecting node $r'$ with $r$ and $l_{E}(P)$ counts the number of edges in $E$ traversed by $P$. 
From the definition, $d_E$ satisfies the axioms of a distance.

The optimal transport cost under such a cost function has the properties of a distance between probability distributions and is called the ($L^1$) Wasserstein distance, which is defined as
\begin{align}
    \label{def Wasserstein}
    \calW_E(p,q):=\min_{\Pi\in\calU(p,q)}\sum_{r,r'}d_E(r,r')\Pi(r,r').
\end{align}
Here, $\calU(p,q)$ is defined as the set of transport matrices given by $\calU(p,q):=\{\Pi\ |\ \Pi(r,r')\geq0, \sum_{r}\Pi(r,r')=q(r'), \sum_{r'}\Pi(r,r')=p(r)\}$, also known as the transportation polytope from $q$ to $p$.

The Wasserstein distance can be bounded from below using the total variation distance $\calT(p,q):= \sum_r |p(r) - q(r)| / 2$, that is, $\calW_E(p, q) \geq \calT(p, q)$ holds \cite{Van-Saito-unification2023}.
In particular, when the graph $G(N, E)$ is fully connected, the equality is achieved.
Whereas the total variation distance focuses only on the differences in the values of the probability distributions, the Wasserstein distance also takes into account the configurational differences of these distributions.

\subsection{Review: Minimal Thermodynamic Costs of the Total System}
\label{subsec: review discrete total}
We next review the minimization of the thermodynamic cost of the total system using optimal transport theory following \ccite{dechant2022minimum}, which plays a central role in our study.
Activity and EP can be interpreted as the thermodynamic costs associated with the time evolution of the probability distribution \cite{Shiraishi-Funo-Saito2018speed,Ito-Dechant2020info}.
This perspective is closely related to optimal transport theory, which discusses the cost of transporting probability distributions.
In fact, as we will see below, the minimum values of activity and EP can be expressed using the Wasserstein distance.

\subsubsection{Minimal activity}
\label{subsubsec: Minimal activity}
First, we consider minimizing the activity rate in the infinitesimal time evolution $\ptt \rightarrow \pdd$.
The initial distribution $\ptt$ and the final distribution $\pdd$ are fixed.
There are numerous transition rates that can realize this time evolution.
Here, we optimize the transition rates to minimize the activity rate $\dot{A}_{XY} := \dot{A}_X + \dot{A}_Y$ of the total system.
This minimum value can be expressed using the Wasserstein distance \eqref{def Wasserstein} as \cite{dechant2022minimum}
\begin{align}
    \label{min AXY rate}
    \min_{R:\ptt\rightarrow\pdd}\dot{A}_{XY}
    &=\frac{\calW_{E}(\pdd,\ptt)}{dt}\nonumber\\
    &=:\dot{\calL}_{E}.
\end{align}
The optimization variable $R$ denotes the transition rates defined on $E$ that give the time evolution of \eref{master eq}.
We note that $\dot{\calL}_E$ can be interpreted as the speed of time evolution measured by the Wasserstein distance.

We next consider minimizing the activity over the finite time evolution $\{\ptt\}_{0 \leq t \leq \tau}$.
Here, $\{\ptt\}_{0 \leq t \leq \tau}$ represents a (continuous) trajectory of probability distributions, specifying the distribution at each moment.
For this trajectory, the infinitesimal time procedure can be repeated at each moment, and the minimum value of the activity is given by $\calL_E := \int_0^\tau dt \, \dot{\calL}_E$.
This represents the length of the trajectory measured by the Wasserstein distance.

Finally, we consider minimizing the activity for the finite-time evolution $\poo \rightarrow \pff$.
In this case, only the initial distribution $\poo$ and the final distribution $\pff$ are fixed, and the trajectory of probability distributions during the intermediate time is also optimized.
Given the fact that the Wasserstein distance satisfies the triangle inequality, the minimum value of the activity obtained when optimizing the trajectory is
\begin{align}
    \label{min AXY}
    \min_{\{R\}_{0\leq t\leq \tau}:\poo\rightarrow\pff}{A}_{XY}
    =\calW_{E}(\pff,\poo).
\end{align}
The optimization variable $\{R\}_{0 \leq t \leq \tau}$ denotes a protocol consisting of transition rates defined on $E$, which realizes the time evolution $\poo \rightarrow \pff$ according to \eref{master eq}.

The trajectory that provides the minimum value in \eref{min AXY} corresponds to the geodesic under the Wasserstein distance.
However, this geodesic is not necessarily unique in general.
One possible geodesic is the linear interpolation between the initial and final distributions, given by $\ptt = (1 - t/\tau)\poo + (t/\tau)\pff$.

\subsubsection{Minimal entropy production}
\label{subsubsec: Minimal EP}
First, we consider minimizing the EP rate in the infinitesimal time evolution $\ptt \rightarrow \pdd$.
We optimize the transition rates that realize this time evolution to minimize the EP rate $\dot{\Sigma}_{XY} := \dot{\Sigma}_X + \dot{\Sigma}_Y$ of the total system.
To keep the system's time scale finite, we impose an upper bound on the total activity rate $\dot{A}_{XY}$.
According to \eref{min AXY rate}, this upper bound must be set to at least $\dot{\calL}_E$.
In this case, the minimum value of the total EP rate is given by \cite{dechant2022minimum}
\begin{align}
    \label{min EP rate discrete}
    \min_{\substack{R:\ptt\rightarrow\pdd\\ \dot{A}_{XY} \textrm{ fixed}}}\dot{\Sigma}_{XY}=2\dot{\calL}_{E}\tanh^{-1}\left(\frac{\dot{\calL}_{E}}{\dot{A}_{XY}}\right).
\end{align}

\begin{figure}
    \centering
    \includegraphics[width=0.55\linewidth]{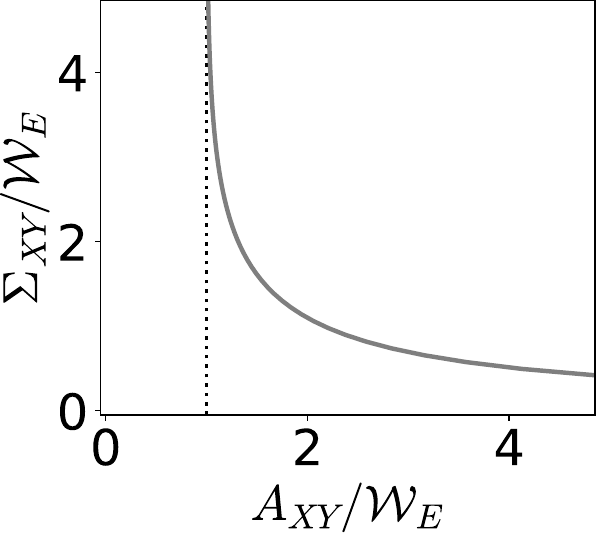}
    \caption{\label{fig: A-EP}
    The tradeoff relation between activity and EP (\eref{min EP}).
    }
\end{figure}

Next, we consider minimizing the EP over the finite-time evolution $\{\ptt\}_{0 \leq t \leq \tau}$.
Let $\calL_E$ be the length of this trajectory measured by the Wasserstein distance.
To keep the system's time scale finite, we impose an upper bound on the total activity $A_{XY}$.
In this case, the minimum value of the EP is given by $2\calL_{E}\tanh^{-1}({\calL_{E}}/{A_{XY}})$.

Furthermore, we optimize the trajectory to minimize the EP in the finite-time evolution $\poo\rightarrow\pff$.
Considering that $x\tanh^{-1}(x)\ (x>0)$ is monotonically increasing, it is clear that when optimizing the trajectory, one should select the trajectory that provides the geodesic distance $\calW_E(\pff, \poo)$.
Therefore, the minimum value of the EP is given by
\begin{align}
    \label{min EP}
    \min_{\substack{\{R\}_{0\leq t\leq \tau}:\poo\rightarrow\pff\\ {A}_{XY} \textrm{ fixed}}}{\Sigma}_{XY}=2{\calW}_{E}\tanh^{-1}\left(\frac{\calW_{E}}{{A}_{XY}}\right),
\end{align}
where $\calW_{E}=\calW_{E}(\pff,\poo)$, and the upper bound of the activity must be chosen to satisfy $A_{XY}\geq\calW_{E}(\pff,\poo)$.
Similar to the case of minimizing the activity, the trajectory that provides the minimum EP is a geodesic in the sense of the Wasserstein distance (see the optimal protocol in Supplemental Material).
However, unlike the continuous case \cite{Nakazato-Ito2021}, it is not necessary to maintain a constant evolution speed $\dot{\calL}_E$.

As is clear from \eref{min EP} and \fref{fig: A-EP}, there is a tradeoff relation between activity and EP, which can be expressed solely in terms of the thermodynamic force $F=2\tanh^{-1}({\calW}_{E}/{A}_{XY})$ (see Supplemental Material for details).
When $F \rightarrow 0$, the transport becomes fully bidirectional, corresponding to the quasistatic limit.
In this case, the dissipation asymptotically approaches zero, whereas the activity diverges.
Conversely, when $F \rightarrow \infty$, the transport becomes completely unidirectional, corresponding to the strong nonequilibrium limit.
In this limit, the dissipation logarithmically diverges, and the activity asymptotically approaches its minimum value ${\calW}_{E}$.

\subsubsection{Thermodynamic speed limit}
Equality \eqref{min EP} can also be interpreted as a speed limit for the time evolution of the total system.
Let us define the average activity rate as $\bar{A}_{XY}:=A_{XY}/\tau$, which provides a scale for the inverse of the relaxation time.
Furthermore, it is possible to construct a protocol that realizes \eref{min EP} and satisfies $\bar{A}_{XY}=\dot{A}_{XY}$.
The operation time $\tau$ must then satisfy
\begin{align}
    \label{speed limit total discrete}
    \tau\geq\tau_{XY}:=\frac{\calW_E}{\bar{A}_{XY}}\coth\left(\frac{\Sigma_{XY}}{2\calW_E}\right).
\end{align}
Therefore, to achieve the desired time evolution in a shorter duration, a higher thermodynamic cost must be paid.
For processes with large dissipation, where $\Sigma_{XY}\gg\calW_E$ and $\coth(x)\simeq 1$, the activity rate $\bar{A}_{XY}$ dominantly determines the operation time \cite{Van-Hasegawa2022unifiedTKUR}.

By using the inequality $\coth(x)\geq 1/x\ (x>0)$ and $\calW_E(\pff,\poo)\geq\calT(\pff,\poo)$ in \eref{speed limit total discrete}, the speed limit derived by \ccite{Shiraishi-Funo-Saito2018speed} can be reproduced.
However, since $\calW_E$ reflects the structure of the graph, \eref{speed limit total discrete} is tighter in general, and there exists a protocol that achieves the equality for any distributions $\poo$ and $\pff$.
This tightness becomes more pronounced as the system size increases \cite{Van-Saito2023topological}.

\begin{figure*}
    \centering
    \includegraphics[width=0.9\linewidth]{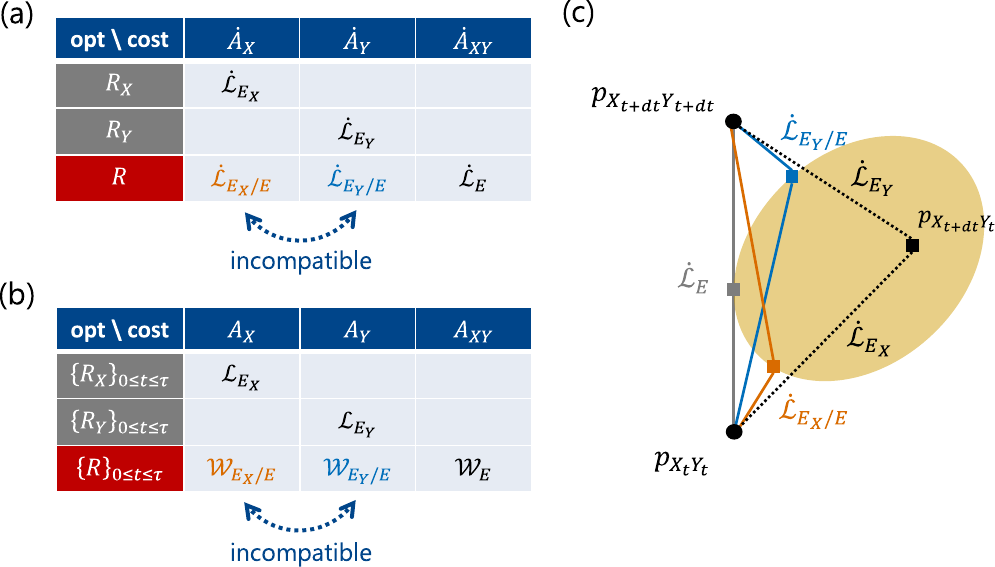}
    \caption{\label{fig: tradeoff_within}
    (a)(b) Differences in the optimization target and method for infinitesimal time evolution $\ptt \rightarrow \pdd$ and finite time evolution $\poo \rightarrow \pff$.
    The gray cells represent local optimization, optimizing only the transition rates related to subsystem transitions while keeping the partial time evolution $\pdt$ fixed.
    The red cells represent global optimization, where the entire transition rates are optimized.
    In this case, the partial time evolution is also optimized.
    In the local optimization for finite time evolution, the overall time evolution $\{\ptt\}_{0 \leq t \leq \tau}$ is not modified.
    Therefore, the minimum is not $\calW$, which depends only on the initial and final distributions, but $\calL$, which depends on the trajectory of the distribution.
    The same applies when minimizing the partial EP (rate).
    (c) Differences in partial time evolution for infinitesimal time evolution $\ptt \rightarrow \pdd$.
    This corresponds to the infinitesimal time version of the central schematic diagram in \fref{fig: summary}.
    The shaded area schematically represents the range of $\pdt$, specifically the range where $\dot{\calL}_{E_X} < \infty$ and $\dot{\calL}_{E_Y} < \infty$.
    The gray line represents the optimal transport of the total system (\eref{min AXY rate}), corresponding to the geodesic under the Wasserstein distance.
    For any $\pdt$, the inequality $\dot{\calL}_{E_X} + \dot{\calL}_{E_Y} \geq \dot{\calL}_{E}$ holds (see Supplemental Material for details).
    }
\end{figure*}

\section{Minimal thermodynamic costs of a single subsystem}
\label{sec: opt for single subsystem}
In this section, we consider the extent to which the activity and EP of subsystems can be minimized in finite-time processes.
Here, we optimize the costs of subsystems using two methods with different constraints.

First, we define partial time evolution.
For simplicity, we consider minimization in the case of infinitesimal time evolution $\ptt \rightarrow \pdd$.
In \eref{master eq}, we separate the probability flux for each subsystem and consider the partial time evolution resulting from it:
\begin{align}
    \label{partial evolution: X}
    \pdt(r)&=\ptt(r)+\sum_{x'(\neq x)}J_{xx'}^{y}dt,\\
    \label{partial evolution: Y}
    \ptd(r)&=\ptt(r)+\sum_{y'(\neq y)}J_{x}^{yy'}dt.
\end{align}
Here, $\pdt$ and $\ptd$ are joint distributions shifted by $dt$ in time, capturing the individual time evolution of $X$ and $Y$, respectively.
In the following, quantities of order $o(dt)$ will be neglected.
Given the transition rates of \eref{master eq}, the partial time evolutions (\esref{partial evolution: X}\eqref{partial evolution: Y}) are uniquely determined.
However, when only the infinitesimal time evolution $\ptt \rightarrow \pdd$ of the total system is provided, there is freedom in choosing the partial time evolution.
In local optimization, a single partial time evolution is fixed, whereas in global optimization, the partial time evolution is also optimized (see \fref{fig: tradeoff_within}(a)(b)).

\subsection{Local Optimization}
\label{subsec: Local optimization}
In this subsection, we discuss local optimization, where the transition rates on $E_X$ are optimized to minimize the thermodynamic cost of $X$ (see \fref{fig: tradeoff_within}(a)(b)).
In this approach, the partial time evolutions (see \esref{partial evolution: X}\eqref{partial evolution: Y}) are fixed, and the optimization within each subsystem is independent.
Local optimization is crucial for understanding the subsequent global optimization, and it also explains the origin of the tradeoff relation between thermodynamic costs through its procedure.

First, we consider minimization in the case of infinitesimal time evolution $\ptt \rightarrow \pdd$.
The initial distribution $\ptt$ and final distribution $\pdd$ are fixed.
In local optimization, the partial time evolution $\ptt \rightarrow \pdt$ is also fixed, which in turn determines $\ptd$.
The transitions of $X$ occur on $E_X$ and contribute solely to the time evolution $\ptt \rightarrow \pdt$, meaning that the transitions of $Y$ are not involved in this time evolution.
Therefore, the approach of \sref{subsec: review discrete total} can be applied to the subgraph $G(N,E_X)$ and the time evolution $\ptt \rightarrow \pdt$.
The local minimum of the partial activity rate for $X$ is
\begin{align}
    \label{min AX rate local}
    \min_{R_X:\ptt\rightarrow\pdt}\dot{A}_X
    &=\frac{\calW_{E_X}(\pdt,\ptt)}{dt}\nonumber\\
    &=:\dot{\calL}_{E_X}.
\end{align}
Here, $R_X$ denotes the transition rates on the subset $E_X$ of $E$, satisfying \eref{partial evolution: X}.
However, if there is no path between nodes $r$ and $r'$ that is entirely within the edges of $E_X$, we define $d_{E_X}(r,r') = \infty$.

Next, we locally optimize the partial EP rate.
To keep the time scale of the subsystems finite, we set an upper bound $\dot{A}_X (\geq \dot{\calL}_{E_X})$ on the partial activity rate.
Similar to the case of the partial activity, we can apply the approach from \sref{subsec: review discrete total}.
As a result, the local minimum of the partial EP rate for $X$ is
\begin{align}
    \label{min EP_X rate local}
    \min_{\substack{R_X:\ptt\rightarrow\pdt\\ \dot{A}_{X} \textrm{ fixed}}}\dot{\Sigma}_{X}=2\dot{\calL}_{E_X}\tanh^{-1}\left(\frac{\dot{\calL}_{E_X}}{\dot{A}_{X}}\right).
\end{align}

Extending to the finite time evolution $\poo \rightarrow \pff$ is straightforward.
In this case, the trajectory of the probability distribution $\{\ptt\}_{0\leq t\leq \tau}$ is fixed.
Furthermore, by fixing the partial time evolution at each moment, local optimization becomes possible.
The minimum partial activity of $X$ is given by $\calL_{E_X}:=\int_0^\tau dt\dot{\calL}_{E_X}$.
Moreover, by fixing the upper bound of the partial activity $A_X(\geq\calL_{E_X})$, the minimum partial EP is given by $2\calL_{E_X}\tanh^{-1}(\calL_{E_X}/A_X)$.

In local optimization, the partial time evolutions $\pdt$ and $\ptd$ are fixed.
Since information flow is expressed using $\pdt$ and $\ptd$, this can be considered as fixing the information flow (see \esref{I_X rate discrete}\eqref{I_Y rate discrete}).

The optimization given in \esref{min AX rate local}\eqref{min EP_X rate local} applies similarly to the subsystem $Y$.
Since $R_X$ and $R_Y$ can be optimized independently, the local minima of the thermodynamic costs for the subsystems can be achieved simultaneously.
However, for the overall time evolution $\ptt \rightarrow \pdd$ to be possible, the time evolution $\ptt \rightarrow \pdt$ must be realized solely by $R_X$, and the time evolution $\pdt \rightarrow \pdd$ (which is the same as $\ptt \rightarrow \ptd$) must be realized solely by $R_Y$.
In other words, $\pdt$ (or $\ptd$) must be chosen such that $\dot{\calL}_{E_X} < \infty$ and $\dot{\calL}_{E_Y} < \infty$ are satisfied (see \fref{fig: tradeoff_within}(c)).

The transition rates that achieve the minimum value in \eref{min AX rate local} can be expressed using the optimal transport matrix concerning $d_{E_X}$ within $\calU(\pdt,\ptt)$.
Similarly, the transition rates $R_X$ that achieve the minimum value in \eref{min EP_X rate local} can be expressed using this optimal transport matrix and the thermodynamic force on $E_X$, $F_X = 2 \tanh^{-1}(\dot{\calL}_{E_X}/\dot{A}_{X})$ (see Supplemental Material for details).
In the case of finite time evolution, $F_X$ needs to be kept constant at each time step.
From the transport perspective, the thermodynamic force $F_X$ characterizes the unidirectionality of transport on the edges $E_X$.
There is a tradeoff relation between the activity rate and the EP rate for $X$, which can be represented solely by the parameter $F_X$.


\subsection{Global Optimization}
\label{subsec: Global optimization}
In this subsection, we discuss global optimization, where the overall transition rates are optimized to minimize the thermodynamic cost of $X$.
In this approach, the previously fixed partial time evolutions are also optimized.
This allows for further minimization of the costs of subsystem $X$ beyond what local optimization (\esref{min AX rate local}\eqref{min EP_X rate local}) can achieve.
To this end, we introduce a generalization of the Wasserstein distance.
Unlike local optimization, this approach does not treat the optimization within each subsystem independently.
Physically, this can be interpreted as optimizing the cost of $X$ under the condition that $X$ and $Y$ can interact in any manner.
The same procedure can be applied to optimize $Y$ as well, while we focus only on X in the following.

\subsubsection{Minimal partial activity}
\label{subsec: Minimal partial activity}
First, consider minimizing the partial activity rate in the case of infinitesimal time evolution $\ptt \rightarrow \pdd$.
The initial distribution $\ptt$ and final distribution $\pdd$ are fixed.
In \sref{subsec: Local optimization}, we performed local optimization by fixing the partial time evolution $\ptt \rightarrow \pdt$.
In this subsection, we perform global optimization by also allowing $\pdt$ to vary.
In this optimization, under the condition that the entire transition rates realize the time evolution $\ptt \rightarrow \pdd$, we aim to minimize the transitions carried by the $X$ component of the transition rates as much as possible.
Specifically, we select $\pdt$ such that $\dot{\calL}_{E_Y} < \infty$ and $\dot{\calL}_{E_X}$ is minimized.

To find the partial time evolution that minimizes the transitions on $E_X$, we consider the cost function $d_{E_X/E}$ defined as
\begin{align}
\label{def of d_EX/E}
    d_{E_X/E}(r,r'):=\min_{\substack{P:r'\rightarrow r\\ \textrm{on }E}} l_{E_X}(P).
\end{align}
Although the cost function $d_E$ (\eref{def of d_E}) counts all transports on $E$, $d_{E_X/E}$ counts only the transports on $E_X$ (see \fref{fig: cost_func}(b)).
The optimal transport cost concerning this cost function $d_{E_X/E}$ is expressed as
\begin{align}
    \label{Ex Wasserstein}
    \calW_{E_X/E}(p,q):=\min_{\Pi\in\calU(p,q)}\sum_{r,r'}d_{E_X/E}(r,r')\Pi(r,r'),
\end{align}
which is regarded as a generalized Wasserstein distance as explained below.
The quantity $\calW_{E_X/E}(p,q)$ represents the minimum transport cost related to $E_X$ for transporting the probability distribution $q$ to $p$.
Intuitively, in the optimal transport matrix concerning $d_{E_X/E}$, all transport that can be performed on $E_Y$ is carried out on $E_Y$.

Similar to $\calW_{E}$ and $\calW_{E_X}$, $\calW_{E_X/E}$ satisfies the properties of symmetry and the triangle inequality, that is, $\calW_{E_X/E}(p,q)=\calW_{E_X/E}(q,p)$ and $\calW_{E_X/E}(p,r)\leq\calW_{E_X/E}(p,q)+\calW_{E_X/E}(q,r)$.
The triangle inequality allows this optimization to be extended to the time-integrated case.
However, since $\calW_{E_X/E}(p,q) = 0$ when the transport can be performed solely on $E_Y$, $\calW_{E_X/E}$ is a pseudo-distance rather than a true distance.
Here, we refer to $\calW_{E_X/E}$ as the Wasserstein pseudo-distance.
The same applies to $\calW_{E_Y/E}$.

Using this Wasserstein pseudo-distance, the minimum partial activity rate of $X$ for the infinitesimal time evolution $\ptt \rightarrow \pdd$ can be expressed as:
\begin{align}
    \label{min AX rate global}
    \min_{R:\ptt\rightarrow\pdd}\dot{A}_X
    &=\frac{\calW_{E_X/E}(\pdd,\ptt)}{dt}\nonumber\\
    &=:\dot{\calL}_{E_X/E}.
\end{align}
By using \eref{Ex Wasserstein}, we perform both the local optimization and the optimization of partial time evolution.
The condition $\Pi \in \calU(\pdd, \ptt)$ during the optimization ensures that $\dot{\calL}_{E_X} < \infty$ and $\dot{\calL}_{E_Y} < \infty$ are satisfied.
By using the cost function $d_{E_X/E}$, we minimize the partial activity rate of $X$.
In local optimization (\eref{min AX rate local}), we optimized the $X$ component of the transition rates, $R_X$.
By contrast, in global optimization (\eref{min AX rate global}), we optimize the entire transition rates, $R$. 
The proof of \eref{min AX rate global} is provided in Supplemental Material.


Next, we consider minimizing the partial activity for the finite time evolution $\poo \rightarrow \pff$.
Only the initial distribution $\poo$ and the final distribution $\pff$ are fixed, and the trajectory of the probability distributions corresponding to the intermediate time evolution is also optimized.
Since $\calW_{E_X/E}$ satisfies the triangle inequality, similar to \eref{min AXY}, the minimum partial activity is given by:
\begin{align}
    \label{min AX global}
    \min_{\{R\}_{0\leq t\leq \tau}:\poo\rightarrow\pff}{A}_{X}
    =\calW_{E_X/E}(\pff,\poo).
\end{align}
In this case, the trajectory forms a geodesic under the Wasserstein pseudo-distance, which generally differs from the geodesic under the standard Wasserstein distance.
However, the linear interpolation between the initial and final distributions $\ptt = (1 - t/\tau)\poo + (t/\tau)\pff$ can be used as a common geodesic.

The optimization in \eref{min AX rate global} can also be performed similarly for the subsystem $Y$.
However, it should be noted that, in general, the transition rates $R$ (and the resulting partial time evolutions $\pdt$ and $\ptd$) that minimize the partial activity rates for $X$ and $Y$ may be different.
This difference arises because the optimal transport matrices concerning $d_{E_X/E}$ and $d_{E_Y/E}$ within $\calU(\pdd, \ptt)$ are different.
Therefore, in general, it is not possible to simultaneously achieve the globally minimal partial activity rates $\dot{\calL}_{E_X/E}$ and $\dot{\calL}_{E_Y/E}$ for both $X$ and $Y$ (see \fref{fig: tradeoff_within}(c)).
This implies a tradeoff relation where reducing the activity rate of one subsystem will increase the activity rate of the other subsystem.
Naturally, the global minimum of the total activity rate $\dot{\calL}_{E}$ is given by the optimal transport of the total system.
The same applies to the time-integrated case.
The bounds of the tradeoff relation in partial activities will be discussed in the next section.

The transition rates that achieve \eref{min AX rate global} can be constructed similarly to those in \eref{min AXY rate}.
However, when converting the optimal transport matrix into transition rates, it is necessary to select paths that minimize the use of edges in $E_X$ as much as possible (see Supplemental Material for details).

\subsubsection{Minimal partial entropy production}
\label{subsubsec: min EP_X EP_Y}
First, we consider minimizing the partial EP rate in the overall infinitesimal time evolution $\ptt \rightarrow \pdd$.
To keep the timescale of the subsystem $X$ finite, we set an upper bound $\dot{A}_X (\geq \dot{\calL}_{E_X/E})$ on the activity rate of $X$.
In the global optimization of the partial EP rate, in addition to the local optimization in the previous section (\eref{min EP_X rate local}), the partial time evolution is also optimized.
This optimal time evolution is determined by the minimization of the partial activity rate (\eref{min AX rate global}).
Therefore, the global minimum partial EP rate for $X$ is given by
\begin{align}
    \label{min EP_X rate global}
    \min_{\substack{R:\ptt\rightarrow\pdd\\ \dot{A}_{X} \textrm{ fixed}}}
    \dot{\Sigma}_X=2\dot{\calL}_{E_X/E}\tanh^{-1}\left(\frac{\dot{\calL}_{E_X/E}}{\dot{A}_{X}}\right).
\end{align}


Next, we minimize the partial EP for finite time evolution $\poo \rightarrow \pff$.
To keep the timescale of the subsystem $X$ finite, we set an upper bound $A_X (\geq \calW_{E_X/E})$ on the partial activity.
As in the case of the total system (\eref{min EP}), we optimize the trajectory of the time evolution.
In global optimization, we further optimize the partial time evolution within this trajectory.
This optimal time evolution is determined by the minimization of the partial activity (\eref{min AX global}).
Therefore, the global minimum of the partial EP for finite time evolution is given by
\begin{align}
    \label{min EP_X global}
    \min_{\substack{\{R\}_{0\leq t\leq \tau}:\poo\rightarrow\pff\\ {A}_{X} \textrm{ fixed}}}
    \Sigma_X=2{\calW}_{E_X/E}\tanh^{-1}\left(\frac{{\calW}_{E_X/E}}{{A}_{X}}\right).
\end{align}


The optimization in \esref{min EP_X rate global}\eqref{min EP_X global} can also be performed similarly for the subsystem $Y$.
However, similarly to the partial activity, it is in general not possible to simultaneously achieve the global minimum of the partial EP for both $X$ and $Y$.
This is because the optimality of the partial evolution, which needs to be specified to calculate thermodynamic costs of $X$ and $Y$, is different for $X$ and $Y$ (see \fref{fig: tradeoff_within}(c)).
Therefore, there exists a tradeoff relation where reducing the EP of one subsystem increases the EP of the other subsystem.

Similar to the case of \eref{min EP_X rate local}, the transition rates $R$ that achieve the minimum value in \eref{min EP_X rate global} can be constructed.
However, since the thermodynamic cost of $Y$ is not the optimization target, the thermodynamic force $F_Y$ on $E_Y$ can be chosen arbitrarily.
The same applies to the time-integrated case.
We note that the entropy productions of subsystems can be lower-bounded through marginalization (see Supplemental Material for details).

\subsubsection{Thermodynamic speed limit}
\label{subsubsec: speed limit}
Equality \eqref{min EP_X global} and its counterpart for $Y$ can also be interpreted as the speed limits for the time evolution.
Let us define the average partial activity rates as $\bar{A}_{X}:=A_{X}/\tau$ and $\bar{A}_{Y}:=A_{Y}/\tau$.
This gives the scale of the inverse of the relaxation time for each subsystem.
Combined with \eref{speed limit total discrete}, the operation time $\tau$ satisfies the following speed limit:
\begin{align}
    \label{speed limit all discrete}
    \tau&\geq\max\{\tau_X,\tau_Y,\tau_{XY}\},\\
    \label{speed limit X discrete}
    &\tau_X:=\frac{\calW_{E_X/E}}{\bar{A}_{X}}\coth\left(\frac{\Sigma_{X}}{2\calW_{E_X/E}}\right),\\
    \label{speed limit Y discrete}
    &\tau_Y:=\frac{\calW_{E_Y/E}}{\bar{A}_{Y}}\coth\left(\frac{\Sigma_{Y}}{2\calW_{E_Y/E}}\right).
\end{align}

These speed limits can be interpreted as follows.
We want to evolve the system from the initial distribution $\poo$ to the final distribution $\pff$ over the graph $G(N,E=E_X\sqcup E_Y)$ by manipulating the transition rates.
During the evolution, we are allowed to dissipate in each subsystem up to $\Sigma_X$ and $\Sigma_Y$, respectively.
The operation speed of each subsystem is finite, and the partial activity rates cannot exceed $\bar{A}_X$ and $\bar{A}_Y$, respectively.
These parameters can be set arbitrarily.
The speed limit \eqref{speed limit all discrete} indicates that we need an operation time of at least $\max\{\tau_X,\tau_Y,\tau_{XY}\}$.
Therefore, to achieve the desired time evolution in a shorter period, it is necessary to increase the dissipation for both $X$ and $Y$.


\subsubsection{Several remarks}
\label{subsubsec: remarks for global}
The protocol that provides the global minimum of the partial activity and partial EP can be expressed using the optimal transport matrix concerning $d_{E_X/E}$ or $d_{E_Y/E}$ within $\calU(\pff,\poo)$.
However, in general, such ($L^1$) optimal transport matrices are not uniquely determined \cite{dechant2022minimum,sudakov1979geometric}.
This is because there may be multiple paths on $G(N,E)$ that give the same optimal transport cost.
Moreover, even if one selects an optimal path, there is freedom in how to perform the transport within that path.
These generate different trajectories that yield distinct geodesics.
For any trajectory, the partial activity and partial EP remain unchanged if the same thermodynamic force is applied, but the time-integrated information flow may vary.
This point is discussed in \sref{sec: Optimal information processing}.
Furthermore, there is also freedom in determining the time dependency of the evolution speeds $\dot{\calL}_{E_X}$ and $\dot{\calL}_{E_Y}$.
Partial activity, partial EP, and the time-integrated information flow do not depend on this speed.
In contrast, for continuous systems, this speed needs to be constant to minimize the dissipation \cite{Nakazato-Ito2021}.


\section{Tradeoff relations in the thermodynamic costs of subsystems}
\label{sec: Tradeoff relations in the thermodynamic costs of subsystems}

In this section, we consider the framework to identify the tradeoff relation between the thermodynamic costs of $X$ and $Y$ in finite-time processes.
Hereafter, when the initial and final distributions are clear from the context, we will omit these variables of the Wasserstein distance.

\subsection{Pareto Front}
\label{subsec: Pareto front}
\begin{figure}
    \centering
    \includegraphics[width=0.85\linewidth]{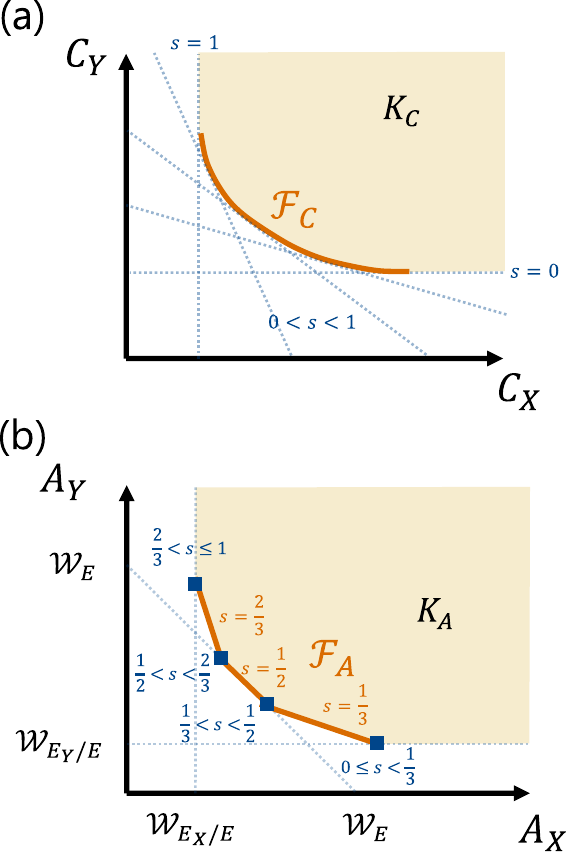}
    \caption{\label{fig: activity_Pareto}
    (a) A schematic diagram of the Pareto front $\calF_C$ for incompatible cost functions $C_X$ and $C_Y$.
    The yellow-shaded area represents the feasible region $K_C$ for $(C_X, C_Y)$. To obtain the Pareto front $\calF_C$, solve \eref{min CX and CY convex} for $0 \leq s \leq 1$.
    Specifically, for each $s$, find the tangent line to $K_C$ with a slope of $-s/(1-s)$ (blue dashed line in the figure).
    (b) A typical example of the Pareto front $\calF_A$ of partial activities in the finite-time evolution $\poo \rightarrow \pff$.
    This figure illustrates that $\calF_A$ consists of vertices (blue squares) and edges (orange lines).
    For instance, the vertex for $1/2 < s < 2/3$ indicates that $sA_X + (1-s)A_Y$ is minimized at this point for values of $s$ in this range.
    Conversely, the edge for $s = 2/3$ indicates that $sA_X + (1-s)A_Y$ is minimized along this edge for this value of $s$.
    }
\end{figure}

As a tool to express the tradeoff relations between conflicting cost functions, we adopt the Pareto front \cite{ngatchou2005pareto,coello2007evolutionary}.
Figure \ref{fig: activity_Pareto}(a) represents a general schematics of the Pareto front.
The feasible region $K_C (\neq \emptyset)$ consists of all pairs of feasible cost functions $C = (C_X, C_Y)$.
The Pareto front $\calF_C$ is the set of all optimal pairs within $K_C$ (see the precise definition in Supplemental Material).
In other words, the Pareto front corresponds to finding the minimum $C_X$ for a fixed value of $C_Y$.
We note that the Pareto front has been applied in thermodynamics to characterize the tradeoff relations in thermodynamic machines, such as those between power and efficiency, or dissipation and precision \cite{Benenti-2017Fundamental,solon-holowitz2018,Ashida-Sagawa2021}.
We here introduce the concept of the Pareto front for thermodynamic costs of subsystems $C$ and provide a general method for constructing it.

The explicit method to construct the Pareto front $\calF_C$ is explained as follows \cite{marler2010weightedsum}.
Take $0 \leq s \leq 1$ and perform the following procedure (see also \fref{fig: activity_Pareto}(a)):
First, among all protocols $\{R\}_{0 \leq t \leq \tau}$ that realize the time evolution $\poo \rightarrow \pff$, minimize the weighted cost function $sC_X + (1-s)C_Y$, i.e., solve
\begin{align}
    \label{min CX and CY convex}
    \min_{\{R\}_{0\leq t\leq \tau}:\poo\rightarrow\pff}[sC_X+(1-s)C_Y].
\end{align}
Then, plot all points $(C_X, C_Y)$ corresponding to the protocols that give the minimum value.
After sweeping  $s$ over the range $0 \leq s \leq 1$, the set of optimal points obtained by this method forms the Pareto front if $K_C$ is a convex set \cite{marler2010weightedsum}. 

\subsection{Pareto Front of the Partial Activities}
\label{subsec: partial activity Pareto}

First, we consider the Pareto front $\calF_A$ of partial activities $A_X, A_Y$.
It can be shown that the feasible region for partial activity, $K_A$, is convex (see Supplemental Material for proof). 
Furthermore, such optimization can be reduced to optimal transport theory in the case of partial activity, which will be discussed in the following.
With $C=A$ and $s=1$, \eref{min CX and CY convex} corresponds to the global minimization of the activity of $X$, and its minimum value is given by \eref{min AX global}.
To find this minimum value, we introduced the cost function $d_{E_X/E}$ of the optimal transport problem.
Similarly, by designing the cost function appropriately, the minimum value of \eref{min CX and CY convex} for other values of $s$ can also be expressed using the Wasserstein distance.

To find the minimum value of \eref{min CX and CY convex}, we introduce the following cost function.
Define $d_{E_X:E_Y}^s\ (0\leq s\leq1)$ as 
\begin{align}
\label{def of d_EXEY}
    d_{E_X:E_Y}^s(r,r'):=\min_{\substack{P:r'\rightarrow r\\ \textrm{on }E}} [sl_{E_X}(P)+(1-s)l_{E_Y}(P)].
\end{align}
The optimal transport cost regarding this cost function is then given by
\begin{align}
    \label{s discrete Wasserstein}
    \calW_{E_X:E_Y}^s(p,q):=\min_{\Pi\in\calU(p,q)}\sum_{r,r'}d_{E_X:E_Y}^s(r,r')\Pi(r,r').
\end{align}
Although $\calW_{E_X/E}$ represents the minimization of the transport cost related to $E_X$, $\calW_{E_X:E_Y}^s$ represents the minimization of a convex combination of the transport costs for $E_X$ and $E_Y$.
Similar to $\calW_{E_X/E}$, $\calW_{E_X:E_Y}^s$ satisfies symmetry and the triangle inequality.
Moreover, for $0 < s < 1$, non-degeneracy holds, making $\calW_{E_X:E_Y}^s$ a distance.

Following a similar argument as in \eref{min AX global}, the minimum value of \eref{min CX and CY convex} for $C=A$ is given by $\calW_{E_X:E_Y}^s(\pff,\poo)$ (see Supplemental Material for proof).
In general, the optimal transport problem \eqref{s discrete Wasserstein} is a linear programming problem, hence this minimum value can be efficiently determined.
By plotting the optimal pairs $(A_X, A_Y)$ based on \eref{min CX and CY convex}, we can obtain the Pareto front $\calF_A$ of partial activities.
However, when converting the optimal transport matrix to have values only on $E$, it is necessary to select the path that gives the cost function $d_{E_X:E_Y}^s$ (see Supplemental Material for details).

\begin{figure*}
    \centering
    \includegraphics[width=0.95\linewidth]{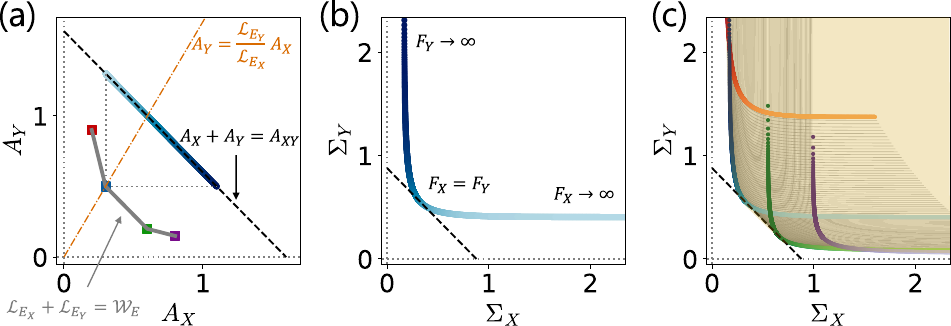}
    \caption{\label{fig: EP_Pareto_general}
    (a) An illustrative example of the Pareto front $\calF_A$ of partial activities.
    $\calF_A$ consists of square vertices and gray edges.
    The vertices, ordered by increasing $A_X$, are $(0.2, 0.9), (0.3, 0.5), (0.6, 0.2), (0.8, 0.15)$.
    (b) For the time evolution corresponding to the blue square in (a), \eref{partial EP tradeoff expression} is plotted.
    The partial activities realized for these EPs correspond to the blue line in (a).
    The black dashed line indicates the minimum total EP.
    When the total EP is minimized, $\calL_{E_X} + \calL_{E_Y} = \calW_E$, $A_X + A_Y = A_{XY}$, and $A_Y = (\calL_{E_Y}/\calL_{E_X})A_X$ are satisfied.
    The partial activity takes the value at the intersection of the black and orange dashed lines in (a).
    (c) The corresponding Pareto front $\calF_\Sigma$ of partial EPs.
    We can obtain $\calF_\Sigma$ by plotting \eref{partial EP tradeoff expression} for all partial time evolutions $(\calL_{E_X}, \calL_{E_Y})$ that constitute $\calF_A$ in (a).
    Here, $\calW_E=0.8$ and $A_{XY}=1.6$.
    }
\end{figure*}

In general, $\calF_A$ forms a convex polyline consisting of a finite number of vertices and edges (see \fref{fig: activity_Pareto}(b)).
Each vertex (blue square) indicates that within the corresponding range of $s$, the pair $(A_X,A_Y)$ that minimizes $sA_X + (1-s)A_Y$ remains the same.
Specifically, the vertices for $s=1$ and $s=0$ give the minimum partial activities $\calW_{E_X/E}(\pff, \poo)$ and $\calW_{E_Y/E}(\pff, \poo)$, respectively.
On the other hand, each edge (orange line segment) indicates that for a specific value of $s$, there are multiple pairs $(A_X,A_Y)$ that minimize $sA_X + (1-s)A_Y$, distributed linearly.
As a result, $\calF_A$ has an edge with a slope of $-s/(1-s)$, showing a singular tradeoff behavior.
Remember that to realize time evolution within a finite operation time $\tau$, the system needs to be controlled at a certain minimum average speed.
On the edges of $\calF_A$, this required operational speed can be adjusted between subsystems.
In particular, the edge with a slope of $-1$ corresponding to $s=1/2$ represents the minimum total activity.
In this case, \eref{min CX and CY convex} reduces to the minimization of half the total activity of the system (see \eref{min AXY}).
This behavior can be derived from the general theory of linear programming (see Supplemental Material for details).

The Pareto front $\calF_A$ of partial activities depends on both the initial and final distributions, but its shape is significantly influenced by the structure of the graph $G(N,E)$.
This will be illustrated with a simple example in \sref{subsection: Pareto example}.

\subsection{Pareto Front of the Partial EPs}
\label{subsec: partial EP Pareto}
We next consider the Pareto front $\calF_\Sigma$ of partial EPs $\Sigma_X,\Sigma_Y$.
To maintain a finite timescale for the system, we set an upper bound on the total activity, ${A}_{XY} (\geq {\mathcal{W}}_{XY})$.
In the context of information processing, this can be seen as imposing a limit on the number of processing steps.
We can obtain $\calF_\Sigma$ by solving \eref{min CX and CY convex} for $C=\Sigma$ under this activity constraint.
In this problem, it is necessary to consider the compatibility of the partial EPs for $X$ and $Y$.
This can be achieved by performing local optimization of the partial EPs (see \sref{subsec: Local optimization}) in each subsystem while ensuring that the activity bound is satisfied:
\begin{align}
    \label{partial EP tradeoff expression}
    \begin{cases}
    {\Sigma}_X=\calL_{E_X}F_X \\
    {\Sigma}_Y=\calL_{E_Y}F_Y \\
    {A}_{XY}=\calL_{E_X}\coth(F_X/2)
    +\calL_{E_Y}\coth(F_Y/2)\\
    \end{cases}.
\end{align}
Here, $\calL_{E_X}$ and $\calL_{E_Y}$ are the local minima of the partial activities when the trajectory of the time evolution $\{\ptt\}_{0 \leq t \leq \tau}$ and the partial time evolution $\{\pdt\}_{0 \leq t \leq \tau}$ are fixed.
On the other hand, $F_X$ and $F_Y$ ($\geq 0$) are the thermodynamic forces when the partial EPs are locally minimized.
In this minimization, the partial activities are given by $\calL_{E_X}\coth(F_X/2)$ and $\calL_{E_Y}\coth(F_Y/2)$, with the thermodynamic forces set to satisfy the activity constraints (see \eref{partial EP tradeoff expression}).

To solve \eref{min CX and CY convex}, the remaining task is to optimize the time evolutions ($\calL_{E_X}$ and $\calL_{E_Y}$) and the thermodynamic forces ($F_X$ and $F_Y$).
However, for $\calL_{E_X}$ and $\calL_{E_Y}$, we only need to consider those on the Pareto front $\calF_A$ of partial activities.
This is because, for $\calL_{E_X}$ and $\calL_{E_Y}$ not on $\calF_A$, either $\Sigma_X$ or $\Sigma_Y$ will become larger (see \esref{min CX and CY convex}\eqref{partial EP tradeoff expression}).
In general, it is not easy to analytically determine the time evolution and thermodynamic forces that minimize \eref{min CX and CY convex}.
Nevertheless, we can obtain $\calF_\Sigma$ by sweeping the parameters as shown in \fref{fig: EP_Pareto_general}.
In \fref{fig: EP_Pareto_general}(a), the gray polyline represents $\calF_A$, and the blue square is one of its vertices.
For the time evolution corresponding to this vertex, the partial EPs are plotted by varying $F_X$ and $F_Y$ in \eref{partial EP tradeoff expression}, shown as the blue line in \fref{fig: EP_Pareto_general}(b).
The corresponding partial activities are represented by the blue line in \fref{fig: EP_Pareto_general}(a), and they satisfy the activity bound $A_{XY}$.
By performing this procedure for all time evolutions on $\calF_A$, we can obtain $\calF_\Sigma$ as shown in \fref{fig: EP_Pareto_general}(c).

\subsubsection{Special cases without parameter sweep}
\label{subsubsec: sweepless}
In the following cases, \eref{min CX and CY convex} can be solved without sweeping the parameters of time evolution and thermodynamic forces.

When $s=1/2$, for any $A_{XY}(\geq \calW_E)$, the optimal values of $\calL_{E_X}$, $\calL_{E_Y}$, $F_X$, and $F_Y$ can be determined as follows.
When $s=1/2$, \eref{min CX and CY convex} reduces to minimizing (half) the EP of the total system (see \eref{min EP}).
First, by fixing $\calL_{E_X}$ and $\calL_{E_Y}$ and optimizing $F_X$ and $F_Y$, we obtain $F_X=F_Y$.
At this point, the partial activity satisfies the relation $A_X/\calL_{E_X}=A_Y/\calL_{E_Y}$.
Then, optimizing the partial time evolution selects those that satisfy $\calL_{E_X} + \calL_{E_Y} = \calW_E$.
Using \fref{fig: EP_Pareto_general}, we can illustrate the above optimization.
The blue square in \fref{fig: EP_Pareto_general}(a) represents one of the time evolutions $(\calL_{E_X}, \calL_{E_Y})$ that yields $\calW_E$.
The blue curve in \fref{fig: EP_Pareto_general}(b) depicts the optimization of thermodynamic forces for this time evolution.
The total dissipation is minimized when $F_X=F_Y$, and this curve is tangent to the black dotted line.
The partial activity at this point satisfies both the orange dotted line $A_X/\calL_{E_X} = A_Y/\calL_{E_Y}$ and the black dotted line $A_X + A_Y = A_{XY}$ simultaneously.
It should be noted that the time evolution that yields $\calW_E$ is not necessarily unique.
When $\calF_A$ has an edge with a slope of $-1$ as in \fref{fig: EP_Pareto_general}(a), minimum dissipation can be achieved with different time evolutions, so $\calF_\Sigma$ also has an edge with the same slope, as shown in \fref{fig: EP_Pareto_general}(c).

Near equilibrium ($A_{XY} \gg \calW_E$), the optimization of \eref{min CX and CY convex} can be performed on \eref{partial EP tradeoff expression}.
This is equivalent to minimizing the following function:
\begin{align}
    &s\calL_{E_X}F_X+(1-s)\calL_{E_Y}F_Y
    \nonumber\\
    &\ +\lambda\left(\calL_{E_X}\coth(F_X/2)-\calL_{E_Y}\coth(F_Y/2)-A_{XY}\right)
    \nonumber\\
    &\simeq s\calL_{E_X}F_X+(1-s)\calL_{E_Y}F_Y\nonumber\\
    &\ +\lambda\left(\frac{2\calL_{E_X}}{F_X}+\frac{2\calL_{E_Y}}{F_Y}-A_{XY}\right),
\end{align}
where $\lambda (\geq 0)$ is the Lagrange multiplier.
Here, near equilibrium, assuming the optimized thermodynamic forces are $F_X \simeq 0$ and $F_Y \simeq 0$, we approximate $\coth(F_X/2) \simeq 2/F_X$ and $\coth(F_Y/2) \simeq 2/F_Y$.
Thus, $\lambda$ becomes $\lambda=2((\sqrt{s}\calL_{E_X}+\sqrt{1-s}\calL_{E_Y})/A_{XY})^2$, and the convex combination of the partial EPs after optimizing the thermodynamic forces is given by
\begin{align}
    \label{near equilibrium convex EP}
    s\Sigma_X+(1-s)\Sigma_Y\simeq
    \frac{2(\sqrt{s}\calL_{E_X}+\sqrt{1-s}\calL_{E_Y})^2}{A_{XY}}.
\end{align}
Therefore, to optimize the partial time evolutions $\calL_{E_X}$ and $\calL_{E_Y}$, one should minimize $\sqrt{s}\calL_{E_X} + \sqrt{1-s}\calL_{E_Y}$.
This is equivalent to the optimization problem of \eref{min CX and CY convex} with the coefficient $s$ replaced by $\sqrt{s}/(\sqrt{s}+\sqrt{1-s})$.
Hence, there is no need to sweep the parameter to find $\calF_\Sigma$.
Furthermore, when $\calF_A$ has an edge corresponding to the coefficient $\sqrt{s}/(\sqrt{s}+\sqrt{1-s})$, $\calF_\Sigma$ also has an edge corresponding to the coefficient $s$ (i.e., the slope is squared).

\subsubsection{Several remarks}
When $s=1$, \eref{min CX and CY convex} is analogous to the global minimization of the EP of $X$ (\eref{min EP_X global}), but it is $A_{XY}$ that is fixed, not $A_X$.
Thus, the time required for the transitions of $Y$ must also be considered.
For optimizing the thermodynamic forces, setting $F_Y \rightarrow \infty$ results in $F_X = 2\tanh^{-1}(\calL_{E_X}/(A_{XY} - \calL_{E_Y}))$.
However, the optimization of the partial time evolutions $\calL_{E_X}$ and $\calL_{E_Y}$ is nontrivial and depends on the shape of $\calF_A$.

In this section, when determining the Pareto front of partial EPs, an upper bound $A_{XY}(\geq \calW_E(\pff, \poo))$ was set for the total activity.
The quasistatic limit is obtained by taking ${A}_{XY} \rightarrow \infty$.
This constraint was imposed to maintain finite time scales for both subsystems, but there are other possible methods for setting such constraints.
For example, one could fix the upper bounds of the partial activities $A_X$ and $A_Y$ as in \eref{min EP_X global}.
It is necessary to examine whether these upper bounds can be simultaneously satisfied.
In this case, there still exists the tradeoff relation between $\Sigma_X$ and $\Sigma_Y$ which originates from the degree of freedom to choose the time evolution $(\calL_{E_X}, \calL_{E_Y})$ within the Pareto front of $A_X$ and $A_Y$.
As another constraint, \ccite{FujimotoIto2023} assumes the near-equilibrium condition and analytically derives the Nash equilibrium solution.

\begin{figure}
    \centering
    \includegraphics[width=0.95\linewidth]{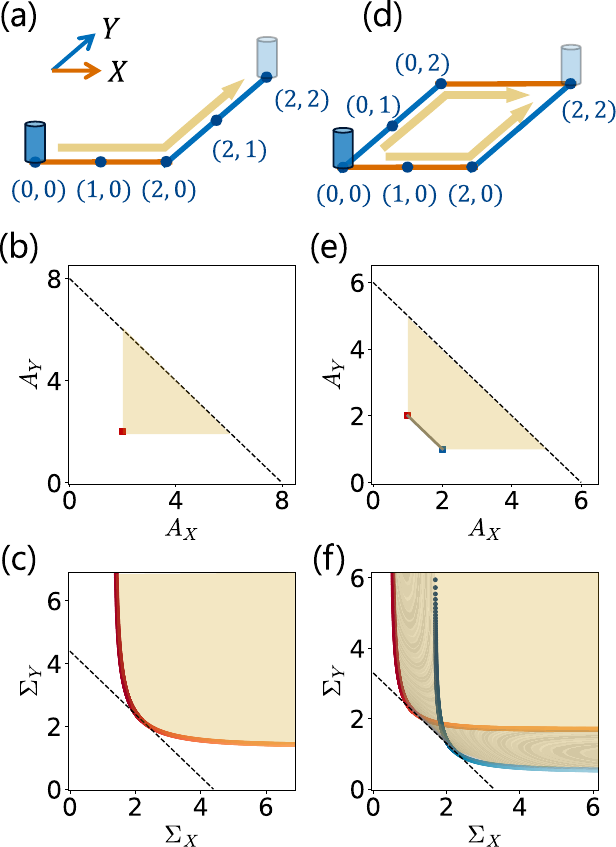}
    \caption{\label{fig: Pareto_example}
    Illustrative examples of the tradeoff relations between thermodynamic costs in finite-time processes.
    Consider the time evolution where probability $1$ is transported from node $(0,0)$ to $(2,2)$. (a) and (d) illustrate the differences in the graph $G(N,E)$ where the dynamics takes place.
    (a) For any $s$, there is only one path that gives the value of the cost function $d_{E_X:E_Y}^s$.
    (d) There are two paths, and the path that gives the value of the cost function $d_{E_X:E_Y}^s$ varies depending on the value of $s$.
    (b) and (e) represent the Pareto front $\calF_A$ of partial activities for the time evolution on graphs (a) and (d), respectively.
    The black dotted line indicates the upper bound $A_{XY}$ imposed on the activity when determining the partial EPs.
    (b) $\calF_A$ consists of a single red square.
    The parameter values are $\calW_{E} = 4$, $\calW_{E_X/E} = 2$, and $\calW_{E_Y/E} = 2$.
    (e) $\calF_A$ is composed of a line segment.
    The parameter values are $\calW_{E} = 3$, $\calW_{E_X/E} = 1$, and $\calW_{E_Y/E} = 1$.
    (c) and (f) illustrate the Pareto front $\calF_\Sigma$ of partial EPs for the time evolution on graphs (a) and (d), respectively.
    The black dotted line indicates the minimum value of the total EP.
    (c) The Pareto front $\calF_\Sigma$ is represented by a red curve, which is tangent to the black dotted line at one point.
    Here, $A_{XY} = 8$.
    (f) The Pareto front $\calF_\Sigma$ consists of parts of the red and blue curves and an edge with a slope of $-1$, overlapping with the black dotted line.
    Here, $A_{XY} = 6$.
    }
\end{figure}

\subsection{Examples}
\label{subsection: Pareto example}
As an illustrative example, consider a graph $G(N,E)$ shown in \fref{fig: Pareto_example}(a).
The time evolution is as follows: at time $t=0$, the system is in state $(0,0)$ with probability 1, and at time $t=\tau$, the system is in state $(2,2)$ with probability 1. 
In this case, the path from node $(0,0)$ to $(2,2)$ that gives the value of the cost function $d_{E_X:E_Y}^s$ is unique for all $s$.
Therefore, the Pareto front $\calF_A$ of partial activities is represented by a single point, as shown in \fref{fig: Pareto_example}(b).

When $\calF_A$ has this form, setting $\calL_{E_X} = \calW_{E_X/E}$ and $\calL_{E_Y} = \calW_{E_Y/E}$ in \eref{partial EP tradeoff expression} and then varying the thermodynamic forces provides the Pareto front $\calF_\Sigma$ of partial EPs (see \fref{fig: Pareto_example}(c)).
The black dotted line represents the minimum value of the total EP, and the Pareto front $\calF_\Sigma$ is tangent to this dotted line at a single point.
From the shape of $\calF_\Sigma$, it can be said that there exists a tradeoff relation between ${\Sigma}_{X}$ and ${\Sigma}_{Y}$ in the nonlinear regime.
However, the cost of reducing the dissipation in one subsystem is not linear; the other subsystem must bear much more dissipation.
This case will also be addressed in \sref{sec: Optimal information processing}.

Next, consider a graph $G(N,E)$ as shown in \fref{fig: Pareto_example}(d).
Similarly to the previous example, the time evolution is as follows: at time $t=0$, the system is in state $(0,0)$ with probability 1, and at time $t=\tau$, the system is in state $(2,2)$ with probability 1.
There are two paths from node $(0,0)$ to $(2,2)$ that do not pass through the same node.
In this case, the Pareto front $\calF_A$ of partial activities is as shown in \fref{fig: Pareto_example}(e).
For $0 \leq s < 1/2$, only the counterclockwise path is optimal (blue square).
Conversely, for $1/2 < s \leq 1$, only the clockwise path is optimal (red square).
When $s = 1/2$, both paths are optimal.
Furthermore, even taking the convex combination of these different paths, the transport cost remains optimal (gray edge).



As stated in \sref{subsec: partial EP Pareto}, since $\calF_A$ has an edge with a slope of $-1$, $\calF_\Sigma$ also has an edge with the same slope (see \fref{fig: Pareto_example}(f)).
This represents the minimum EP of the total system.
Again, there is a tradeoff relation where reducing ${\Sigma}_{X}$ increases ${\Sigma}_{Y}$.
However, unlike the case in \fref{fig: Pareto_example}(c), the optimality of the total EP is always maintained along the edge.
Therefore, the cost of reducing the dissipation in one subsystem is linear, and the other subsystem can simply bear the equivalent amount of increased dissipation.

\section{Optimal information processing}
\label{sec: Optimal information processing}
In this section, we apply the results obtained so far to a simple information processing setup. 
In bipartite systems, we define the information flow, which quantitatively describes the exchange of information and formulate Maxwell's demon.
Specifically, we consider a minimal four-state model of Maxwell's demon that performs measurement and feedback. 
We optimize the thermodynamic costs required for each information processing task and discuss the tradeoff relations between them.
We also implement a protocol to achieve optimal information processing using double quantum dots.

\subsection{Review: Information Flow and Maxwell's Demon}
In a bipartite system, we introduce the concept of information flow.
The mutual information between subsystems $X$ and $Y$ at time $t$ is defined as \cite{cover1999elements}
\begin{align}
    I_{XY}=I_{X_tY_t}:=\sum_{x,y} \ptt(x,y)\ln{\frac{\ptt(x,y)}{\pxt(x)\pyt(y)}}.
\end{align}
In bipartite systems, the time variation of mutual information can be decomposed into contributions from $X$ and $Y$ \cite{infoflow-Holowitz-2014},
\begin{align}
    d_t I_{XY}&=\dot{I}_X+\dot{I}_Y ,\\
    \label{I_X rate discrete}
    \dot{I}_X&=\sum_{x>x',y}J_{xx'}^{y}
    \ln\frac{\pycx(y|x)}{\pycx(y|x')},\\
    \label{I_Y rate discrete}
    \dot{I}_Y&=\sum_{y>y',x}J_{x}^{yy'}
    \ln\frac{\pxcy(x|y)}{\pxcy(x|y')},
\end{align}
where $\pycx(y|x):=\ptt(x,y)/\pxt(x)$ and $\pxcy(x|y):=\ptt(x,y)/\pyt(y)$ are the conditional probability distributions.
The information flow of $X$, i.e., $\dot{I}_X$, corresponds to the change in mutual information due to the time evolution of $X$: $\dot{I}_X = \lim_{dt \rightarrow 0} (I_{X_{t+dt}Y_{t}} - I_{X_{t}Y_{t}}) / dt$. The time integral of $\dot{I}_X$ is denoted as $\Delta I_X$. If $\Delta I_X > 0$, $X$ is acquiring information by measuring $Y$. Conversely, if $\Delta I_X < 0$, $X$ is either receiving feedback based on information about $X$ or simply losing information. The same applies to $\dot{I}_Y$.

We now formulate Maxwell's demon.
Ignoring the correlation with the other subsystem and focusing only on one subsystem may lead to an apparent violation of the second law of thermodynamics, as illustrated by the Szilard engine \cite{leff2002maxwell}.
Ignoring the correlation with $Y$, the apparent EP rate of $X$, $\dot{\Sigma}^{\rm ap}_X$, is given by
\begin{align}
    \dot{\Sigma}^{\rm ap}_X:=
    \sum_{x>x',y}J_{xx'}^{y}\ln\frac{\pxt(x')}{\pxt(x)}
    +\sum_{x>x',y}J_{xx'}^{y}\ln\frac{R_{xx'}^{y}}{R_{x'x}^{y}}.
\end{align}
The first term represents the change in Shannon entropy of $X$, whereas the second term represents the energy dissipation to the heat bath due to the transition of $X$.
The EP rate $\dot{\Sigma}_X$ defined in \eref{partial EP rate} considers the correlation with $Y$, and its relation to the information flow is expressed as
\begin{align}
\dot{\Sigma}_X=\dot{\Sigma}^{\rm ap}_X-\dot{I}_X.
\end{align}
Therefore, if $\dot{I}_X < 0$, it is possible for $\dot{\Sigma}^{\rm ap}_X$ to become negative, exhibiting Maxwell's demon.
By taking into account the ignored correlation, i.e., the change in mutual information $-\dot{I}_X$, the second law is restored \cite{Sagawa-Ueda2008quantumfeedback, Sagawa-Ueda2009erasure, sagawaNJP2013}.


\begin{figure}
    \centering
    \includegraphics[width=0.85\linewidth]{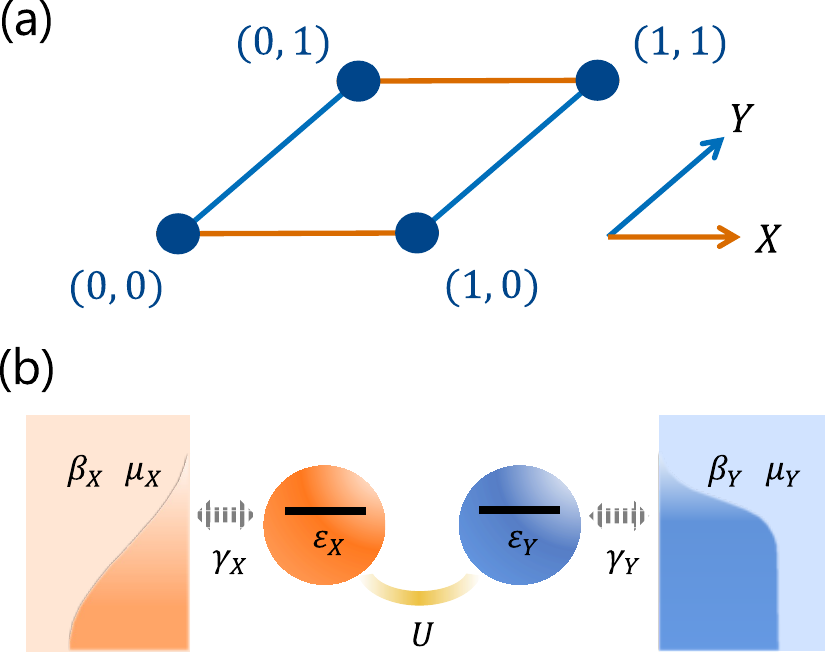}
    \caption{\label{fig: TQD_sketch}
    (a) A four-state model.
    The states of the total system are specified by the states $x=0,1$ of subsystem $X$ and $y=0,1$ of subsystem $Y$.
    There are no transitions where the states of $X$ and $Y$ change simultaneously, ensuring bipartiteness.
    (b) A schematic diagram of a double quantum dot that can be described by the model (a).
    Each dot $X$ and $Y$ is in contact with a thermal bath, with electrons entering and exiting a single energy level.
    In dot $X\ (Y)$, the level being empty is represented by $x=0\ (y=0)$, and being occupied by an electron is represented by $x=1\ (y=1)$.
    The two dots are connected by capacitance, which creates interactions that enable feedback or measurement of the other dot.
    }
\end{figure}

\subsection{Optimal Measurement and Optimal Feedback}
We consider a simple model of Maxwell's demon \cite{Strasberg2013demon,Diana-Esposito2014bipartite}, which consists of four states as shown in \fref{fig: TQD_sketch}(a).
In the corresponding graph $G(N,E)$, $X$ and $Y$ each take on states $0$ or $1$, and only transitions where either $X$ or $Y$ change are allowed.
Also, the upper bound of the total activity is imposed.
Since information processing is realized through transitions, this also limits the number of times information processing can occur.

To consider a process involving measurement and feedback, let the initial distribution be $\poo=[1/2,0,1/2,0]$ and the final distribution be $\pff=[1/2,1/2,0,0]$.
Here, the probabilities of the states $r=(0,0), (0,1), (1,0), (1,1)$ are described in this order.
In the initial distributions, the mutual information between $X$ and $Y$ is zero.
During the time evolution from the initial distribution to the final distribution, the mutual information may increase depending on the intermediate states.
This increase can be interpreted as a measurement taking place.
Since the mutual information of the final distribution is again zero, the mutual information generated by the measurement is consumed through feedback (or loss of information).
Using the method described in \sref{sec: Tradeoff relations in the thermodynamic costs of subsystems}, the bound of the tradeoff relation between dissipation due to measurement and feedback can be determined.

\begin{figure}
    \centering
    \includegraphics[width=0.95\linewidth]{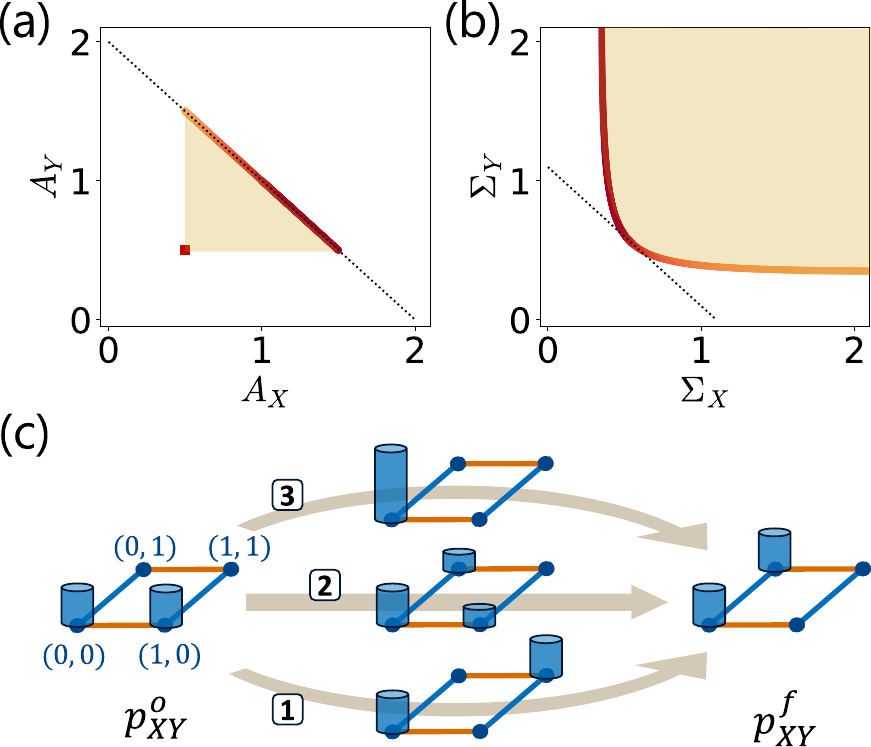}
    \caption{\label{fig: opt_info}
    Pareto front of the partial activity (a) and Pareto front of the partial EP (b) for the finite-time evolution $\poo=[1/2,0,1/2,0]\rightarrow\pff=[1/2,1/2,0,0]$.
    The parameter values are $\calW_{E}=1$, $\calW_{E_X/E}=\calW_{E_Y/E}=1/2$, and $A_{XY}=1.4$.
    (c) Differences in time evolution from the initial distribution $\poo$ to the final distribution $\pff$.
    In method $\bm{1}$, the probability at node $(1,0)$ is first transported to node $(1,1)$ and then to node $(0,1)$.
    Conversely, in method $\bm{3}$, the probability at node $(1,0)$ is first transported to node $(0,0)$ and then to node $(0,1)$.
    In method $\bm{2}$, a small probability is transported from node $(1,0)$ to node $(1,1)$ while simultaneously transporting it to node $(0,1)$.
    The distributions on the arrows illustrate the probability distributions at $t=\tau/2$.
    In all cases, the transport cost remains the same.
    Thus, $\calF_A$ and $\calF_\Sigma$ are represented by (a) and (b), respectively. 
    However, there are differences in how the mutual information changes.
    }
\end{figure}

The Pareto fronts of the partial activity and partial EP are shown in \fref{fig: opt_info}(a)(b).
For any $s$, there are two paths from node $(1,0)$ to $(0,1)$ that give the value of the cost function $d_{E_X:E_Y}^s$, and the transport costs related to $E_X$ and $E_Y$ are the same for both paths.
The shaded area in \fref{fig: opt_info}(a) represents the feasible region $K_A$ for the partial activity, where the activity upper bound $A_{XY}$ is satisfied.
In the protocol giving the Pareto front $\calF_\Sigma$ (the red line in \fref{fig: opt_info}(b)) for partial EP, the partial activity takes the value on the red line in \fref{fig: opt_info}(a).

Since the geodesic of the Wasserstein distance is not necessarily unique, there exist processes with various changes in the mutual information.
However, as long as the same thermodynamic force is applied, the partial activity and partial EP remain unchanged.
In other words, the Pareto fronts are identical in those processes.
In the present model, we illustrate changes in the mutual information due to variations in the geodesic.
To do this, we focus on the two paths that give the optimal transport cost (see \fref{fig: opt_info}(c)).
Either path, or any combination of them, is optimal.
There are numerous geodesics along these paths, among which we explain the three shown in \fref{fig: opt_info}(c).

\subsubsection{Measurement-feedback separation}
\label{subsubsec: calculation of measurement and feedback}
As a geodesic using only the path via node $(1,1)$, consider the following time evolution:
\begin{align}
    \label{time evolution step}
    \ptt=
    \begin{cases}
        \frac{1}{2}[1,0,1-\frac{2t}{\tau},\frac{2t}{\tau}]\ &(0\leq t\leq\frac{\tau}{2})\\
        \frac{1}{2}[1,\frac{2t-\tau}{\tau},0,1-\frac{2t-\tau}{\tau}]\ &(\frac{\tau}{2}<t\leq\tau)
    \end{cases}.
\end{align}
This describes a situation where $Y$ is measured for $0 \leq t \leq \tau/2$, and $X$ is fed back for $\tau/2 < t \leq \tau$.
In this case, $\Delta I_X = -\ln 2$ and $\Delta I_Y = \ln 2$.
Since measurement and feedback are temporally separated, we refer to this time evolution as measurement-feedback separation.
The Pareto fronts of the partial activity and partial EP are represented as in \fref{fig: opt_info}(a)(b).
Therefore, there is a tradeoff relation between the dissipation required for measurement and the dissipation required for feedback.

\begin{figure*}
    \centering
    \includegraphics[width=0.95\linewidth]{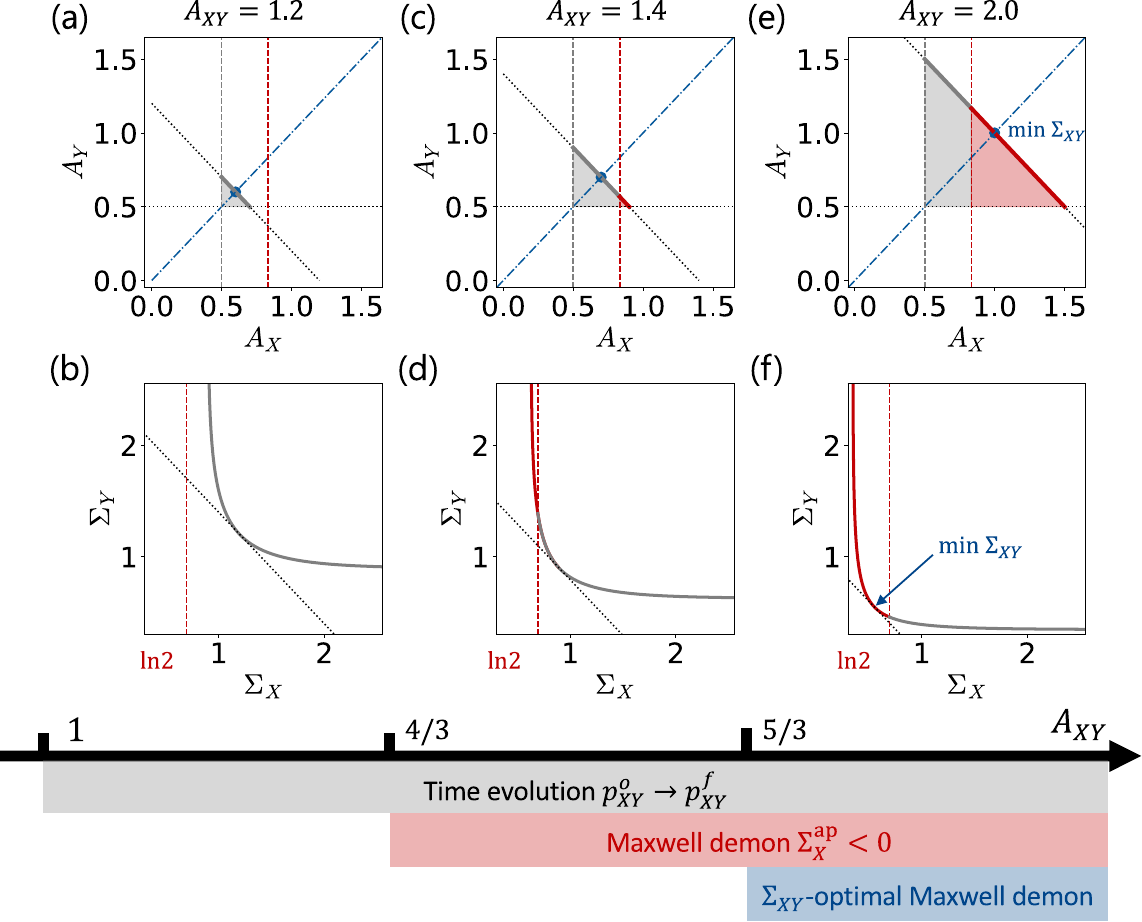}
    \caption{\label{fig: opt_info_AXY}
    Changes in the Pareto fronts $\calF_A$ and $\calF_\Sigma$ when increasing the upper bound of activity $A_{XY}$ during the measurement-feedback separation (see \eref{time evolution step}).
    In (a)(b), $A_{XY}$ is set to $1.2$, in (c)(d) to $1.4$, and in (e)(f) to $2.0$.
    In the $A_X-A_Y$ plane, the black diagonal dashed line represents the upper bound $A_{XY}$, the black horizontal dashed line represents $\calW_{E_Y}$, and the gray vertical dashed line represents $\calW_{E_X}$.
    The red vertical dashed line represents $A_X=\calW_{E_X/E}\coth(-\Delta{I}_X/2\calW_{E_X/E})=5/6$, indicating that Maxwell's demon for $X$ is possible on the right side of this line.
    The orange dashed line represents $A_Y=(\calW_{E_Y/E}/\calW_{E_X/E})A_X$, and the point where this orange line intersects the $A_{XY}$ dashed line is where the total EP is minimized (see also \fref{fig: EP_Pareto_general}).
    In the $\Sigma_X-\Sigma_Y$ plane, the black diagonal dashed line represents the minimum EP of the total system.
    The Pareto front $\calF_\Sigma$ is tangent to this line at one point.
    In addition, the red vertical dashed line represents $-\Delta I_X=\ln 2$, indicating that Maxwell's demon for $X$ is possible on the left side of this line.
    }
\end{figure*}

By increasing $A_{XY}$, which corresponds to the number of information processing steps, we observe changes in the feasibility and optimality of Maxwell's demon (see \fref{fig: opt_info_AXY}).
First, to enable the finite-time evolution $\poo \rightarrow \pff$, it is necessary that $A_{XY} \geq \calW_E$.
Also, since $\Delta I_X = -\ln 2$, Maxwell's demon (${\Sigma}^{\rm ap}_X < 0$) can be realized if $\Sigma_X < \ln 2$.
Now, considering the optimal protocol, we focus on the Pareto front $\calF_\Sigma$ of partial EPs (see \fref{fig: opt_info_AXY}(b)(d)(f)).
When $\calF_\Sigma$ is realized, the partial activity takes values on the red or gray line segments (see \fref{fig: opt_info_AXY}(a)(c)(e)).
On $\calF_\Sigma$, $\Sigma_X$ can be expressed using $A_X$ as $2\calW_{E_X/E} \tanh^{-1}(\calW_{E_X/E}/A_X)$.
Therefore, combined with $\calW_{E_X/E} = 1/2$, $\Sigma_X < -\Delta I_X$ on $\calF_\Sigma$ if $A_X > 5/6$.
When $\Sigma_X$ is minimized, the partial activity of $Y$ is $\calW_{E_Y/E} = 1/2$.
Thus, considering the partial activity of $Y$ as well, the upper bound of the total activity required to realize Maxwell's demon is $A_{XY} > 5/6 + 1/2 = 4/3$.
In the feasible region of the partial activity, Maxwell's demon can be realized in the red-shaded areas of \fref{fig: opt_info_AXY}(c)(e).
When $1 \leq A_{XY} \leq 4/3$ as in \fref{fig: opt_info_AXY}(a)(b), Maxwell's demon cannot be realized at any thermodynamic costs.

Next, we consider the optimality of the total EP when realizing Maxwell's demon.
As stated in \sref{subsubsec: sweepless}, when the total EP is minimized, the partial activity satisfies the relation $A_X/\calW_{E_X/E} = A_Y/\calW_{E_Y/E}$.
The orange dashed lines in \fref{fig: opt_info_AXY}(a)(c)(e) represent this relation, and the orange dots correspond to the partial activity when the total EP is minimized.
Therefore, to minimize the total EP while realizing Maxwell's demon, it is necessary that $A_Y > (\calW_{E_Y/E}/\calW_{E_X/E}) \cdot 5/6 = 5/6$.
Thus, the upper bound of the total activity must be $A_{XY} > 5/6 + 5/6 = 5/3$.
When $4/3 < A_{XY} \leq 5/3$ as in \fref{fig: opt_info_AXY}(c)(d), a protocol that minimizes the total EP cannot realize Maxwell's demon.
Nevertheless, Maxwell's demon can be realized by increasing the dissipation $\Sigma_Y$ in measurement and decreasing the dissipation $\Sigma_X$ in feedback.
When $A_{XY} > 5/3$ as in \fref{fig: opt_info_AXY}(e)(f), it is possible to realize Maxwell's demon while optimizing the total EP simultaneously.

In summary, to enable time evolution, it is necessary to have $A_{XY} \geq 1$, and to realize Maxwell's demon, $A_{XY} > 4/3$ is required.
Furthermore, to minimize the total EP while realizing Maxwell's demon, $A_{XY} > 5/3$ is needed.
By increasing $A_{XY}$, more diverse information processing becomes possible.
This is because the total dissipation decreases, and according to the tradeoff relation, the dissipation in the subsystems can be further reduced.
This is considered as a general property independent of the initial and final distributions or the graph.

\subsubsection{Continuous measurement-feedback}
\label{subsubsec: calculation of continuous process model}
Consider a different geodesic using the same path as in the \sref{subsubsec: calculation of measurement and feedback}:
\begin{align}
    \label{continuous time evolution: XY}
    \ptt&=\left(1-\frac{t}{\tau}\right)\poo+\frac{t}{\tau}\pff
    \nonumber\\
    &=\frac{1}{2}[1-\epsilon,t/\tau,1-t/\tau,\epsilon],\\
    \label{continuous time evolution: X}
    \pdt&=\ptt+\frac{dt}{2\tau}[0,1,0,-1],\\
    \label{continuous time evolution: Y}
    \ptd&=\ptt+\frac{dt}{2\tau}[0,0,-1,1].
\end{align}
Here, the time evolutions of $X$ and $Y$ in the \sref{subsubsec: calculation of measurement and feedback} are performed simultaneously.
However, to prevent the elements of $\pdt$ from becoming negative, we modify the initial distribution to $\poo = [(1-\epsilon)/2, 0, 1/2, \epsilon/2] \ (\epsilon \ll 1)$, ensuring $\epsilon > dt/\tau$ and keeping both values small.
This correction does not change $\calW_E, \calT, \calW_{E_X/E}$, or $ \calW_{E_Y/E}$.
Therefore, the Pareto fronts of the partial activity and partial EP are still represented by \fref{fig: opt_info}(a)(b).

From the partial time evolution (\esref{continuous time evolution: X}\eqref{continuous time evolution: Y}), the information flow can be calculated, yielding
\begin{align}
    \label{cont: dIX}
    \dot{I}_X(t)
    &=\frac{1}{2\tau}\ln\frac{\frac{t}{\tau}(1-\frac{t}{\tau}+\epsilon)}{\epsilon(1+\frac{t}{\tau}-\epsilon)},\\
    \label{cont: dIY}
    \dot{I}_Y(t)
    &=\frac{1}{2\tau}\ln\frac{\epsilon(2-\frac{t}{\tau}-\epsilon)}{(1-\frac{t}{\tau})(\frac{t}{\tau}+\epsilon)}=-\dot{I}_X(\tau-t).
\end{align}
For $t/\tau < \epsilon$ and $t/\tau > 1-\epsilon$, we have $\dot{I}_X < 0$, whereas for $\epsilon < t/\tau < 1-\epsilon$, we have $\dot{I}_X > 0$.
Unlike the measurement-feedback separation case discussed before, measurement and feedback are not temporally separated.
Since measurement and feedback are continuously repeated, we refer to this time evolution as continuous measurement-feedback.

The change in the mutual information due to $X$ and $Y$ is entirely different from that in the measurement-feedback separation case:
\begin{align}
    \label{DeltaI: continuous information processing}
    \Delta I_X&=-\Delta I_Y\nonumber\\
    &=\frac{1}{2}\big(-(2-\epsilon)\ln(2-\epsilon)-1-\ln\epsilon-\epsilon\ln\epsilon\nonumber\\
    &\ \ \ \ \ \ \ \ \ +(1+\epsilon)\ln(1+\epsilon)+(1-\epsilon)\ln(1-\epsilon)\big).
\end{align}
The first and final terms contribute in the same way as the measurement-feedback separation case, which can be seen from simple calculations.
Surprisingly, by making $\epsilon$ smaller, a larger information flow can be achieved.
However, since $\Delta I_X > 0$ and $\Delta I_Y < 0$, the roles are reversed, with $X$ performing the measurement and $Y$ receiving the feedback.
Therefore, when $\epsilon$ is sufficiently small, Maxwell's demon concerning $Y$ can be realized, except in the region where $A_Y \simeq \calW_{E_Y/E}$.

Since both the initial and final distributions are uncorrelated, $\Delta I = \Delta I_X + \Delta I_Y = 0$.
In the limit $\epsilon \rightarrow 0$, the first and second terms in \eref{DeltaI: continuous information processing} converge to finite values, and the fourth and subsequent terms converge to zero.
However, the third term contains $\ln\epsilon$, which diverges logarithmically.
Nevertheless, because the transition rates implementing the time evolution are proportional to $\epsilon^{-1}$ (see \eref{protocol W of cont}), it is practically impossible to make the information flow infinitely large.

\subsubsection{Reset and redistribute}
Finally, as a geodesic using only the path via node $(0,0)$, consider the following time evolution:
\begin{align}
    \label{time evolution reset and redistribute}
    \ptt=
    \begin{cases}
        \frac{1}{2}[1+\frac{2t}{\tau},0,1-\frac{2t}{\tau},0]\ &(0\leq t\leq\frac{\tau}{2})\\
        \frac{1}{2}[1-\frac{2t}{\tau},\frac{2t}{\tau},0,0]\ &(\frac{\tau}{2}<t\leq\tau)
    \end{cases}.
\end{align}
This describes a situation where, for $0 \leq t \leq \tau/2$, the state of $X$ is reset to 0, and for $\tau/2 < t \leq \tau$, the state of $Y$ is redistributed between $0$ and $1$.
In this case, the mutual information does not change at all during the process, with $\Delta I_X = \Delta I_Y = 0$.
Therefore, this is a process without any information processing.

\subsubsection{Several remarks}
From the perspective of transport, in \sref{subsubsec: calculation of measurement and feedback}, the probability $1/2$ was first moved to node $(1,1)$ and then to node $(0,1)$ (see also \fref{fig: opt_info}(c)).
On the other hand, as in \sref{subsubsec: calculation of continuous process model}, it is also possible to move the probability directly to node $(0,1)$.
That is, a small probability can be transported simultaneously from node $(1,0)$ to node $(1,1)$ and from node $(1,1)$ to node $(0,1)$.
Whichever the optimal transport matrix elements are sequentially transported or simultaneously transported, the trajectory of the time evolution remains geodesic.
The difference in these geodesics results in different changes in the mutual information, but this is not due to moving the final distribution as discussed in \ccite{Nagase-Sagawa2023infogain}.

We note that in this example, the Wasserstein distance and the total variation distance $\calT$ may differ because the graph $G(N,E)$ is not fully connected.
In fact, for this initial and final distributions, $\calW_E=1$ and $\calT=1/2$, meaning that the bound using $\calT$ as in \ccite{Lee-Park_highly2022,Salazar2022EPbound} cannot be achieved.

It should be noted that optimization can similarly be performed for general initial and final distributions.
For example, if $\poo = [1, 0, 0, 0]$ and $\pff = [1/2, 0, 0, 1/2]$, there exists a process where $\Delta I_X > 0$ and $\Delta I_Y > 0$. In this case, $X$ and $Y$ mutually measure each other, enhancing their correlation through information processing.
We note that if $\poo = [1/2, 0, 0, 1/2]$ and $\pff = [1/2, 0, 1/2, 0]$, this represents the optimal bit erasure of a memory $Y$ that stores the information of system $X$, reproducing the results of \ccite{Lee-Park_highly2022}.

\subsection{Implementation with Double Quantum Dots}
\label{subsec: install with double QD}
The optimal information processing described above can be implemented using double quantum dots \cite{Strasberg2013demon,Diana-Esposito2014bipartite,Kutvonen-Sagawa2016}.
Consider a system as depicted in \fref{fig: TQD_sketch}(b).
The quantum dots $X$ and $Y$ are each coupled to thermal baths with inverse temperatures $\beta_X$ and $\beta_Y$, and chemical potentials $\mu_X$ and $\mu_Y$, respectively. Electrons can enter and exit the quantum dots from these thermal baths.
We consider the low-temperature and Coulomb blockade regime.
Thus, each dot cannot be occupied by more than one electron, and the dynamics can be treated as a Markov jump process.
Using the state where dot $X$ is empty ($x=0$) as the reference energy, let $\varepsilon_X$ be the energy of the state where $X$ is occupied ($x=1$).
The same applies to dot $Y$.
To introduce interaction between $X$ and $Y$, the quantum dots are connected via capacitance.
As a result, when both $X$ and $Y$ are occupied ($x=y=1$), the total system energy, including the interaction energy $U$, becomes $\varepsilon_X + \varepsilon_Y + U$.

To implement optimal information processing, the transition rates are controlled.
According to Fermi's golden rule, the transition rates are given by
\begin{align}
    \label{Fermi golden}
    \begin{cases}
        &
        R_{xx'}^{y}=\gamma_X f_X\left((x-x')(\varepsilon_X+yU)\right)\\
        &
        R^{yy'}_{x}=\gamma_Y f_Y\left((y-y')(\varepsilon_Y+xU)\right)
    \end{cases},
\end{align}
where $\gamma_X$ and $\gamma_Y$ represent the coupling strengths with the thermal baths,
and $f_X(\varepsilon) := 1/(1 + e^{\beta_X(\varepsilon - \mu_X)})$ and $f_Y(\varepsilon) := 1/(1 + e^{\beta_Y(\varepsilon - \mu_Y)})$ are the Fermi distribution functions.
To implement optimal transport, it is necessary to manipulate the symmetric and antisymmetric parts of the transition rates in a time-dependent manner \cite{dechant2022minimum, Remlein-Seifert2021}.
In our quantum dot setup, this corresponds to adjusting the energy levels, interaction energy, and coupling strengths.

Henceforth, we will consider the foregoing protocols of information processing (\sref{subsubsec: calculation of measurement and feedback}, \ref{subsubsec: calculation of continuous process model}) separately.
However, in each case, the protocol realizes the same Pareto front $\calF_\Sigma$ of partial EPs (\fref{fig: opt_info}(a)(b)).

\subsubsection{Measurement-feedback separation}
\label{subsub: implementation of measurement and feedback}
\begin{figure}
    \centering
    \includegraphics[width=0.95\linewidth]{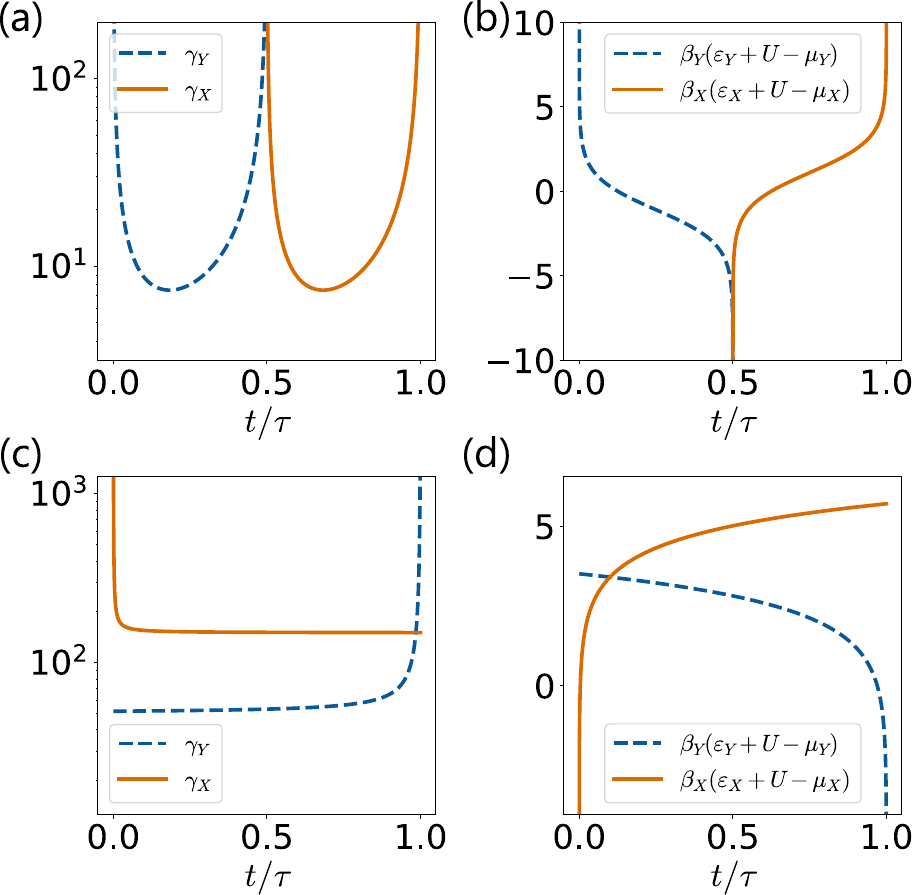}
    \caption{\label{fig: Qdot_protocol}
    Protocol for implementing optimal information processing with a double quantum dot.
    The upper bound of activity is set to $A_{XY}=2.0$.
    The thermodynamic forces are chosen as $F_X=F_Y=2\tanh^{-1}(\calW_E/A_{XY})$ to minimize the EP of the total system.
    Using the optimal transition rates, each parameter of the two quantum dots is calculated according to \eref{Fermi golden}.
    Other parts of $\calF_\Sigma$ can be determined similarly.
    (a)(b) Measurement-feedback separation (see \esref{protocol R_Y of M and F}\eqref{protocol R_X of M and F}).
    (c)(d) Continuous measurement-feedback (see \eref{protocol W of cont}). Here, $\tau=1$ and $\epsilon=10^{-2}$.
    }
\end{figure}

In \textit{measurement-feedback separation}, we implement the protocol that realizes $\calF_\Sigma$ shown in \fref{fig: opt_info}(b).
The time evolution is given by \eref{time evolution step}.
During $0 \leq t \leq \tau/2$, the measurement transports probability $1/2$ to node $(1,1)$, and during $\tau/2 < t \leq \tau$, the feedback transports probability $1/2$ to node $(0,1)$.
The optimal transition rates for the measurement ($0 \leq t \leq \tau/2$) are given by
\begin{align}
    \label{protocol R_Y of M and F}
    R_{1}^{10}=\frac{1}{1-e^{-F_Y}}\frac{1}{\tau}\frac{1}{\frac{1}{2}(1-\frac{2t}{\tau})},
    R_{1}^{01}=\frac{1}{e^{F_Y}-1}\frac{1}{\tau}\frac{1}{\frac{1}{2}\frac{2t}{\tau}}.
\end{align}
The first factor adjusts the reversibility of transitions on $E_Y$.
The second factor is the transport rate per unit time (equal to $\dot{\calL}_{E_Y}$), and the third factor is the reciprocal of the probability of the node from which the transport originates.
In the optimal protocol, other transition rates are set to zero.
This can be achieved by setting $\gamma_X = 0$ and $\beta_Y(\varepsilon_Y-\mu_Y) \gg1$.

Similarly, the optimal transition rates for the feedback ($\tau/2 < t \leq \tau$) are given by
\begin{align}
    \label{protocol R_X of M and F}
    R_{01}^{1}=\frac{1}{1-e^{-F_X}}\frac{1}{\tau}\frac{1}{\frac{1}{2}(1-\frac{2t-\tau}{\tau})},
    R_{10}^{1}=\frac{1}{e^{F_X}-1}\frac{1}{\tau}\frac{1}{\frac{1}{2}\frac{2t-\tau}{\tau}}.
\end{align}
Here again, in the optimal protocol, other transition rates are set to zero.
This can be achieved by setting $\gamma_Y = 0$ and $\beta_X(\varepsilon_X-\mu_X) \gg1$.

Based on \eref{Fermi golden}, the parameters of the quantum dots that implement the optimal protocol are shown in \fref{fig: Qdot_protocol}(a)(b).
The thermodynamic forces $F_X$ and $F_Y$ are chosen to minimize the total EP.
It is necessary to significantly change the parameters according to the changes in the node probabilities.
Specifically, near $t=0$, $\tau/2$, and $\tau$, where the node probabilities become zero, the coupling strengths can become extremely large.
In practice, it is necessary to slightly modify the initial and final distributions to ensure that the node probabilities do not become zero.

\subsubsection{Continuous measurement-feedback}
\label{subsub: implementation of continuous information processing}
Next, in \textit{continuous measurement-feedback}, we implement the protocol that realizes the same $\calF_\Sigma$ as in \sref{subsub: implementation of measurement and feedback}.
The time evolution is given by \eref{continuous time evolution: XY}.
The optimal transition rates are given by
\begin{align}
    \label{protocol W of cont}
    \begin{cases}
        &R_{01}^{1}=\frac{1}{1-e^{-F_X}}\frac{1}{2\tau}\frac{1}{\frac{1}{2}\epsilon},
        R_{10}^{1}=\frac{1}{e^{F_X}-1}\frac{1}{2\tau}\frac{1}{\frac{1}{2}\frac{t}{\tau}}
        \\
        &R_{1}^{10}=\frac{1}{1-e^{-F_Y}}\frac{1}{2\tau}\frac{1}{\frac{1}{2}(1-\frac{t}{\tau})},
        R_{1}^{01}=\frac{1}{e^{F_Y}-1}\frac{1}{2\tau}\frac{1}{\frac{1}{2}\epsilon}
    \end{cases}.
\end{align}
In the optimal protocol, other transition rates are set to zero.
This can be achieved by setting $\beta_X(\varepsilon_X-\mu_X) \gg1,\beta_Y(\varepsilon_Y-\mu_Y) \gg1$. $\varepsilon_X$ and $\varepsilon_Y$ are manipulated satisfying this constraint.
The protocol that minimizes the total EP is shown in \fref{fig: Qdot_protocol}(c)(d).
Since the probability of node $(1,1)$ is small and does not change significantly, the divergence observed near $t=\tau/2$ in \fref{fig: Qdot_protocol}(a)(b) can be avoided.

As described above, the optimal information processing can be implemented using double quantum dots.
In general, when the probability on a node becomes small, such as near $t=0$ and $\tau$, large transition rates are required.
This issue cannot be resolved by adjusting the thermodynamic forces.
Moreover, it cannot be avoided by slowing down the rate of time evolution $\dot{\calL}_E$.
This is a bottleneck for achieving precise transport in discrete systems.

\begin{figure}
    \centering
    \includegraphics[width=0.95\linewidth]{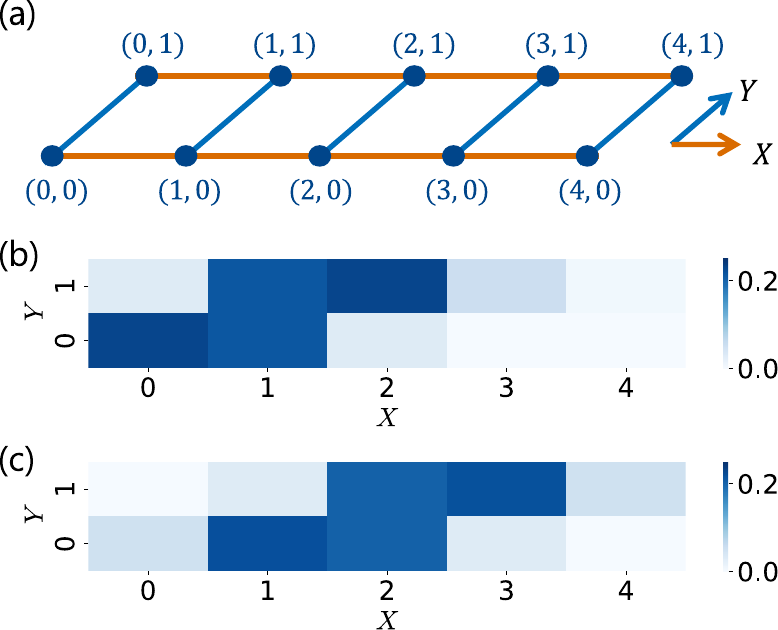}
    \caption{\label{fig: chemo_dist}
    (a) Model of \textit{E. coli} chemotaxis.
    The autonomous transition rates are given as follows \cite{lan-2012ESA}.
    The methylation and demethylation rates are $R_{x+1,x}^0=k_R$, $R_{x,x+1}^0=\chi k_R$, $R_{x,x+1}^1=k_B$, and $R_{x+1,x}^1=\chi k_B$ for $x=0,\dots,3$.
    The activation and deactivation rates satisfy $R_{x}^{01}=R_{x}^{10}e^{2(1-x)+\calC}$ with $\min{R_{x}^{01}, R_{x}^{10}}=k_Y$ for $x=0,\dots,4$.
    The parameter $\chi$ adjusts the reverse reaction, and $\calC$ describes the free energy's dependence on ligand concentration.
    We set $k_R=k_B=1.0$, $k_Y=0.010$, and $\chi=0.10$.
    (b) Steady state for low ligand concentration ($\calC=0$).
    This state is specified as the initial distribution, $\poo$, with the value of $\poo(x,y)$ shown by color in each cell.
    (c) State at time $t=\tau=10$, where the initial distribution $\poo$ at $t=0$ relaxes in response to a high ligand concentration signal ($\calC=2.0$).
    This state is specified as the final distribution, $\pff$, with $\pff(x,y)$ values shown by color in each cell.}
\end{figure}

\section{Application to Multi-state Systems}
\label{sec: multistate}
In Secs.~\ref{sec: Tradeoff relations in the thermodynamic costs of subsystems},\ref{sec: Optimal information processing}, we examined prototypical few-state systems to illustrate our results. 
However, our framework is not restricted to such situation but also immediately applicable to larger scale systems.  
In this section, we demonstrate its application to a multi-state model of chemotaxis.

We consider a model of \textit{E. coli} chemotaxis \cite{lan-2012ESA,sartori-2014sensory} as shown in \fref{fig: chemo_dist}(a).
The state of $X$ represents the receptor's methylation level, while the state of $Y$ represents the kinase activity, where $y=0$ and $y=1$ correspond to the inactive and active states, respectively.
The average value of $X$ varies according to the environmental ligand concentration.
Specifically, $X$ tends to take smaller (larger) values at lower (higher) ligand concentrations (see \fref{fig: chemo_dist}(b)(c)).
The change in $Y$ occurs much faster than in $X$, and in the (nonequilibrium) steady states, $Y$ takes $y=0$ and $y=1$ with nearly equal probabilities, regardless of ligand concentration.
\begin{figure*}
    \centering
    \includegraphics[width=0.95\linewidth]{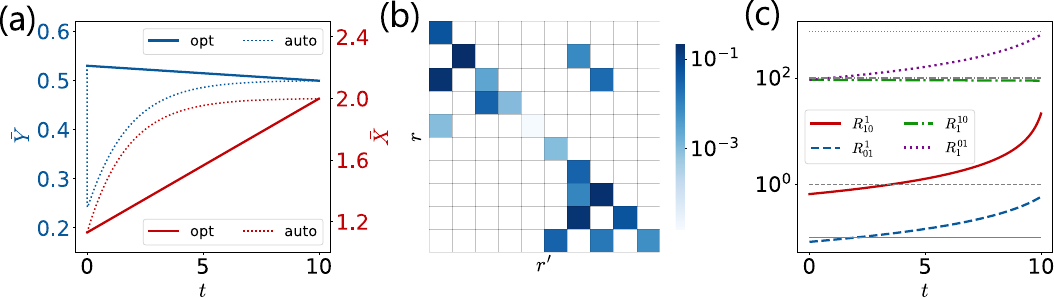}
    \caption{\label{fig: chemo_protocol}
    (a) Time evolution of the average values of $X$ and $Y$.
    The blue (red) solid line represents $\bar{Y}$ ($\bar{X}$) in the optimal time evolution.
    The blue (red) dotted curve shows $\bar{Y}$ ($\bar{X}$) in the time evolution driven by autonomous transition rates (see \fref{fig: chemo_dist}).
    In the autonomous process, $\bar{Y}$ responds rapidly to the ligand signal and returns to approximately $\bar{Y}=0.5$, indicating adaptation.
    (b) Optimal transport matrix $\Pi^*$ within $\calU(\pff,\poo)$ with respect to $d_E$.
    The values of $\Pi^*(r,r')$ are shown by color in each cell.
    Nodes $(0,0), (0,1), \dots, (1,4)$ are labeled sequentially from top to bottom for $r$ and from left to right for $r'$.
    (c) Protocol for implementing the optimal time evolution.
    Gray lines indicate the values of the autonomous transition rates. 
    For ease of viewing, not all rates are shown.}
\end{figure*}

Consider the following adaptation situation.
The system initially relaxes to the less methylated steady state (\fref{fig: chemo_dist}(b)), after which the ligand concentration suddenly increases.
The subsystem $Y$ responds rapidly to this signal, while subsystem $X$ gradually relaxes to the highly methylated steady state (\fref{fig: chemo_dist}(c)).
Eventually, the average value of $Y$, denoted as $\bar{Y}$, returns to a level near its original value prior to the signal (see also the dotted line in \fref{fig: chemo_protocol}(a)).
This adaptation is governed by autonomous transition rates that depend on both the methylation level and the ligand concentration (see the caption of \fref{fig: chemo_dist} for details).
These rates violate the detailed balance condition to maintain $\bar{Y} \simeq 0.5$, and thereby, chemical energy is continuously supplied to the system \cite{lan-2012ESA}.
In this process ($0 \leq t \leq \tau$), the partial activities and entropy productions (EPs) for the autonomous transition rates are computed as $A_X = 11$, $A_Y = 1.5 \times 10^3$, $\Sigma_X^{\rm{auto}} = 30$, and $\Sigma_Y^{\rm{auto}} = 0.33$.

We examine the minimal cost required for the time evolution from the less methylated state $\poo$ (\fref{fig: chemo_dist}(b)) to the highly methylated state $\pff$ (\fref{fig: chemo_dist}(c)).
The optimal time evolution is obtained by applying \eref{partial EP tradeoff expression}.
Using the Python Optimal Transport library \cite{flamary2021pot}, we compute the optimal transport matrix, as shown in \fref{fig: chemo_protocol}(b), and the Wasserstein distances as $\calW_E=0.90$, $\calW_{E_X/E}=0.87$, and $\calW_{E_Y/E}=0.030$.
The Pareto fronts exhibit forms similar to those in \fref{fig: Pareto_example}(b)(c).
Here, we consider the speed difference in the dynamics of $X$ and $Y$, setting the thermodynamic forces and activity bounds so that the partial activities match their autonomous values, $A_X=11$ and $A_Y=1.5\times10^3$.
The optimal time evolution is achieved by linearly interpolating between $\poo$ and $\pff$, and thus the average values of $X$ and $Y$ also interpolates between its initial and final values (see \fref{fig: chemo_protocol}(a)).
The optimal protocol is determined from the optimal transport matrix (see \fref{fig: chemo_protocol}(b)(c)).
The optimal partial entropy productions (EPs) are $\Sigma_X=0.14$ and $\Sigma_Y=1.2\times10^{-6}$.
The EP for $X$ is reduced by a factor of more than $200$ compared to the autonomous case, $\Sigma_X^{\rm{auto}}$.
The minimal cost for the time evolution itself is given by $\Sigma_X$, while $\Sigma_X^{\rm{auto}}$ accounts for the cost of implementing this evolution autonomously between steady states.
Notably, the optimal time evolution does not involve a tradeoff between EP and adaptation accuracy \cite{lan-2012ESA}.
On the other hand, the EP for $Y$ is nearly zero in the optimal protocol, indicating that $Y$ changes so rapidly that it remains almost equilibrated throughout the optimal time evolution.

Before concluding this section, we remark on the applicability to larger systems.
We remember that the Wasserstein distance can be computed exactly for the examples discussed in this paper.
For distributions with $|N|$ states, the exact calculation requires $\tilde{\calO}(|N|^3)$ time steps \cite{Pele-Werman2009EMD}, which is practically computable up to $|N|\simeq 10^4$.
In even larger systems, the Wasserstein distance can be approximated with $\tilde{\calO}(|N|^2/\epsilon)$ time steps using the Sinkhorn algorithm \cite{cuturi2013sinkhorn, Pham-2020sinkhorn}, where $\epsilon$ denotes the desired accuracy of the solution.

\section{Conclusion and Discussion}
In this paper, we have introduced the concept of Pareto front for thermodynamic costs, revealing tradeoff relations between the costs associated with finite-time information processing (\esref{s discrete Wasserstein}\eqref{partial EP tradeoff expression}). 
The primary focus of this paper is on discrete Markov jump processes, where we identified the optimal thermodynamic costs for each subsystem (\esref{min AX global}\eqref{min EP_X global}).
To this end, we extended the Wasserstein distance to measure only the transport costs of subsystems such as the memory and the engine (\eref{Ex Wasserstein}).
We developed a general method for constructing the Pareto fronts for thermodynamic costs (\fref{fig: activity_Pareto}, \ref{fig: EP_Pareto_general}) and showed that the Pareto front of partial EPs is determined by the Pareto front of partial activities.
We found that the non-uniqueness of the optimal transport matrix significantly affects the shapes of the Pareto fronts (\fref{fig: Pareto_example}).
Our results imply that in the process of measurement and feedback, there is a tradeoff relation between the dissipation of measurement and that of feedback (\fref{fig: opt_info}).
Increasing the activity enables more diverse information processing such as Maxwell's demon (\fref{fig: opt_info_AXY}).
We also demonstrated that our results are applicable to the optimization of multi-state systems (\fref{fig: chemo_protocol}).
Below, we address some future perspectives.

\textit{Continuous systems.---}
In continuous systems such as Langevin systems, the activity is replaced by the diffusion constant, which typically cannot be controlled.
Consequently, the tradeoff relation for thermodynamic costs in continuous systems fundamentally differs from those in discrete systems.
For continuous systems, another recent paper \cite{kamijima-2024continuous} derived the Pareto front of partial EPs using the $L^2$ Wasserstein distance and revealed the existence of the tradeoff relation between EPs.
On the other hand, taking the continuous limit of \eref{min AX global} would yield an $L^1$ inequality, which cannot be achieved in general \cite{dechant2022minimum,Van-Saito-unification2023}.

\textit{More practical situations.---}
Optimizing the thermodynamic costs of information processing has various potential applications.
For instance, current industrial computational devices exhibit thermal dissipation much larger than the fundamental bound.
It is thus crucial to understand the extent to which computational costs can be fundamentally reduced in such devices \cite{Wolpert2019computation,Freitas-2021Circuits}.
Applications to biological systems are also highly intriguing.
In cells, various chemical reactions occur to maintain life, consuming free energy in the process.
These reactions and processes are subject to thermodynamic constraints due to thermal fluctuations in the environment \cite{Rao-Esposito2016CRN}.
Additionally, biological constraints introduce various tradeoff relations between functionality and cost, such as in sensory adaptation \cite{Lan-2012energy-speed-accuracy,Ito-Sgawa2015signaltransduction}, kinetic proofreading \cite{Arvind-2012speed-proofreading,Sartori-Pigolotti2015Correction}, and biochemical copying \cite{Ouldridge-2017Copying-Biochemical}.

The minimal costs and tradeoff relations obtained through our method can be considered as an important step toward optimization and functional understanding in such systems.
However, optimization using optimal transport theory assumes full control, requiring the ability to arbitrarily manipulate transition rates.
In realistic systems, however, the time dependence of transition rates can only be partially controlled.
Therefore, to establish more relevant bounds to those situations, it would be necessary to explore the optimization of thermodynamic costs under limited control \cite{Kolchinsky-Wolpert2021Constraints,Zhong-DeWeese2022limited}.

\begin{acknowledgments}
We thank Sosuke Ito for the fruitful discussion.
This work is supported by JST ERATO Grant No. JPMJER2302, Japan.
T.K. is supported by World-leading Innovative Graduate Study Program for Materials Research, Information, and Technology (MERIT-WINGS) of the University of Tokyo.
T.K. is also supported by JSPS KAKENHI Grant No. JP24KJ0611.
K.F. acknowledges support from JSPS KAKENHI (Grant Nos. JP23K13036 and JP24H00831).
T.S. is supported by JSPS KAKENHI Grant No. JP19H05796 and JST CREST Grant No. JPMJCR20C1. 
T.S. is also supported by Institute of AI and Beyond of the University of Tokyo.
\end{acknowledgments}

%

\clearpage

\end{document}


\setcounter{section}{0}
\setcounter{equation}{0}
\setcounter{figure}{0}
\setcounter{table}{0}
\setcounter{page}{1}
\renewcommand{\thesection}{S\arabic{section}}
\renewcommand{\theequation}{S\arabic{equation}}
\renewcommand{\thefigure}{S\arabic{figure}}
\renewcommand{\thetable}{S\arabic{table}}
\renewcommand{\bibnumfmt}[1]{[S#1]}
\renewcommand{\citenumfont}[1]{S#1}
\begin{center}
	\Large
	\textbf{Supplemental Material for ``Finite-time thermodynamic bounds and tradeoff relations for information processing''}
\end{center}


\maketitle

\section{Supplemental to Section III}
\label{sec: supplement of single subsystem}
Here, we supplement the proofs and discussions omitted in Sec. \blue{III}.

\subsection{Correspondence between Transition Rates and Transport Matrices}
\label{subsec: transition and transport}
We consider an infinitesimal time evolution from $\ptt$ to $\pdd$, where $\calW_E(\pdd,\ptt)=\calO(dt)$ and terms of $o(dt)$ are neglected.
The transition rates $R:\ptt\rightarrow\pdd$ that realize this time evolution satisfy Eq.~(\blue{2}).
These transition rates have nonzero values only on $E$.
In this subsection, we explain the procedure for establishing a correspondence between the transport matrix within $\calU(\pdd,\ptt)$ and the transition rates that realize the time evolution $\ptt\rightarrow\pdd$ following \ccite{dechant2022minimum}.
As will be shown later, this correspondence allows for the optimization of the transition rates to be equivalent to the optimization of the transport matrices when minimizing the partial activity rate.
Here, we use the cost function $d_E$, but the same applies when using $d_{E_X}, d_{E_X/E}$, and $d_{E_X:E_Y}^s$.

First, we construct the transition rates from the transport matrix.
We choose an arbitrary transport matrix $\hat{\Pi}$ that is an element of $\calU(\pdd,\ptt)$, where the off-diagonal elements are $\calO(dt)$.
Given that $\calW_E(\pdd,\ptt) = \calO(dt)$, such a transport matrix always exists.
In general, $\hat{\Pi}$ has nonzero values for elements that are not on $E$, making it unsuitable for direct conversion to transition rates.
Therefore, we transform the transport matrix so that it belongs to $\calU(\pdd,\ptt)$ and retains the transport cost concerning $d_E$, while only having nonzero values on $E$ and the diagonal elements.
To achieve this, we repeatedly apply the following procedures for all $\hat{\Pi}(r, r') > 0$ where $[r, r'] \not\in E$ (i.e., $d_E(r, r') > 1$).
0. Select a path $r' = r_0 \rightarrow r_1 \rightarrow \cdots \rightarrow r_m = r$ that connects node $r'$ to $r$ and gives the minimal $l_E=m$.
\begin{align}
    &1.\ \hat{\Pi}(r_{i+1},r_i)\rightarrow\hat{\Pi}(r_{i+1},r_i)+\hat{\Pi}(r,r')\ (i=0,\cdots,m-1)\nonumber.\\
    &2.\ \hat{\Pi}(r_i,r_i)\rightarrow\hat{\Pi}(r_i,r_i)-\hat{\Pi}(r,r')\ (i=1,\cdots,m-1)\nonumber.\\
    &3.\ \hat{\Pi}(r,r')\rightarrow0 \nonumber.
\end{align}

In procedure (1), the transport $\hat{\Pi}(r,r')>0$ that cannot be performed on the edge is divided along the shortest path from node $r'$ to $r$. Since $d_E$ is linear, the transport cost does not change in this procedure. 
In procedure (2), at nodes along the shortest path, the probability of not transporting (diagonal elements) is reduced by the amount transported to the next node.
The transport matrix transformed by such procedures is denoted as $\Pi$, which satisfies $\Pi \in \calU(\pdd,\ptt)$. Moreover, the transport cost also remains unchanged: $\sum_{r,r'}d_E(r,r')\hat{\Pi}(r,r')=\sum_{r,r'}d_E(r,r')\Pi(r,r')$.

In general, the shortest path between nodes is not unique, so $\Pi$ generated from $\hat{\Pi}$ is also not necessarily unique.
Moreover, in procedure (2), if the diagonal element of an intermediate node is originally zero, this may become negative after the transformation.
In this case, a small amount $\epsilon$ is in advance allocated to the probability of the intermediate node.
Since the off-diagonal elements are $\calO(dt)$, the non-negativity of the transport matrix is preserved as $dt \rightarrow 0$.
Simultaneously, by maintaining $\epsilon \gg dt$ while letting $\epsilon \rightarrow 0$, the Wasserstein distance remains unaffected.
When transforming a general transport matrix, the transport amount $\hat{\Pi}(r, r')$ is not $\calO(dt)$, so this method cannot be used, and the diagonal elements may become negative.
Therefore, to avoid this problem in the case of finite-time evolution, the transformation should be applied to the transport matrix for the infinitesimal time evolution at each time step.

Given that $\Pi$ has nonzero values only on the edges of $E$, we can construct the transition rates $R: \ptt \rightarrow \pdd$ based on this:
\begin{align}
\label{P to W}
    R(r,r')=
    \begin{cases}
     \frac{\Pi(r,r')}{\ptt(r')dt}\ &([r,r']\in E) \\
     0\ &([r,r']\not\in E)  \\
    \end{cases}.
\end{align}
By doing so, $R$ realizes the time evolution from $\ptt$ to $\pdd$.


Conversely, we now construct the transport matrix from the transition rate. Choose arbitrary transition rates $R: \ptt \rightarrow \pdd$. Define the transport matrix $\Pi$ as follows:
\begin{align}
    \label{W to P}
    \Pi(r,r')=
    \begin{cases}
     R(r,r')\ptt(r')dt\ &([r,r']\in E) \\
     0\ &([r,r']\not\in E) \\
     (1-\sum_{r''(\neq r)}W(r'',r)dt)\ptt(r)\ &(r=r')
    \end{cases}.
\end{align}
As $dt \rightarrow 0$, the diagonal elements are guaranteed to be nonnegative, ensuring that $\Pi \in \calU(\pdd, \ptt)$.
We note that the transport matrix $\Pi$ constructed in this manner has nonzero values only on $E$ and the diagonal elements.

\subsection{Explicit Form of Optimal Protocols}
\label{subsec: apdx explicit form}
Here, we explicitly provide the optimal protocols discussed in the main text.
We only show the case of the total system (Sec. \blue{II C}) for simplicity following \ccite{dechant2022minimum}, but the protocols for subsystems (Sec. \blue{III}, \blue{IV}) can be constructed in a similar way.

The transition rates $R$ that realize the minimum value in Eq.~(\blue{7}) can be expressed using the optimal transport matrix $\Pi^* \in \calU(\pdd, \ptt)$ with respect to $d_E$ as $R(r, r') \ptt(r') dt = \Pi^*(r, r')$ for $[r, r'] \in E$.
However, in general, $\Pi^*$ obtained by solving the optimal transport problem may have nonzero values on edges not included in $E$.
Therefore, it is necessary to transform $\Pi^*$ so that it only has values on $E$, allowing it to be converted into the transition rates (see \sref{subsec: transition and transport}).
For the time-integrated case (Eq.~(\blue{8})), the optimal protocol $\{R\}_{0 \leq t \leq \tau}$ can be expressed using the optimal transport matrix within $\calU(\pff, \poo)$ with respect to $d_E$.

The transition rates $R$ that achieve the minimum value of Eq.~(\blue{9}) can be expressed using the optimal transport matrix $\Pi^*\in\calU(\pdd,\ptt)$ with respect to $d_E$ as follows:
\begin{align}
    \label{W for min EP rate}
    R(r,r')\ptt(r')dt=
    \begin{cases}
        \frac{1}{1-e^{-F}}\Pi^*(r,r')\ \ (\Pi^*(r,r')\geq0, \Pi^*(r',r)=0)\\
        \frac{1}{e^F-1}\Pi^*(r',r)\ \ (\Pi^*(r,r')=0, \Pi^*(r',r)>0)
    \end{cases}.
\end{align}
Note that due to the unidirectional nature of optimal transport, it is never the case that both $\Pi^*(r,r')>0$ and $\Pi^*(r',r)>0$.
Here, $F(\geq0)$ is the thermodynamic force driving the transition on edge $E$, given by $F=2\tanh^{-1}(\dot{\calL}_{E}/\dot{A}_{XY})$.
From the perspective of transport, the thermodynamic force characterizes the unidirectionality of transport on edge $E$.
For the time-integrated case (Eq.~(\blue{10})), the optimal transport matrix concerning $d_E$ within $\calU(\pff,\poo)$ is used.
In this case, the thermodynamic force should be taken to be a constant value $F=2\tanh^{-1}({\calW}_{E}/{A}_{XY})$ at each time step.

\subsection{Remarks on the Activity Bound}
Let us provide some additional remarks regarding the upper bounds on activity.
These remarks are also applicable to the minimization in Sec. \blue{III}.
In Sec. \blue{II C 2}, we set an upper bound on the activity (rate) to minimize the EP (rate).
In the absence of such constraints, it is known that EP can be reduced to zero in discrete systems \cite{Muratore2013heat}.
However, this does not imply that dissipation can be reduced to zero in finite time.
In discrete systems, if there is no limit on activity, the transition rates can be increased indefinitely, effectively achieving a quasistatic limit.
Therefore, imposing an upper bound on activity allows us to consider the dynamics over a finite timescale.

On the other hand, it is possible to impose other physical constraints that compete with EP.
In \ccite{Van-Saito-unification2023}, the dynamical state mobility is fixed (although the optimal protocol is equivalent to that of activity).
In continuous systems, since the diffusion coefficient, which provides the timescale, is typically not controlled, there exists a finite lower bound on EP even without other constraints \cite{Aurell2011Wasserstein,Aurell2012refined}.
\ccite{proesmans2023precision} further imposes a constraint that reduces the deviation from a target trajectory in continuous systems.

Moreover, if the realized activity does not reach its upper bound in a protocol that provides the desired time evolution, it is possible to increase the activity to reduce the EP.
Therefore, when discussing the minimum EP, it can be assumed that the upper bound of activity is achieved.
For simplicity, we abuse the notation and use ${A}_{XY} (\dot{A}_{XY})$ to refer to both the upper bound of activity (rate) and the activity (rate) itself.

\subsection{Decomposition of Time Evolution}
The master equation (\blue{2}) examines the probability distribution of $X$ and $Y$ at the same time, but it is also possible to consider probability distributions where only $X$ or $Y$ has partially evolved.
Utilizing the bipartite nature, the probability $\pdt(x,y)$ for $X_{t+dt}=x, Y_{t}=y$ can be expressed as
\begin{align}
\label{apdx: partial evolution of X}
    \pdt(x,y)&=\left(1-\sum_{x'(\neq x)}R_{x'x}^{y}dt-\sum_{y'(\neq y)}R_{x}^{y'y}dt\right)\ptt(x,y)+\sum_{x'(\neq x)}R_{xx'}^{y}\ptt(x',y)dt
    +\sum_{y'(\neq y)}R_{x}^{y'y}\ptt(x,y)dt\nonumber\\
    &=\ptt(x,y)+\sum_{x'(\neq x)}J_{x'x}^{y}dt.
\end{align}
Hereafter, we will ignore the $o(dt)$ terms.
Similarly, for $\ptd(x,y)$, we have
\begin{align}
\label{apdx: partial evolution of Y}
\ptd(x,y)=\ptt(x,y)+\sum_{y'(\neq y)}J_{x}^{y'y}dt.
\end{align}
Combining these with the master equation (\blue{2}), we get
\begin{align}
    \label{decomp of time evolution}
    {\pdd(r)-\pdt(r)}={\ptd(r)-\ptt(r)}.
\end{align}
This indicates that when the infinitesimal time evolution $\ptt\rightarrow\pdd$ is given and we determine one of the partial time evolutions $\pdt$ or $\ptd$, the other is automatically determined.
If we specify transition rates that realize the infinitesimal time evolution $\ptt\rightarrow\pdd$, the partial time evolutions \esref{apdx: partial evolution of X}\eqref{apdx: partial evolution of Y} are determined.
However, from $\ptt\rightarrow\pdd$ alone, these cannot be determined.

Furthermore, for any subset $E'$ of $E$ and any probability distributions $p_1, q_1, p_2, q_2$ on $N$, the following holds:
\begin{align}
    \label{Wasserstein equivalence general}
    \begin{cases}
        \calW_{E'}(p_1,q_1)=\calO(dt),\calW_{E'}(p_2,q_2)=\calO(dt),\\
        p_1(r)-q_1(r)=p_2(r)-q_2(r)\ (\forall r\in N)
    \end{cases}\Longrightarrow\calW_{E'}(p_1,q_1)=\calW_{E'}(p_2,q_2).
\end{align}
This can be shown as follows.
Let $\Pi^*$ be the optimal transport matrix concerning $d_{E'}$ within $\calU(p_1, q_1)$. 
We define a new transport matrix $\Pi$ as
\begin{align}
    \Pi(r,r')=
    \begin{cases}
    \Pi^*(r,r')\ &(r\neq r')\\
    q_2(r)-\sum_{r''(\neq r)}\Pi^*(r'',r)
    \ &(r=r')
    \end{cases}.
\end{align}
Since the transport amount on the edge is $\calO(dt)$, the diagonal elements are nonnegative.
Thus, we have
\begin{align*}
    \sum_{r'}\Pi(r,r')&=
    \sum_{r'(\neq r)}\Pi^*(r,r')
    +q_2(r)-\sum_{r''(\neq r)}\Pi^*(r'',r)\\
    &=\left(p_1(r)-\Pi^*(r,r)\right)+q_2(r)+\left(\Pi^*(r,r)-q_1(r)\right)\\
    &=p_2(r).
\end{align*}
In the second equality, we used the fact that $\Pi^* \in \calU(p_1, q_1)$, and in the third equality, we used $p_1(r) - q_1(r) = p_2(r) - q_2(r)$.
Therefore, $\Pi \in \calU(p_2, q_2)$, but it is not necessarily optimal with respect to $d_{E'}$. Thus we get
\begin{align*}
    \calW_{E'}(p_2,q_2)\leq
    \sum_{r,r'}d_{E'}(r,r')\Pi(r,r')=\sum_{r,r'}d_{E'}(r,r')\Pi^*(r,r')=\calW_{E'}(p_1,q_1).
\end{align*}
The reverse inequality can be shown in exactly the same way, thus establishing \eref{Wasserstein equivalence general}.
Combining this with \eref{decomp of time evolution}, we obtain
\begin{align}
    \label{Wasserstein equivalence partial}
    \calW_{E_X}(\pdt,\ptt)=\calW_{E_X}(\pdd,\ptd),\calW_{E_Y}(\ptd,\ptt)=\calW_{E_Y}(\pdd,\pdt).
\end{align}

Now, in addition to the overall time evolution $\ptt\rightarrow\pdd$, we fix one partial time evolution $\ptt\rightarrow\pdt$ such that $\calW_{E_X}(\pdt,\ptt) = \calO(dt)$ and $\calW_{E_Y}(\ptd,\ptt) = \calO(dt)$.
Regarding these Wasserstein distances, the following inequality holds:
\begin{align}
    \label{EX EY triangle}
    \calW_{E_X}(\pdt,\ptt)+\calW_{E_Y}(\ptd,\ptt)\geq\calW_{E}(\pdd,\ptt).
\end{align}
The equality holds when the optimal transport matrix concerning $d_E$ in $\calU(\pdd, \ptt)$ is converted into transition rates, and $\pdt$ and $\ptd$ are given by the partial time evolutions based on their $X$ and $Y$ components.

To prove \eref{EX EY triangle}, we start by considering the optimal transport matrix concerning $d_{E_X}$ in $\calU(\pdt,\ptt)$, transformed to have nonzero values on the edges in $E_X$.
We denote this matrix as $\Pi^\star_X$, where the elements on $E_Y$ are zero.
Similarly, we denote the corresponding matrix for $Y$ as $\Pi^\star_Y$.
From these transport matrices, we construct the overall transport matrix as
\begin{align}
    \label{Pi from Pi_X and Pi_Y}
    \Pi(r,r')=
    \begin{cases}
    \Pi^\star_X(r,r')\ &([r,r']\in E_X)\\
    \Pi^\star_Y(r,r')\ &([r,r']\in E_Y)\\
    \ptt(r)-\sum_{r''(\neq r)}\Pi^\star_X(r'',r) &\\
    -\sum_{r''(\neq r)}\Pi^\star_Y(r'',r)
    \ &(r=r')\\
    0\ &({\rm otherwise})
    \end{cases}.
\end{align}
Since the transport amount on the edges is $\calO(dt)$, the diagonal elements are nonnegative.
From \eref{decomp of time evolution}, we see that $\Pi \in \calU(\pdd,\ptt)$, but it is not necessarily optimal with respect to $d_E$.
Therefore, we have
\begin{align*}
    \calW_{E_X}(\pdt,\ptt)+\calW_{E_Y}(\ptd,\ptt)&=\sum_{r,r'}d_{E_X}(r,r')\Pi^\star_X(r,r')
    +\sum_{r,r'}d_{E_Y}(r,r')\Pi^\star_Y(r,r')\\
    &=\sum_{r,r'}d_{E}(r,r')\Pi(r,r')\\
    &\geq \calW_E(\pdd,\ptt).
\end{align*}
In the second equality, we used the fact that when there is a transport on the edges in $E_X$ ($E_Y$), $d_E = d_{E_X} = 1$ ($d_E = d_{E_Y} = 1$).
The equality holds only when $\Pi$ is the optimal transport matrix with respect to $d_E$.
Hence, if we denote the partial time evolutions induced by this optimal transport matrix as $\ptt \rightarrow \pdt^*, \ptd^*$, the following equality
\begin{align}
    \calW_{E_X}(\pdt^*,\ptt)+\calW_{E_Y}(\ptd^*,\ptt)=\calW_{E}(\pdd,\ptt)
\end{align}
holds.
In general, the optimal transport matrix is not unique, and $X$ and $Y$ components of the transport cost, $\calW_{E_X}(\pdt^*,\ptt)$ and $\calW_{E_Y}(\ptd^*,\ptt)$, can vary.
However, their sum always matches the overall optimal transport cost (see Fig.~\blue{6}(b)).

\subsection{Proof of Eqs.~(\blue{14})(\blue{18})}
\label{subsec: proof of min AX rate}
In addition to the overall infinitesimal time evolution $\ptt \rightarrow \pdd$, the partial time evolution $\ptt \rightarrow \pdt$ is fixed.
For any transition rates $R_X: \ptt \rightarrow \pdt$, we use the transformation \eqref{W to P} to convert it into a transport matrix $\Pi_X \in \calU(\pdt, \ptt)$.
The activity rate of $X$ can be expressed as
\begin{align}
    \label{activity rate equivalence}
    \dot{A}_{X}=\sum_{x\neq x',y}R_{xx'}^{y}\ptt(x',y)=\frac{1}{dt}\sum_{r\neq r'}\Pi_X(r,r')=\frac{1}{dt}\sum_{r,r'}d_{E_X}(r,r')\Pi_X(r,r').
\end{align}
In this transformation, we utilized the fact that $\Pi_X$ only takes nonzero values on $E_X$ (where $d_{E_X}=1$) or on the diagonal elements (where $d_{E_X}=0$).
On the other hand, any transport matrix in $\calU(\pdd, \ptt)$ that has off-diagonal elements of $\calO(dt)$ (as matrices without this property cannot be optimal transport matrices) can be converted into transition rates that realize $\ptt \rightarrow \pdt$ using the transformation \eqref{P to W}.
Thus, \eref{activity rate equivalence} holds.
Therefore, the local minimum of the partial activity rate is given by
\begin{align}
    \label{min AX local proof}
    \min_{R_X:\ptt\rightarrow\pdt}\dot{A}_{X}=\min_{\Pi\in\calU(\pdt,\ptt)}\frac{1}{dt}\sum_{r,r'}d_{E_X}(r,r')\Pi_X(r,r')=\frac{1}{dt}\calW_{E_X}(\pdt,\ptt),
\end{align}
demonstrating Eq.~(\blue{14}).

Next, we prove Eq.~(\blue{18}).
We will show that for a general $\pdt$ satisfying $\dot{\calL}_{E_X} < \infty$ and $\dot{\calL}_{E_Y} < \infty$, the inequality $\dot{\calL}_{E_X/E} \leq \dot{\calL}_{E_X}$ holds, and that equality can be achieved with a suitable choice of $\pdt$.
Let $\Pi^\star_X$ be the optimal transport matrix concerning $d_{E_X}$ in $\calU(\pdt, \ptt)$.
Also, let $\Pi_Y$ be a transport matrix in $\calU(\ptd, \ptt)$ that can be transformed on the edges only in $E_Y$ (though it is not necessarily optimal concerning $d_{E_Y}$).
The matrices $\Pi^\star_X$ and $\Pi_Y$ are transformed on the edges in $E_X$ and $E_Y$, respectively.
We construct the overall transport matrix $\Pi \in \calU(\pdd, \ptt)$ from these transport matrices in the same manner as \eref{Pi from Pi_X and Pi_Y}.
Thus, we have
\begin{align}
    \calW_{E_X}(\pdt,\ptt)=&
    \sum_{[r,r']\in E_X}d_{E_X}(r,r')\Pi^\star_X(r,r')\nonumber\\
    =&\sum_{[r,r']\in E_X}d_{E_X/E}(r,r')\Pi^\star_X(r,r')+\sum_{[r,r']\in E_Y}d_{E_X/E}(r,r')\Pi_Y(r,r')\nonumber\\
    =&\sum_{r,r'}d_{E_X/E}(r,r')\Pi(r,r')\nonumber\\
    \geq&\calW_{E_X/E}(\pdd,\ptt).
\end{align}
In the second equality, we used the fact that $d_{E_X/E} = 0$ for $[r, r'] \in E_Y$.
This inequality holds for any $\pdt$ satisfying $\dot{\calL}_{E_X} < \infty$ and $\dot{\calL}_{E_Y} < \infty$.
Equality is achieved if and only if $\Pi$ is the optimal transport matrix concerning $d_{E_X/E}$ in $\calU(\pdd, \ptt)$.
Combining this with \eref{min AX local proof}, when this optimal transport matrix is converted into transition rates and the system is evolved partially in time according to the $X$ and $Y$ components, the global minimum of the activity rate of $X$ is realized (as stated in Eq.~(\blue{18})).

\subsection{Marginalization}
\label{subsec: marginalization}
\begin{figure}
    \centering
    \includegraphics[width=0.5\linewidth]{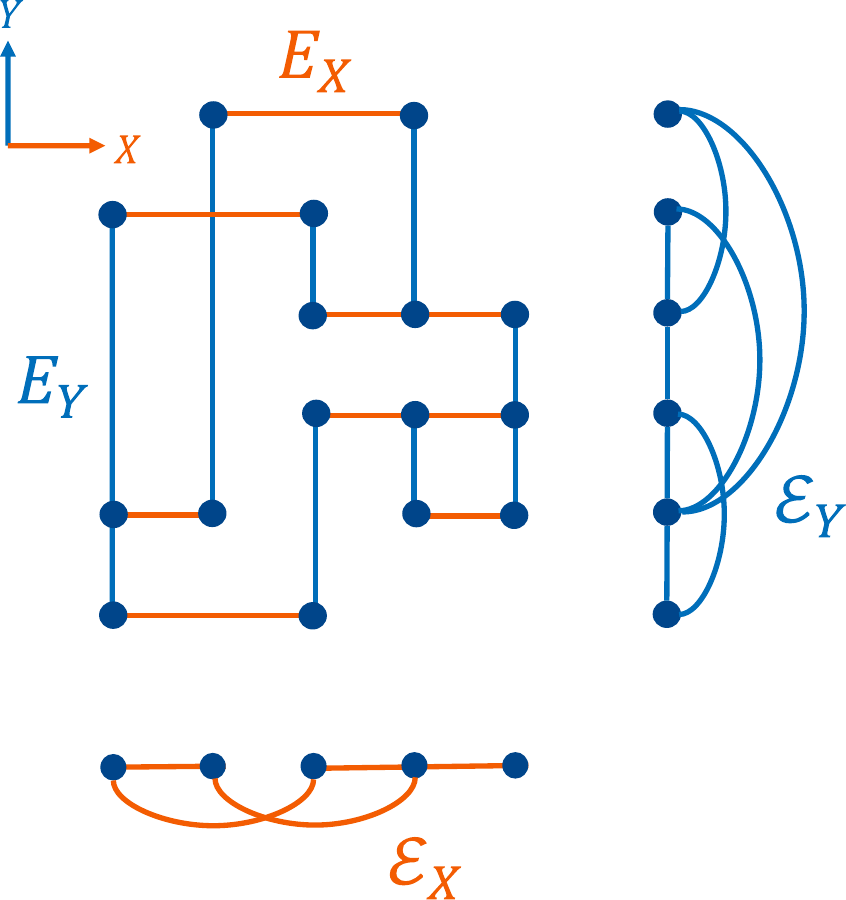}
    \caption{\label{fig: marginal_graph}
    Marginalization of the graph.
    }
\end{figure}
Here, we consider the dynamics of only one subsystem by tracing out the information of the other subsystem.
We observe that the lower bounds on activity and EP obtained in this manner are no longer tight and cannot achieve equality in general.
In this section, we primarily address the marginalization to $X$, but the same applies when marginalizing to $Y$.

As a preparation, we construct the marginalized graph $G(\calN_X,\calE_X)$ from the original graph $G(N,E)$ as shown in \fref{fig: marginal_graph}.
Formally, the set of nodes is defined as $\calN_X := \{x | \exists y \ \text{s.t.} \ (x,y) \in N \}$, and the set of edges is defined as $\calE_X := \{[x,x'] | \exists y \ \text{s.t.} \ [(x,y),(x',y)] \in E_X \}$.
From the time evolution of the total system $\ptt \rightarrow \pdd$, the time evolution of the subsystem $\pxt \rightarrow \pxd$ is uniquely determined.
This is realized by the transition rates on $\calE_X$.

On the other hand, let us fix a partial time evolution $\ptt \rightarrow \pdt$.
Denote the optimal transport matrix concerning $d_{E_X}$ within $\calU(\pdt, \ptt)$, which has nonzero values on the edges $E_X$, as $\Pi^\star_X$.
The transport matrix obtained by tracing out the $Y$ component of $\Pi^\star_X$ is denoted as $\pi_X$.
Specifically, this transport matrix is defined as $\pi_X(x, x') := \sum_y \Pi^\star_X((x, y), (x', y))$.
A simple calculation shows that $\pi_X \in \calU(\pxd, \pxt)$, but it is not necessarily optimal with respect to the cost function $d_{\calE_X}$.
Therefore,
\begin{align}
    \label{marginal Wasserstein ineq}
    \calW_{E_X}(\pdt,\ptt)&=\sum_{r,r'}d_{E_X}(r,r')\Pi_X^\star(r,r')\nonumber\\
    &=\sum_{x,x',y}d_{E_X}((x,y),(x',y))\Pi^\star_X((x,y),(x',y)) \nonumber\\
    &=\sum_{x,x'}d_{\calE_X}(x,x')\pi_X(x,x')\nonumber\\
    &\geq\calW_{\calE_X}(\pxd,\pxt).
\end{align}
In the third equality, we used the fact that the summand is nonzero only when $[(x, y), (x', y)] \in E_X$, at which point $d_{E_X}((x, y), (x', y)) = d_{\calE_X}(x, x') = 1$.
The equality in \eref{marginal Wasserstein ineq} holds if and only if $\pi_X$ is the optimal transport matrix.
However, the constraints after marginalization do not need to consider the correlation with $Y$, so in general, there exists a transport matrix more optimal than $\pi_X$ concerning $d_{\calE_X}$.
Therefore, the equality generally cannot be achieved.

The inequality \eqref{marginal Wasserstein ineq} holds for any $\pdt$ that satisfies $\dot{\calL}_{E_X} < \infty$ and $\dot{\calL}_{E_Y} < \infty$.
This includes the partial time evolution that gives $\dot{\calL}_{E_X/E}$.
Therefore, denoting $\dot{\calL}_{\calE_X} := \calW_{\calE_X}(\pxd, \pxt) / dt$, we have $\dot{\calL}_{E_X/E} \geq \dot{\calL}_{\calE_X}$.
Combining this with Eq.~(\blue{20}), for an infinitesimal time evolution $\ptt \rightarrow \pdd$, when the upper bound of the partial activity rate of $X$ is $\dot{A}_X$, the lower bound of the minimum partial EP rate is given by:
\begin{align}
    \label{mg EP rate lower}
    \min_{\substack{R:\ptt\rightarrow\pdd\\ \dot{A}_{X} \textrm{ fixed}}}
    \dot{\Sigma}_X\geq
    2\dot{\calL}_{\calE_X} \tanh^{-1}\left(\frac{\dot{\calL}_{\calE_X}}{\dot{A}_X}\right).
\end{align}
This lower bound does not use information about the correlation with $Y$.

In addition, this lower bound can be obtained even when considering the marginalized dynamics from the beginning.
Using the transition rates $R$ that realize the infinitesimal time evolution $\ptt \rightarrow \pdd$, we define the marginalized transition rates for $X$ as $\calR_{xx'} := \sum_y R_{xx'}^{y} \ptt(x', y) / \pxt(x')$.
These transition rates realize the infinitesimal time evolution $\pxt \rightarrow \pxd$.
Furthermore, its activity rate matches the partial activity rate of $X$ given by the original $R$: $\dot{A}_X = \sum_{x \neq x'} \calR_{xx'} \pxt(x')$.
Regarding the partial EP rate, by the log-sum inequality, we have
\begin{align}
    \dot{\Sigma}_X&=\sum_{x>x',y}\left(R_{xx'}^{y}\ptt(x',y)-R_{x'x}^{y}\ptt(x,y)\right)\ln\frac{R_{xx'}^{y}\ptt(x',y)}{R_{x'x}^{y}\ptt(x,y)}\nonumber\\
    &\geq\sum_{x>x'}\left(\calR_{xx'}\pxt(x')-\calR_{x'x}\pxt(x)\right)\ln\frac{\calR_{xx'}\pxt(x')}{\calR_{x'x}\pxt(x)}.
\end{align}
The rightmost term represents the EP rate generated by the transition rates $\calR$ during the infinitesimal time evolution $\pxt \rightarrow \pxd$.
Applying the discussion from Sec. \blue{II C} to this time evolution, the minimum value of the rightmost term is given by the right-hand side of \eref{mg EP rate lower}.

\section{Supplemental to Section {IV}}
\label{sec: supplement of tradeoffs}
\subsection{Definition of the Pareto Front}
\label{subsec: def of Pareto front}
We define the Pareto front $\calF_C$ for the feasible region $K_C (\neq \emptyset)$, which is the set of all feasible cost function pairs $C = (C_X, C_Y)$.
A pair $(C_X, C_Y)$ is said to dominate another pair $(C_X', C_Y')$ if either $C_X < C_X', C_Y \leq C_Y'$, or $C_X \leq C_X', C_Y < C_Y'$.
This is denoted as $(C_X, C_Y) \prec (C_X', C_Y')$. 
Conversely, if $(C_X', C_Y')$ is not dominated by $(C_X, C_Y)$, it is denoted as $(C_X, C_Y) \not\prec (C_X', C_Y')$.
The Pareto front $\calF_C$ is the set of elements in $K_C$ that are not dominated by any other element in $K_C$. Specifically, it is defined as:
\begin{align}
    \label{def of Pareto front}
    \calF_C:=\{&(C_X,C_Y)\in K_C|\forall(C_X',C_Y')\in K_C,
    (C_X',C_Y')\not\prec(C_X,C_Y)
    \}
\end{align}
(see Fig.~\blue{6}(a)).
When comparing an element on this front with any other element in $K_C$, if one cost can be decreased, the other cost must inevitably increase.
The same applies to sets of three or more dimensions.

\subsection{Minimization of Convex Combination of Partial Activities}
\label{subsec: Minimal partial activity convex}
The infinitesimal time evolution $\ptt \rightarrow \pdd$ of the total system is fixed.
For any transition rates $R: \ptt \rightarrow \pdt$, we use the transformation \eqref{W to P} to convert it into a transport matrix $\Pi \in \calU(\pdd, \ptt)$.
The convex combination of the partial activity rates can be expressed as:
\begin{align}
    \label{convex activity rate equivalence}
    s\dot{A}_{X}+(1-s)\dot{A}_{Y}
    &=s\sum_{x\neq x',y}R_{xx'}^{y}\ptt(x',y)
    +(1-s)\sum_{y\neq y',x}R_{x}^{yy'}\ptt(x,y)\nonumber\\
    &=s\frac{1}{dt}\sum_{[r,r']\in E_X}\Pi(r,r')
    +(1-s)\frac{1}{dt}\sum_{[r,r']\in E_Y}\Pi(r,r')\nonumber\\
    &=\frac{1}{dt}\sum_{r,r'}d_{E_X:E_Y}^s(r,r')\Pi(r,r').
\end{align}
In this transformation, we used the fact that $\Pi$ takes nonzero values only on $E_X$ (where $d_{E_X:E_Y}^s=s$), $E_Y$ (where $d_{E_X:E_Y}^s=1-s$), and diagonal elements (where $d_{E_X:E_Y}^s=0$).
On the other hand, any transport matrix in $\calU(\pdd, \ptt)$, which has off-diagonal elements of order $\calO(dt)$, can be converted back into the transition rates $R$ that realize $\ptt \rightarrow \pdt$ using the transformation \eqref{P to W}. Thus, \eref{convex activity rate equivalence} holds.
Therefore, the minimum value of the convex combination of the partial activity rates is given by:
\begin{align}
    \label{min sAX+(1-s)AY proof}
    \min_{R:\ptt\rightarrow\pdd}[s\dot{A}_{X}+(1-s)\dot{A}_{Y}]
    &=\min_{\Pi\in\calU(\pdt,\ptt)}\frac{1}{dt}\sum_{r,r'}d_{E_X:E_Y}^s(r,r')\Pi(r,r')\nonumber\\
    &=\frac{1}{dt}\calW_{E_X:E_Y}^s(\pdd,\ptt).
\end{align}
Since $\calW_{E_X:E_Y}^s$ satisfies the triangle inequality, the minimum value of $s {A}_X + (1-s) {A}_Y$ is $\calW_{E_X:E_Y}^s(\pff, \poo)$ (cf. Eq.~(\blue{19})).

\subsection{Shape of $\calF_A$}
\label{subsec: shape of FA}
Similar to \eref{W for min EP rate}, the protocols that give $\calF_A$ can be constructed from the optimal transport matrices concerning $d_{E_X:E_Y}^s$ within $\calU(\pff, \poo)$.
Equation (\blue{27}) is a linear programming problem, and since the transport polytope $\calU(p, q)$ is bounded, an optimal solution always exists.
Furthermore, this solution can be expressed as a vertex of the transport polytope \cite{nocedal1999numerical}.
If the vertex solution has values outside $E$ (and diagonal elements), it can be transformed into a vertex solution that only has values on $E$.
Therefore, when seeking $\calF_A$, which considers the components of the transport cost, we only need to consider the vertex solutions that have values solely on $E$.

The shape of $\calF_A$ can be understood from the perspective of vertex solutions to the optimal transport problem.
For the interval of $s$ corresponding to a vertex (excluding the endpoints), the optimal transport problem's vertex solution remains the same, or there are multiple vertex solutions with the same transport cost components for $X$ and $Y$.
Conversely, for $s$ corresponding to an edge, multiple vertex solutions with different transport cost components exist, providing the vertices at both ends of the edge.
The convex combination of these vertex solutions also forms an optimal solution, covering all points on the edge.
Different vertices correspond to different vertex solutions.
The number of vertices is at most the number of vertices of the transport polytope (at most $|N|^{2(|N|-1)}$ \cite{dyer2003random}).
Therefore, $\calF_A$ will not have curves consisting of infinitely many edges and vertices.

\subsection{Convexity of $K_A$}
\label{subsec: Convexity of KA}
In the case of infinitesimal time evolution, we consider the feasible set of partial activity rates $K_{\dot{A}}$.
The same can be derived for finite time.
By lowering the partial EP rate, the partial activity rates for $X$ and $Y$ can be independently increased without bound, so we only need to focus on the lower left part of $K_{\dot{A}}$.
Consider any two points from $K_{\dot{A}}$.
These points can be represented as $(\dot{\calL}_{E_X}(\pdt), \dot{\calL}_{E_Y}(\pdt))$ and $(\dot{\calL}_{E_X}(\pdt'), \dot{\calL}_{E_Y}(\pdt'))$ because, by construction, for any point $(\dot{A}_X, \dot{A}_Y)$ in $K_{\dot{A}}$, there exist $\pdt, F_X \geq 0, F_Y \geq 0$ such that $\dot{A}_X = \coth(F_X/2)\dot{\calL}_{E_X}(\pdt)$ and $\dot{A}_Y = \coth(F_Y/2)\dot{\calL}_{E_Y}(\pdt)$.
The convex combination using $(\dot{\calL}_{E_X}(\pdt), \dot{\calL}_{E_Y}(\pdt))$ will dominate the convex combination using $(\dot{A}_X, \dot{A}_Y)$.
Therefore, when seeking the optimal pair of partial activity rates, we only need to consider $(\dot{\calL}_{E_X}(\pdt), \dot{\calL}_{E_Y}(\pdt))$.

To show that $K_{\dot{A}}$ is convex, we need to demonstrate that for any $\lambda \ (0 \leq \lambda \leq 1)$,
\begin{align}
    \label{activity rate convex condition}
    \left(\lambda\dot{\calL}_{E_X}(\pdt)+(1-\lambda)\dot{\calL}_{E_X}(\pdt'),\lambda\dot{\calL}_{E_Y}(\pdt)+(1-\lambda)\dot{\calL}_{E_Y}(\pdt')\right)\in K_{\dot{A}}.
\end{align}
The optimal transport matrices concerning $d_{E_X}$ in $\calU(\pdt, \ptt)$ and $\calU(\pdt', \ptt)$ are taken to yield a convex combination with coefficient $\lambda$.
This will be an element of $\calU(\lambda \pdt + (1 - \lambda) \pdt', \ptt)$, and the transport cost will be $\lambda \dot{\calL}_{E_X}(\pdt) + (1 - \lambda) \dot{\calL}_{E_X}(\pdt')$.
However, this is not necessarily optimal concerning $d_{E_X}$.
Therefore,
\begin{align}
    \label{activity rate X convexity}
    \lambda\dot{\calL}_{E_X}(\pdt)+(1-\lambda)\dot{\calL}_{E_X}(\pdt')\geq \dot{\calL}_{E_X}(\lambda\pdt+(1-\lambda)\pdt').
\end{align}
The same reasoning applies for $d_{E_Y}$, so
\begin{align}
    \label{activity rate Y convexity}
    \lambda\dot{\calL}_{E_Y}(\pdt)+(1-\lambda)\dot{\calL}_{E_Y}(\pdt')\geq \dot{\calL}_{E_Y}(\lambda\pdt+(1-\lambda)\pdt')
\end{align}
holds.
Thus, by adjusting the thermodynamic force, we can achieve \eref{activity rate convex condition}.

We focus on the edge of the Pareto front with the slope $-s/(1-s)$.
There exist multiple optimal transportation matrices for $\calW_{E_X:E_Y}^s(\pdd,\ptt)$ where the $X$ and $Y$ components of the transportation cost differ.
By selecting two of these transportation matrices and converting them into transition rates, we obtain probability distributions $\pdt$ and $\pdt'$ which represent the partial time evolution of $X$.
For any $\lambda$ such that $0 \leq \lambda \leq 1$, the equality conditions of \eref{activity rate X convexity} and \eqref{activity rate Y convexity} hold.
If they did not hold, it would imply that $s\dot{A}_X + (1-s)\dot{A}_Y$ could be made smaller than $\calW_{E_X:E_Y}^s(\pdd,\ptt)/dt$, which is a contradiction.
Therefore, by taking a convex combination in this manner, any point on the edge of the Pareto front can be realized.

\section{Nomenclature}
Here, we list and briefly explain the symbols that appear in this paper (see Table \ref{tab: nomenclature}).
\begin{table*}[htbp]
\caption{\label{tab: nomenclature}
Nomenclature. 
}
\begin{ruledtabular}
\begin{tabular}{c|l}
$	A_{XY},A_{X},A_{Y}	$	&	activity and upper bound of activity	\\
$	C=(C_X,C_Y)	$	&	cost function in multiobjective optimization	\\
$	\calC	$	&	free energy's dependence on ligand concentration	\\
$	d_{E},d_{E_X/E},d_{E_X:E_Y}^s	$	&	cost function in optimal transport problem	\\
$	E, E_{X}, E_{Y}	$	&	edge set	\\
$	\calE_{X},\calE_{Y}	$	&	marginalized edge set	\\
$	f_{X},f_{Y} 	$	&	Fermi distribution	\\
$	F, F_{X},F_{Y} 	$	&	thermodynamic force	\\
$	\calF_{C},\calF_{A},\calF_{\Sigma}	$	&	Pareto front	\\
$	G	$	&	graph	\\
$	I_{XY}	$	&	mutual information	\\
$	\dot{I}_{X},\dot{I}_{Y}	$	&	information flow	\\
$	J_{xx'}^{y},J_{x}^{yy'}	$	&	probability current	\\
$	k_B,k_R,k_Y	$	&	reaction rates in \textit{E.coli} chemotaxis	\\
$	K_{C},K_{A},K_{\Sigma}	$	&	feasible region	\\
$l_{E}(P)$  &	the number of edges in $E$ traversed by $P$\\
$	\calL_{E},\calL_{E_X},\calL_{E_Y},\calL_{E_X/E},\calL_{E_Y/E}	$	&	trajectory length measured by the Wasserstein distance	\\
$	N	$	&	node set	\\
$	\calN_{X}, \calN_{Y}	$	&	marginalized node set	\\
$	p,q	$	&	probability distribution	\\
$	\ptt,\pdd,\poo,\pff	$	&	probability distribution of the total system	\\
$	\pdt,\ptd	$	&	partially evolved probability distribution	\\
$	\pxt,\pxd,\pyt,\pyd	$	&	probability distribution of the subsystem	\\
$	\{\ptt\}_{0\leq t\leq\tau},\{\pdt\}_{0\leq t\leq\tau}	$	&	trajectory of time evolution	\\
$P$    &    path on graph\\
$	r	$	&	state of the total system	\\
$	R, R_{X}	$	&	transition rates	\\
$	\{R\}_{0\leq t\leq\tau}, \{R_{X}\}_{0\leq t\leq\tau}	$	&	protocol	\\
$	\calR	$	&	marginalized transition rate	\\
$	s	$	&	ratio in convex combination	\\
$	t	$	&	time	\\
$	\calT	$	&	total variation distance	\\
$	U	$	&	interaction energy between the quantum dots	\\
$	\calU	$	&	set of transport matrix	\\
$	\calW_{E}	$	&	Wasserstein distance (the optimal transport cost regarding $d_{E}$)	\\
$	\calW_{E_X/E}	$	&	Wasserstein pseudo distance (the optimal transport cost regarding $d_{E_X/E}$)	\\
$	\calW_{E_X:E_Y}^s	$	&	the optimal transport cost regarding $d_{E_X:E_Y}^s$	\\
$	x,y	$	&	state of the subsystem	\\
$	X,Y	$	&	subsystem	\\
$	\beta_{X},\beta_{Y}	$	&	inverse temperature of reservoir	\\
$	\gamma_{X},\gamma_{Y}	$	&	coupling strength to reservoir	\\
$	\epsilon	$	&	infinitesimal amount	\\
$	\varepsilon_{X}, \varepsilon_{Y}	$	&	energy level of quantum dot	\\
$	\lambda	$	&	Lagrange multiplier	\\
$	\mu_{X},\mu_{Y}	$	&	chemical potential of reservoir	\\
$	\pi	$	&	marginalized transport matrix	\\
$	\Pi	$	&	transport matrix	\\
$	\Sigma_{XY},\Sigma_{X},\Sigma_{Y}	$	&	entropy production	\\
$	{\Sigma}^{\rm ap}_{X},{\Sigma}^{\rm ap}_{Y}	$	&	apparent entropy production	\\
$	\tau	$	&	operation time of the process	\\
$	\chi	$	&	bias in methylation and demethylation reactions	\\
\end{tabular}
\end{ruledtabular}
\end{table*}

%

%% file: commands.tex
\newcommand{\todayd}{\the\year/\the\month/\the\day}

\newcommand{\bel}{\begin{easylist}}
\newcommand{\eel}{\end{easylist}}

\newcommand{\eref}[1]{Eq.~\eqref{#1}}
\newcommand{\esref}[1]{Eqs.~\eqref{#1}}

\newcommand{\fref}[1]{Fig.~\ref{#1}}
\newcommand{\sref}[1]{Sec.~\ref{#1}}
\newcommand{\ccite}[1]{Ref.~\cite{#1}}

\newcommand{\sumtwo}[2]%
{\mathop{\sum_{#1}}_{#2}}
\newcommand{\sumthree}[3]%
{\mathop{\mathop{\sum_{#1}}_{#2}}_{#3}}
\newcommand{\sumfour}[4]%
{\mathop{\mathop{\mathop{\sum_{#1}}_{#2}}_{#3}}_{#4}} 
\newcommand{\prodtwo}[2]%
{\mathop{\prod_{#1}}_{#2}}
\newcommand{\mintwo}[2]%
{\mathop{\min_{#1}}_{#2}}
\newcommand{\maxtwo}[2]%
{\mathop{\max_{#1}}_{#2}}
\newcommand{\maxthree}[3]%
{\mathop{\mathop{\max_{#1}}_{#2}}_{#3}}
\newcommand{\limtwo}[2]%
{\mathop{\lim_{#1}}_{#2}}
\newcommand{\suptwo}[2]%
{\mathop{\sup_{#1}}_{#2}}
\newcommand{\supthree}[3]%
{\mathop{\mathop{\sup_{#1}}_{#2}}_{#3}}
\newcommand{\supfour}[4]%
{\mathop{\mathop{\mathop{\sup_{#1}}_{#2}}_{#3}}_{#4}} 
\newcommand{\inftwo}[2]%
{\mathop{\inf_{#1}}_{#2}}
\newcommand{\infthree}[3]%
{\mathop{\mathop{\inf_{#1}}_{#2}}_{#3}}
\newcommand{\inffour}[4]%
{\mathop{\mathop{\mathop{\inf_{#1}}_{#2}}_{#3}}_{#4}} 

\newcommand\calC{{\mathcal C}}

\newcommand\calF{{\mathcal F}}

\newcommand\calL{{\mathcal L}}

\newcommand\calO{{\mathcal O}}

\newcommand\calT{{\mathcal T}}
\newcommand\calU{{\mathcal U}}

\newcommand\calW{{\mathcal W}}

\newcommand{\ptt}{p_{X_{t}Y_{t}}}
\newcommand{\pdt}{p_{X_{t+dt}Y_{t}}}
\newcommand{\ptd}{p_{X_{t}Y_{t+dt}}}
\newcommand{\pdd}{p_{X_{t+dt}Y_{t+dt}}}
\newcommand{\poo}{p^o_{XY}}
\newcommand{\pff}{p^f_{XY}}

\newcommand{\pxcy}{p_{X_{t}|Y_{t}}}
\newcommand{\pycx}{p_{Y_{t}|X_{t}}}
\newcommand{\pxt}{p_{X_{t}}}
\newcommand{\pyt}{p_{Y_{t}}}


%% file: main.bbl
\begin{thebibliography}{102}%
\makeatletter
\providecommand \@ifxundefined [1]{%
 \@ifx{#1\undefined}
}%
\providecommand \@ifnum [1]{%
 \ifnum #1\expandafter \@firstoftwo
 \else \expandafter \@secondoftwo
 \fi
}%
\providecommand \@ifx [1]{%
 \ifx #1\expandafter \@firstoftwo
 \else \expandafter \@secondoftwo
 \fi
}%
\providecommand \natexlab [1]{#1}%
\providecommand \enquote  [1]{``#1''}%
\providecommand \bibnamefont  [1]{#1}%
\providecommand \bibfnamefont [1]{#1}%
\providecommand \citenamefont [1]{#1}%
\providecommand \href@noop [0]{\@secondoftwo}%
\providecommand \href [0]{\begingroup \@sanitize@url \@href}%
\providecommand \@href[1]{\@@startlink{#1}\@@href}%
\providecommand \@@href[1]{\endgroup#1\@@endlink}%
\providecommand \@sanitize@url [0]{\catcode `\\12\catcode `\$12\catcode `\&12\catcode `\#12\catcode `\^12\catcode `\_12\catcode `\%12\relax}%
\providecommand \@@startlink[1]{}%
\providecommand \@@endlink[0]{}%
\providecommand \url  [0]{\begingroup\@sanitize@url \@url }%
\providecommand \@url [1]{\endgroup\@href {#1}{\urlprefix }}%
\providecommand \urlprefix  [0]{URL }%
\providecommand \Eprint [0]{\href }%
\providecommand \doibase [0]{https://doi.org/}%
\providecommand \selectlanguage [0]{\@gobble}%
\providecommand \bibinfo  [0]{\@secondoftwo}%
\providecommand \bibfield  [0]{\@secondoftwo}%
\providecommand \translation [1]{[#1]}%
\providecommand \BibitemOpen [0]{}%
\providecommand \bibitemStop [0]{}%
\providecommand \bibitemNoStop [0]{.\EOS\space}%
\providecommand \EOS [0]{\spacefactor3000\relax}%
\providecommand \BibitemShut  [1]{\csname bibitem#1\endcsname}%
\let\auto@bib@innerbib\@empty
\bibitem [{\citenamefont {Sekimoto}(2010)}]{sekimoto2010stochastic}%
  \BibitemOpen
  \bibfield  {author} {\bibinfo {author} {\bibfnamefont {K.}~\bibnamefont {Sekimoto}},\ }\href@noop {} {\emph {\bibinfo {title} {Stochastic energetics}}},\ Vol.\ \bibinfo {volume} {799}\ (\bibinfo  {publisher} {Springer},\ \bibinfo {year} {2010})\BibitemShut {NoStop}%
\bibitem [{\citenamefont {Seifert}(2012)}]{Seifert2012}%
  \BibitemOpen
  \bibfield  {author} {\bibinfo {author} {\bibfnamefont {U.}~\bibnamefont {Seifert}},\ }\bibfield  {title} {\bibinfo {title} {Stochastic thermodynamics, fluctuation theorems and molecular machines},\ }\href {https://doi.org/10.1088/0034-4885/75/12/126001} {\bibfield  {journal} {\bibinfo  {journal} {Rep. Prog. Phys.}\ }\textbf {\bibinfo {volume} {75}},\ \bibinfo {pages} {126001} (\bibinfo {year} {2012})}\BibitemShut {NoStop}%
\bibitem [{\citenamefont {Peliti}\ and\ \citenamefont {Pigolotti}(2021)}]{peliti2021stochastic}%
  \BibitemOpen
  \bibfield  {author} {\bibinfo {author} {\bibfnamefont {L.}~\bibnamefont {Peliti}}\ and\ \bibinfo {author} {\bibfnamefont {S.}~\bibnamefont {Pigolotti}},\ }\href@noop {} {\emph {\bibinfo {title} {Stochastic thermodynamics: an introduction}}}\ (\bibinfo  {publisher} {Princeton University Press},\ \bibinfo {year} {2021})\BibitemShut {NoStop}%
\bibitem [{\citenamefont {Ciliberto}(2017)}]{Ciliberto2017experiment-history}%
  \BibitemOpen
  \bibfield  {author} {\bibinfo {author} {\bibfnamefont {S.}~\bibnamefont {Ciliberto}},\ }\bibfield  {title} {\bibinfo {title} {Experiments in stochastic thermodynamics: Short history and perspectives},\ }\href {https://doi.org/10.1103/PhysRevX.7.021051} {\bibfield  {journal} {\bibinfo  {journal} {Phys. Rev. X}\ }\textbf {\bibinfo {volume} {7}},\ \bibinfo {pages} {021051} (\bibinfo {year} {2017})}\BibitemShut {NoStop}%
\bibitem [{\citenamefont {Curzon}\ and\ \citenamefont {Ahlborn}(1975)}]{Curzon-Ahlborn1975}%
  \BibitemOpen
  \bibfield  {author} {\bibinfo {author} {\bibfnamefont {F.~L.}\ \bibnamefont {Curzon}}\ and\ \bibinfo {author} {\bibfnamefont {B.}~\bibnamefont {Ahlborn}},\ }\bibfield  {title} {\bibinfo {title} {Efficiency of a carnot engine at maximum power output},\ }\href {https://doi.org/10.1119/1.10023} {\bibfield  {journal} {\bibinfo  {journal} {Am. J. Phys.}\ }\textbf {\bibinfo {volume} {43}},\ \bibinfo {pages} {22} (\bibinfo {year} {1975})}\BibitemShut {NoStop}%
\bibitem [{\citenamefont {van~den Broeck}(2005)}]{Broeck2005EMP}%
  \BibitemOpen
  \bibfield  {author} {\bibinfo {author} {\bibfnamefont {C.}~\bibnamefont {van~den Broeck}},\ }\bibfield  {title} {\bibinfo {title} {Thermodynamic efficiency at maximum power},\ }\href {https://doi.org/10.1103/PhysRevLett.95.190602} {\bibfield  {journal} {\bibinfo  {journal} {Phys. Rev. Lett.}\ }\textbf {\bibinfo {volume} {95}},\ \bibinfo {pages} {190602} (\bibinfo {year} {2005})}\BibitemShut {NoStop}%
\bibitem [{\citenamefont {Schmiedl}\ and\ \citenamefont {Seifert}(2007)}]{Schmiedl-Seifert2007optimal}%
  \BibitemOpen
  \bibfield  {author} {\bibinfo {author} {\bibfnamefont {T.}~\bibnamefont {Schmiedl}}\ and\ \bibinfo {author} {\bibfnamefont {U.}~\bibnamefont {Seifert}},\ }\bibfield  {title} {\bibinfo {title} {Optimal finite-time processes in stochastic thermodynamics},\ }\href {https://doi.org/10.1103/PhysRevLett.98.108301} {\bibfield  {journal} {\bibinfo  {journal} {Phys. Rev. Lett.}\ }\textbf {\bibinfo {volume} {98}},\ \bibinfo {pages} {108301} (\bibinfo {year} {2007})}\BibitemShut {NoStop}%
\bibitem [{\citenamefont {Crooks}(2007)}]{Crooks2007length}%
  \BibitemOpen
  \bibfield  {author} {\bibinfo {author} {\bibfnamefont {G.~E.}\ \bibnamefont {Crooks}},\ }\bibfield  {title} {\bibinfo {title} {Measuring thermodynamic length},\ }\href {https://doi.org/10.1103/PhysRevLett.99.100602} {\bibfield  {journal} {\bibinfo  {journal} {Phys. Rev. Lett.}\ }\textbf {\bibinfo {volume} {99}},\ \bibinfo {pages} {100602} (\bibinfo {year} {2007})}\BibitemShut {NoStop}%
\bibitem [{\citenamefont {Esposito}\ \emph {et~al.}(2010)\citenamefont {Esposito}, \citenamefont {Kawai}, \citenamefont {Lindenberg},\ and\ \citenamefont {Van~den Broeck}}]{Esposito-2010EMP-low}%
  \BibitemOpen
  \bibfield  {author} {\bibinfo {author} {\bibfnamefont {M.}~\bibnamefont {Esposito}}, \bibinfo {author} {\bibfnamefont {R.}~\bibnamefont {Kawai}}, \bibinfo {author} {\bibfnamefont {K.}~\bibnamefont {Lindenberg}},\ and\ \bibinfo {author} {\bibfnamefont {C.}~\bibnamefont {Van~den Broeck}},\ }\bibfield  {title} {\bibinfo {title} {Efficiency at maximum power of low-dissipation carnot engines},\ }\href {https://doi.org/10.1103/PhysRevLett.105.150603} {\bibfield  {journal} {\bibinfo  {journal} {Phys. Rev. Lett.}\ }\textbf {\bibinfo {volume} {105}},\ \bibinfo {pages} {150603} (\bibinfo {year} {2010})}\BibitemShut {NoStop}%
\bibitem [{\citenamefont {Sivak}\ and\ \citenamefont {Crooks}(2012)}]{Sivak-Crooks2012metrics}%
  \BibitemOpen
  \bibfield  {author} {\bibinfo {author} {\bibfnamefont {D.~A.}\ \bibnamefont {Sivak}}\ and\ \bibinfo {author} {\bibfnamefont {G.~E.}\ \bibnamefont {Crooks}},\ }\bibfield  {title} {\bibinfo {title} {Thermodynamic metrics and optimal paths},\ }\href {https://doi.org/10.1103/PhysRevLett.108.190602} {\bibfield  {journal} {\bibinfo  {journal} {Phys. Rev. Lett.}\ }\textbf {\bibinfo {volume} {108}},\ \bibinfo {pages} {190602} (\bibinfo {year} {2012})}\BibitemShut {NoStop}%
\bibitem [{\citenamefont {Blaber}\ and\ \citenamefont {Sivak}(2023)}]{Blaber-Sivak2023optimal}%
  \BibitemOpen
  \bibfield  {author} {\bibinfo {author} {\bibfnamefont {S.}~\bibnamefont {Blaber}}\ and\ \bibinfo {author} {\bibfnamefont {D.~A.}\ \bibnamefont {Sivak}},\ }\bibfield  {title} {\bibinfo {title} {Optimal control in stochastic thermodynamics},\ }\href@noop {} {\bibfield  {journal} {\bibinfo  {journal} {Journal of Physics Communications}\ }\textbf {\bibinfo {volume} {7}},\ \bibinfo {pages} {033001} (\bibinfo {year} {2023})}\BibitemShut {NoStop}%
\bibitem [{\citenamefont {Ma}\ \emph {et~al.}(2020)\citenamefont {Ma}, \citenamefont {Zhai}, \citenamefont {Chen}, \citenamefont {Sun},\ and\ \citenamefont {Dong}}]{Ma-2020exp-scaling}%
  \BibitemOpen
  \bibfield  {author} {\bibinfo {author} {\bibfnamefont {Y.-H.}\ \bibnamefont {Ma}}, \bibinfo {author} {\bibfnamefont {R.-X.}\ \bibnamefont {Zhai}}, \bibinfo {author} {\bibfnamefont {J.}~\bibnamefont {Chen}}, \bibinfo {author} {\bibfnamefont {C.~P.}\ \bibnamefont {Sun}},\ and\ \bibinfo {author} {\bibfnamefont {H.}~\bibnamefont {Dong}},\ }\bibfield  {title} {\bibinfo {title} {Experimental test of the $1/\ensuremath{\tau}$-scaling entropy generation in finite-time thermodynamics},\ }\href {https://doi.org/10.1103/PhysRevLett.125.210601} {\bibfield  {journal} {\bibinfo  {journal} {Phys. Rev. Lett.}\ }\textbf {\bibinfo {volume} {125}},\ \bibinfo {pages} {210601} (\bibinfo {year} {2020})}\BibitemShut {NoStop}%
\bibitem [{\citenamefont {Shiraishi}\ \emph {et~al.}(2018)\citenamefont {Shiraishi}, \citenamefont {Funo},\ and\ \citenamefont {Saito}}]{Shiraishi-Funo-Saito2018speed}%
  \BibitemOpen
  \bibfield  {author} {\bibinfo {author} {\bibfnamefont {N.}~\bibnamefont {Shiraishi}}, \bibinfo {author} {\bibfnamefont {K.}~\bibnamefont {Funo}},\ and\ \bibinfo {author} {\bibfnamefont {K.}~\bibnamefont {Saito}},\ }\bibfield  {title} {\bibinfo {title} {Speed limit for classical stochastic processes},\ }\href {https://doi.org/10.1103/PhysRevLett.121.070601} {\bibfield  {journal} {\bibinfo  {journal} {Phys. Rev. Lett.}\ }\textbf {\bibinfo {volume} {121}},\ \bibinfo {pages} {070601} (\bibinfo {year} {2018})}\BibitemShut {NoStop}%
\bibitem [{\citenamefont {Ito}(2018)}]{Ito2018geometry}%
  \BibitemOpen
  \bibfield  {author} {\bibinfo {author} {\bibfnamefont {S.}~\bibnamefont {Ito}},\ }\bibfield  {title} {\bibinfo {title} {Stochastic thermodynamic interpretation of information geometry},\ }\href {https://doi.org/10.1103/PhysRevLett.121.030605} {\bibfield  {journal} {\bibinfo  {journal} {Phys. Rev. Lett.}\ }\textbf {\bibinfo {volume} {121}},\ \bibinfo {pages} {030605} (\bibinfo {year} {2018})}\BibitemShut {NoStop}%
\bibitem [{\citenamefont {Ito}\ and\ \citenamefont {Dechant}(2020)}]{Ito-Dechant2020info}%
  \BibitemOpen
  \bibfield  {author} {\bibinfo {author} {\bibfnamefont {S.}~\bibnamefont {Ito}}\ and\ \bibinfo {author} {\bibfnamefont {A.}~\bibnamefont {Dechant}},\ }\bibfield  {title} {\bibinfo {title} {Stochastic time evolution, information geometry, and the cram\'er-rao bound},\ }\href {https://doi.org/10.1103/PhysRevX.10.021056} {\bibfield  {journal} {\bibinfo  {journal} {Phys. Rev. X}\ }\textbf {\bibinfo {volume} {10}},\ \bibinfo {pages} {021056} (\bibinfo {year} {2020})}\BibitemShut {NoStop}%
\bibitem [{\citenamefont {Falasco}\ and\ \citenamefont {Esposito}(2020)}]{Falasco-Esposito2020dissipation-time}%
  \BibitemOpen
  \bibfield  {author} {\bibinfo {author} {\bibfnamefont {G.}~\bibnamefont {Falasco}}\ and\ \bibinfo {author} {\bibfnamefont {M.}~\bibnamefont {Esposito}},\ }\bibfield  {title} {\bibinfo {title} {Dissipation-time uncertainty relation},\ }\href {https://doi.org/10.1103/PhysRevLett.125.120604} {\bibfield  {journal} {\bibinfo  {journal} {Phys. Rev. Lett.}\ }\textbf {\bibinfo {volume} {125}},\ \bibinfo {pages} {120604} (\bibinfo {year} {2020})}\BibitemShut {NoStop}%
\bibitem [{\citenamefont {Hamazaki}(2022)}]{Hamazaki2022speed}%
  \BibitemOpen
  \bibfield  {author} {\bibinfo {author} {\bibfnamefont {R.}~\bibnamefont {Hamazaki}},\ }\bibfield  {title} {\bibinfo {title} {Speed limits for macroscopic transitions},\ }\href {https://doi.org/10.1103/PRXQuantum.3.020319} {\bibfield  {journal} {\bibinfo  {journal} {PRX Quantum}\ }\textbf {\bibinfo {volume} {3}},\ \bibinfo {pages} {020319} (\bibinfo {year} {2022})}\BibitemShut {NoStop}%
\bibitem [{\citenamefont {Barato}\ and\ \citenamefont {Seifert}(2015)}]{Barato-Seifert2015}%
  \BibitemOpen
  \bibfield  {author} {\bibinfo {author} {\bibfnamefont {A.~C.}\ \bibnamefont {Barato}}\ and\ \bibinfo {author} {\bibfnamefont {U.}~\bibnamefont {Seifert}},\ }\bibfield  {title} {\bibinfo {title} {Thermodynamic uncertainty relation for biomolecular processes},\ }\href {https://doi.org/10.1103/PhysRevLett.114.158101} {\bibfield  {journal} {\bibinfo  {journal} {Phys. Rev. Lett.}\ }\textbf {\bibinfo {volume} {114}},\ \bibinfo {pages} {158101} (\bibinfo {year} {2015})}\BibitemShut {NoStop}%
\bibitem [{\citenamefont {Gingrich}\ \emph {et~al.}(2016)\citenamefont {Gingrich}, \citenamefont {Horowitz}, \citenamefont {Perunov},\ and\ \citenamefont {England}}]{Gingrich-Horowitz2016}%
  \BibitemOpen
  \bibfield  {author} {\bibinfo {author} {\bibfnamefont {T.~R.}\ \bibnamefont {Gingrich}}, \bibinfo {author} {\bibfnamefont {J.~M.}\ \bibnamefont {Horowitz}}, \bibinfo {author} {\bibfnamefont {N.}~\bibnamefont {Perunov}},\ and\ \bibinfo {author} {\bibfnamefont {J.~L.}\ \bibnamefont {England}},\ }\bibfield  {title} {\bibinfo {title} {Dissipation bounds all steady-state current fluctuations},\ }\href@noop {} {\bibfield  {journal} {\bibinfo  {journal} {Phys. Rev. Lett.}\ }\textbf {\bibinfo {volume} {116}},\ \bibinfo {pages} {120601} (\bibinfo {year} {2016})}\BibitemShut {NoStop}%
\bibitem [{\citenamefont {Pietzonka}\ \emph {et~al.}()\citenamefont {Pietzonka}, \citenamefont {Barato},\ and\ \citenamefont {Seifert}}]{pietzonka-Barato-Seifert2016}%
  \BibitemOpen
  \bibfield  {author} {\bibinfo {author} {\bibfnamefont {P.}~\bibnamefont {Pietzonka}}, \bibinfo {author} {\bibfnamefont {A.~C.}\ \bibnamefont {Barato}},\ and\ \bibinfo {author} {\bibfnamefont {U.}~\bibnamefont {Seifert}},\ }\bibfield  {title} {\bibinfo {title} {Universal bound on the efficiency of molecular motors},\ }\href@noop {} {\bibinfo  {journal} {J. Stat. Mech. (2016) 124004}\ }\BibitemShut {NoStop}%
\bibitem [{\citenamefont {Shiraishi}\ \emph {et~al.}(2016)\citenamefont {Shiraishi}, \citenamefont {Saito},\ and\ \citenamefont {Tasaki}}]{Shiraishi-Saito-Tasaki2016heatengine}%
  \BibitemOpen
\bibfield  {journal} {  }\bibfield  {author} {\bibinfo {author} {\bibfnamefont {N.}~\bibnamefont {Shiraishi}}, \bibinfo {author} {\bibfnamefont {K.}~\bibnamefont {Saito}},\ and\ \bibinfo {author} {\bibfnamefont {H.}~\bibnamefont {Tasaki}},\ }\bibfield  {title} {\bibinfo {title} {Universal trade-off relation between power and efficiency for heat engines},\ }\href {https://doi.org/10.1103/PhysRevLett.117.190601} {\bibfield  {journal} {\bibinfo  {journal} {Phys. Rev. Lett.}\ }\textbf {\bibinfo {volume} {117}},\ \bibinfo {pages} {190601} (\bibinfo {year} {2016})}\BibitemShut {NoStop}%
\bibitem [{\citenamefont {Proesmans}\ and\ \citenamefont {Van~den Broeck}(2017)}]{proesmans2017discrete}%
  \BibitemOpen
  \bibfield  {author} {\bibinfo {author} {\bibfnamefont {K.}~\bibnamefont {Proesmans}}\ and\ \bibinfo {author} {\bibfnamefont {C.}~\bibnamefont {Van~den Broeck}},\ }\bibfield  {title} {\bibinfo {title} {Discrete-time thermodynamic uncertainty relation},\ }\href@noop {} {\bibfield  {journal} {\bibinfo  {journal} {EPL}\ }\textbf {\bibinfo {volume} {119}},\ \bibinfo {pages} {20001} (\bibinfo {year} {2017})}\BibitemShut {NoStop}%
\bibitem [{\citenamefont {Maes}(2017)}]{Maes2017Frenetic}%
  \BibitemOpen
  \bibfield  {author} {\bibinfo {author} {\bibfnamefont {C.}~\bibnamefont {Maes}},\ }\bibfield  {title} {\bibinfo {title} {Frenetic bounds on the entropy production},\ }\href {https://doi.org/10.1103/PhysRevLett.119.160601} {\bibfield  {journal} {\bibinfo  {journal} {Phys. Rev. Lett.}\ }\textbf {\bibinfo {volume} {119}},\ \bibinfo {pages} {160601} (\bibinfo {year} {2017})}\BibitemShut {NoStop}%
\bibitem [{\citenamefont {Dechant}(2018)}]{dechant2018multidimensional}%
  \BibitemOpen
  \bibfield  {author} {\bibinfo {author} {\bibfnamefont {A.}~\bibnamefont {Dechant}},\ }\bibfield  {title} {\bibinfo {title} {Multidimensional thermodynamic uncertainty relations},\ }\href@noop {} {\bibfield  {journal} {\bibinfo  {journal} {J. Phys. A Math. Theor.}\ }\textbf {\bibinfo {volume} {52}},\ \bibinfo {pages} {035001} (\bibinfo {year} {2018})}\BibitemShut {NoStop}%
\bibitem [{\citenamefont {Brandner}\ \emph {et~al.}(2018)\citenamefont {Brandner}, \citenamefont {Hanazato},\ and\ \citenamefont {Saito}}]{Brandner-Hanazato-Saito2018}%
  \BibitemOpen
  \bibfield  {author} {\bibinfo {author} {\bibfnamefont {K.}~\bibnamefont {Brandner}}, \bibinfo {author} {\bibfnamefont {T.}~\bibnamefont {Hanazato}},\ and\ \bibinfo {author} {\bibfnamefont {K.}~\bibnamefont {Saito}},\ }\bibfield  {title} {\bibinfo {title} {Thermodynamic bounds on precision in ballistic multiterminal transport},\ }\href {https://doi.org/10.1103/PhysRevLett.120.090601} {\bibfield  {journal} {\bibinfo  {journal} {Phys. Rev. Lett.}\ }\textbf {\bibinfo {volume} {120}},\ \bibinfo {pages} {090601} (\bibinfo {year} {2018})}\BibitemShut {NoStop}%
\bibitem [{\citenamefont {Pietzonka}\ and\ \citenamefont {Seifert}(2018)}]{Pietzonka-Seifert2018}%
  \BibitemOpen
  \bibfield  {author} {\bibinfo {author} {\bibfnamefont {P.}~\bibnamefont {Pietzonka}}\ and\ \bibinfo {author} {\bibfnamefont {U.}~\bibnamefont {Seifert}},\ }\bibfield  {title} {\bibinfo {title} {Universal trade-off between power, efficiency, and constancy in steady-state heat engines},\ }\href {https://doi.org/10.1103/PhysRevLett.120.190602} {\bibfield  {journal} {\bibinfo  {journal} {Phys. Rev. Lett.}\ }\textbf {\bibinfo {volume} {120}},\ \bibinfo {pages} {190602} (\bibinfo {year} {2018})}\BibitemShut {NoStop}%
\bibitem [{\citenamefont {Hasegawa}\ and\ \citenamefont {Van~Vu}(2019)}]{Hasegawa-Van2019}%
  \BibitemOpen
  \bibfield  {author} {\bibinfo {author} {\bibfnamefont {Y.}~\bibnamefont {Hasegawa}}\ and\ \bibinfo {author} {\bibfnamefont {T.}~\bibnamefont {Van~Vu}},\ }\bibfield  {title} {\bibinfo {title} {Uncertainty relations in stochastic processes: An information inequality approach},\ }\href {https://doi.org/10.1103/PhysRevE.99.062126} {\bibfield  {journal} {\bibinfo  {journal} {Phys. Rev. E}\ }\textbf {\bibinfo {volume} {99}},\ \bibinfo {pages} {062126} (\bibinfo {year} {2019})}\BibitemShut {NoStop}%
\bibitem [{\citenamefont {Koyuk}\ and\ \citenamefont {Seifert}(2020)}]{koyuk2020thermodynamic}%
  \BibitemOpen
  \bibfield  {author} {\bibinfo {author} {\bibfnamefont {T.}~\bibnamefont {Koyuk}}\ and\ \bibinfo {author} {\bibfnamefont {U.}~\bibnamefont {Seifert}},\ }\bibfield  {title} {\bibinfo {title} {Thermodynamic uncertainty relation for time-dependent driving},\ }\href@noop {} {\bibfield  {journal} {\bibinfo  {journal} {Phys. Rev. Lett.}\ }\textbf {\bibinfo {volume} {125}},\ \bibinfo {pages} {260604} (\bibinfo {year} {2020})}\BibitemShut {NoStop}%
\bibitem [{\citenamefont {Liu}\ \emph {et~al.}(2020)\citenamefont {Liu}, \citenamefont {Gong},\ and\ \citenamefont {Ueda}}]{Liu-Gong-Ueda2020}%
  \BibitemOpen
  \bibfield  {author} {\bibinfo {author} {\bibfnamefont {K.}~\bibnamefont {Liu}}, \bibinfo {author} {\bibfnamefont {Z.}~\bibnamefont {Gong}},\ and\ \bibinfo {author} {\bibfnamefont {M.}~\bibnamefont {Ueda}},\ }\bibfield  {title} {\bibinfo {title} {Thermodynamic uncertainty relation for arbitrary initial states},\ }\href@noop {} {\bibfield  {journal} {\bibinfo  {journal} {Phys. Rev. Lett.}\ }\textbf {\bibinfo {volume} {125}},\ \bibinfo {pages} {140602} (\bibinfo {year} {2020})}\BibitemShut {NoStop}%
\bibitem [{\citenamefont {Otsubo}\ \emph {et~al.}(2020)\citenamefont {Otsubo}, \citenamefont {Ito}, \citenamefont {Dechant},\ and\ \citenamefont {Sagawa}}]{Otsubo-Ito-Dechant-Sagawa2020}%
  \BibitemOpen
  \bibfield  {author} {\bibinfo {author} {\bibfnamefont {S.}~\bibnamefont {Otsubo}}, \bibinfo {author} {\bibfnamefont {S.}~\bibnamefont {Ito}}, \bibinfo {author} {\bibfnamefont {A.}~\bibnamefont {Dechant}},\ and\ \bibinfo {author} {\bibfnamefont {T.}~\bibnamefont {Sagawa}},\ }\bibfield  {title} {\bibinfo {title} {Estimating entropy production by machine learning of short-time fluctuating currents},\ }\href {https://doi.org/10.1103/PhysRevE.101.062106} {\bibfield  {journal} {\bibinfo  {journal} {Phys. Rev. E}\ }\textbf {\bibinfo {volume} {101}},\ \bibinfo {pages} {062106} (\bibinfo {year} {2020})}\BibitemShut {NoStop}%
\bibitem [{\citenamefont {Villani}(2009)}]{villani2009}%
  \BibitemOpen
  \bibfield  {author} {\bibinfo {author} {\bibfnamefont {C.}~\bibnamefont {Villani}},\ }\href@noop {} {\emph {\bibinfo {title} {Optimal transport: old and new}}},\ Vol.\ \bibinfo {volume} {338}\ (\bibinfo  {publisher} {Springer},\ \bibinfo {year} {2009})\BibitemShut {NoStop}%
\bibitem [{\citenamefont {Peyr{\'e}}\ and\ \citenamefont {Cuturi}(2019)}]{peyre2019computational}%
  \BibitemOpen
  \bibfield  {author} {\bibinfo {author} {\bibfnamefont {G.}~\bibnamefont {Peyr{\'e}}}\ and\ \bibinfo {author} {\bibfnamefont {M.}~\bibnamefont {Cuturi}},\ }\bibfield  {title} {\bibinfo {title} {Computational optimal transport: With applications to data science},\ }\href@noop {} {\bibfield  {journal} {\bibinfo  {journal} {Foundations and Trends{\textregistered} in Machine Learning}\ }\textbf {\bibinfo {volume} {11}},\ \bibinfo {pages} {355} (\bibinfo {year} {2019})}\BibitemShut {NoStop}%
\bibitem [{\citenamefont {Aurell}\ \emph {et~al.}(2011)\citenamefont {Aurell}, \citenamefont {Mej\'{\i}a-Monasterio},\ and\ \citenamefont {Muratore-Ginanneschi}}]{Aurell2011Wasserstein}%
  \BibitemOpen
  \bibfield  {author} {\bibinfo {author} {\bibfnamefont {E.}~\bibnamefont {Aurell}}, \bibinfo {author} {\bibfnamefont {C.}~\bibnamefont {Mej\'{\i}a-Monasterio}},\ and\ \bibinfo {author} {\bibfnamefont {P.}~\bibnamefont {Muratore-Ginanneschi}},\ }\bibfield  {title} {\bibinfo {title} {Optimal protocols and optimal transport in stochastic thermodynamics},\ }\href {https://doi.org/10.1103/PhysRevLett.106.250601} {\bibfield  {journal} {\bibinfo  {journal} {Phys. Rev. Lett.}\ }\textbf {\bibinfo {volume} {106}},\ \bibinfo {pages} {250601} (\bibinfo {year} {2011})}\BibitemShut {NoStop}%
\bibitem [{\citenamefont {Aurell}\ \emph {et~al.}(2012)\citenamefont {Aurell}, \citenamefont {Gaw\c{e}dzki}, \citenamefont {Mej\'{\i}a-Monasterio}, \citenamefont {Mohayaee},\ and\ \citenamefont {Muratore-Ginanneschi}}]{Aurell2012refined}%
  \BibitemOpen
  \bibfield  {author} {\bibinfo {author} {\bibfnamefont {E.}~\bibnamefont {Aurell}}, \bibinfo {author} {\bibfnamefont {K.}~\bibnamefont {Gaw\c{e}dzki}}, \bibinfo {author} {\bibfnamefont {C.}~\bibnamefont {Mej\'{\i}a-Monasterio}}, \bibinfo {author} {\bibfnamefont {R.}~\bibnamefont {Mohayaee}},\ and\ \bibinfo {author} {\bibfnamefont {P.}~\bibnamefont {Muratore-Ginanneschi}},\ }\bibfield  {title} {\bibinfo {title} {Refined second law of thermodynamics for fast random processes},\ }\href@noop {} {\bibfield  {journal} {\bibinfo  {journal} {Journal of statistical physics}\ }\textbf {\bibinfo {volume} {147}},\ \bibinfo {pages} {487} (\bibinfo {year} {2012})}\BibitemShut {NoStop}%
\bibitem [{\citenamefont {Dechant}\ and\ \citenamefont {Sakurai}(2019)}]{Dechant-Sakurai2019}%
  \BibitemOpen
  \bibfield  {author} {\bibinfo {author} {\bibfnamefont {A.}~\bibnamefont {Dechant}}\ and\ \bibinfo {author} {\bibfnamefont {Y.}~\bibnamefont {Sakurai}},\ }\bibfield  {title} {\bibinfo {title} {Thermodynamic interpretation of wasserstein distance},\ }\href@noop {} {\bibfield  {journal} {\bibinfo  {journal} {arXiv preprint arXiv:1912.08405}\ } (\bibinfo {year} {2019})}\BibitemShut {NoStop}%
\bibitem [{\citenamefont {Nakazato}\ and\ \citenamefont {Ito}(2021)}]{Nakazato-Ito2021}%
  \BibitemOpen
  \bibfield  {author} {\bibinfo {author} {\bibfnamefont {M.}~\bibnamefont {Nakazato}}\ and\ \bibinfo {author} {\bibfnamefont {S.}~\bibnamefont {Ito}},\ }\bibfield  {title} {\bibinfo {title} {Geometrical aspects of entropy production in stochastic thermodynamics based on wasserstein distance},\ }\href {https://doi.org/10.1103/PhysRevResearch.3.043093} {\bibfield  {journal} {\bibinfo  {journal} {Phys. Rev. Res.}\ }\textbf {\bibinfo {volume} {3}},\ \bibinfo {pages} {043093} (\bibinfo {year} {2021})}\BibitemShut {NoStop}%
\bibitem [{\citenamefont {Muratore-Ginanneschi}\ \emph {et~al.}(2013)\citenamefont {Muratore-Ginanneschi}, \citenamefont {Mej{\'\i}a-Monasterio},\ and\ \citenamefont {Peliti}}]{Muratore2013heat}%
  \BibitemOpen
  \bibfield  {author} {\bibinfo {author} {\bibfnamefont {P.}~\bibnamefont {Muratore-Ginanneschi}}, \bibinfo {author} {\bibfnamefont {C.}~\bibnamefont {Mej{\'\i}a-Monasterio}},\ and\ \bibinfo {author} {\bibfnamefont {L.}~\bibnamefont {Peliti}},\ }\bibfield  {title} {\bibinfo {title} {Heat release by controlled continuous-time markov jump processes},\ }\href@noop {} {\bibfield  {journal} {\bibinfo  {journal} {Journal of Statistical Physics}\ }\textbf {\bibinfo {volume} {150}},\ \bibinfo {pages} {181} (\bibinfo {year} {2013})}\BibitemShut {NoStop}%
\bibitem [{\citenamefont {Dechant}(2022)}]{dechant2022minimum}%
  \BibitemOpen
  \bibfield  {author} {\bibinfo {author} {\bibfnamefont {A.}~\bibnamefont {Dechant}},\ }\bibfield  {title} {\bibinfo {title} {Minimum entropy production, detailed balance and wasserstein distance for continuous-time markov processes},\ }\href@noop {} {\bibfield  {journal} {\bibinfo  {journal} {J. Phys. A}\ }\textbf {\bibinfo {volume} {55}},\ \bibinfo {pages} {094001} (\bibinfo {year} {2022})}\BibitemShut {NoStop}%
\bibitem [{\citenamefont {Van~Vu}\ and\ \citenamefont {Saito}(2023{\natexlab{a}})}]{Van-Saito-unification2023}%
  \BibitemOpen
  \bibfield  {author} {\bibinfo {author} {\bibfnamefont {T.}~\bibnamefont {Van~Vu}}\ and\ \bibinfo {author} {\bibfnamefont {K.}~\bibnamefont {Saito}},\ }\bibfield  {title} {\bibinfo {title} {Thermodynamic unification of optimal transport: Thermodynamic uncertainty relation, minimum dissipation, and thermodynamic speed limits},\ }\href {https://doi.org/10.1103/PhysRevX.13.011013} {\bibfield  {journal} {\bibinfo  {journal} {Phys. Rev. X}\ }\textbf {\bibinfo {volume} {13}},\ \bibinfo {pages} {011013} (\bibinfo {year} {2023}{\natexlab{a}})}\BibitemShut {NoStop}%
\bibitem [{\citenamefont {Van~Vu}\ and\ \citenamefont {Saito}(2023{\natexlab{b}})}]{Van-Saito2023topological}%
  \BibitemOpen
  \bibfield  {author} {\bibinfo {author} {\bibfnamefont {T.}~\bibnamefont {Van~Vu}}\ and\ \bibinfo {author} {\bibfnamefont {K.}~\bibnamefont {Saito}},\ }\bibfield  {title} {\bibinfo {title} {Topological speed limit},\ }\href {https://doi.org/10.1103/PhysRevLett.130.010402} {\bibfield  {journal} {\bibinfo  {journal} {Phys. Rev. Lett.}\ }\textbf {\bibinfo {volume} {130}},\ \bibinfo {pages} {010402} (\bibinfo {year} {2023}{\natexlab{b}})}\BibitemShut {NoStop}%
\bibitem [{\citenamefont {Landauer}(1991)}]{Landauer1991information}%
  \BibitemOpen
  \bibfield  {author} {\bibinfo {author} {\bibfnamefont {R.}~\bibnamefont {Landauer}},\ }\bibfield  {title} {\bibinfo {title} {Information is physical},\ }\href@noop {} {\bibfield  {journal} {\bibinfo  {journal} {Physics Today}\ }\textbf {\bibinfo {volume} {44}},\ \bibinfo {pages} {23} (\bibinfo {year} {1991})}\BibitemShut {NoStop}%
\bibitem [{\citenamefont {Esposito}\ and\ \citenamefont {Van~den Broeck}(2011)}]{Esposito2011Landauer}%
  \BibitemOpen
  \bibfield  {author} {\bibinfo {author} {\bibfnamefont {M.}~\bibnamefont {Esposito}}\ and\ \bibinfo {author} {\bibfnamefont {C.}~\bibnamefont {Van~den Broeck}},\ }\bibfield  {title} {\bibinfo {title} {Second law and landauer principle far from equilibrium},\ }\href@noop {} {\bibfield  {journal} {\bibinfo  {journal} {Europhysics Letters}\ }\textbf {\bibinfo {volume} {95}},\ \bibinfo {pages} {40004} (\bibinfo {year} {2011})}\BibitemShut {NoStop}%
\bibitem [{\citenamefont {B{\'e}rut}\ \emph {et~al.}(2012)\citenamefont {B{\'e}rut}, \citenamefont {Arakelyan}, \citenamefont {Petrosyan}, \citenamefont {Ciliberto}, \citenamefont {Dillenschneider},\ and\ \citenamefont {Lutz}}]{berut2012Landauer_exp}%
  \BibitemOpen
  \bibfield  {author} {\bibinfo {author} {\bibfnamefont {A.}~\bibnamefont {B{\'e}rut}}, \bibinfo {author} {\bibfnamefont {A.}~\bibnamefont {Arakelyan}}, \bibinfo {author} {\bibfnamefont {A.}~\bibnamefont {Petrosyan}}, \bibinfo {author} {\bibfnamefont {S.}~\bibnamefont {Ciliberto}}, \bibinfo {author} {\bibfnamefont {R.}~\bibnamefont {Dillenschneider}},\ and\ \bibinfo {author} {\bibfnamefont {E.}~\bibnamefont {Lutz}},\ }\bibfield  {title} {\bibinfo {title} {Experimental verification of landauer's principle linking information and thermodynamics},\ }\href@noop {} {\bibfield  {journal} {\bibinfo  {journal} {Nature}\ }\textbf {\bibinfo {volume} {483}},\ \bibinfo {pages} {187} (\bibinfo {year} {2012})}\BibitemShut {NoStop}%
\bibitem [{\citenamefont {Jun}\ \emph {et~al.}(2014)\citenamefont {Jun}, \citenamefont {Gavrilov},\ and\ \citenamefont {Bechhoefer}}]{Jun-Gavrilov-Bechhoefer2014highprecision}%
  \BibitemOpen
  \bibfield  {author} {\bibinfo {author} {\bibfnamefont {Y.}~\bibnamefont {Jun}}, \bibinfo {author} {\bibfnamefont {M.~c.~v.}\ \bibnamefont {Gavrilov}},\ and\ \bibinfo {author} {\bibfnamefont {J.}~\bibnamefont {Bechhoefer}},\ }\bibfield  {title} {\bibinfo {title} {High-precision test of landauer's principle in a feedback trap},\ }\href {https://doi.org/10.1103/PhysRevLett.113.190601} {\bibfield  {journal} {\bibinfo  {journal} {Phys. Rev. Lett.}\ }\textbf {\bibinfo {volume} {113}},\ \bibinfo {pages} {190601} (\bibinfo {year} {2014})}\BibitemShut {NoStop}%
\bibitem [{\citenamefont {Hong}\ \emph {et~al.}(2016)\citenamefont {Hong}, \citenamefont {Lambson}, \citenamefont {Dhuey},\ and\ \citenamefont {Bokor}}]{Jeongmin2016Landauernanomagnetic}%
  \BibitemOpen
  \bibfield  {author} {\bibinfo {author} {\bibfnamefont {J.}~\bibnamefont {Hong}}, \bibinfo {author} {\bibfnamefont {B.}~\bibnamefont {Lambson}}, \bibinfo {author} {\bibfnamefont {S.}~\bibnamefont {Dhuey}},\ and\ \bibinfo {author} {\bibfnamefont {J.}~\bibnamefont {Bokor}},\ }\bibfield  {title} {\bibinfo {title} {Experimental test of landauer's principle in single-bit operations on nanomagnetic memory bits},\ }\href {https://doi.org/10.1126/sciadv.1501492} {\bibfield  {journal} {\bibinfo  {journal} {Science Advances}\ }\textbf {\bibinfo {volume} {2}},\ \bibinfo {pages} {e1501492} (\bibinfo {year} {2016})}\BibitemShut {NoStop}%
\bibitem [{\citenamefont {Dago}\ \emph {et~al.}(2021)\citenamefont {Dago}, \citenamefont {Pereda}, \citenamefont {Barros}, \citenamefont {Ciliberto},\ and\ \citenamefont {Bellon}}]{Dago-2021exp-Landauer-underdamped}%
  \BibitemOpen
  \bibfield  {author} {\bibinfo {author} {\bibfnamefont {S.}~\bibnamefont {Dago}}, \bibinfo {author} {\bibfnamefont {J.}~\bibnamefont {Pereda}}, \bibinfo {author} {\bibfnamefont {N.}~\bibnamefont {Barros}}, \bibinfo {author} {\bibfnamefont {S.}~\bibnamefont {Ciliberto}},\ and\ \bibinfo {author} {\bibfnamefont {L.}~\bibnamefont {Bellon}},\ }\bibfield  {title} {\bibinfo {title} {Information and thermodynamics: Fast and precise approach to landauer's bound in an underdamped micromechanical oscillator},\ }\href {https://doi.org/10.1103/PhysRevLett.126.170601} {\bibfield  {journal} {\bibinfo  {journal} {Phys. Rev. Lett.}\ }\textbf {\bibinfo {volume} {126}},\ \bibinfo {pages} {170601} (\bibinfo {year} {2021})}\BibitemShut {NoStop}%
\bibitem [{\citenamefont {Proesmans}\ \emph {et~al.}(2020)\citenamefont {Proesmans}, \citenamefont {Ehrich},\ and\ \citenamefont {Bechhoefer}}]{Proesmans2020finiteLandauer}%
  \BibitemOpen
  \bibfield  {author} {\bibinfo {author} {\bibfnamefont {K.}~\bibnamefont {Proesmans}}, \bibinfo {author} {\bibfnamefont {J.}~\bibnamefont {Ehrich}},\ and\ \bibinfo {author} {\bibfnamefont {J.}~\bibnamefont {Bechhoefer}},\ }\bibfield  {title} {\bibinfo {title} {Finite-time landauer principle},\ }\href {https://doi.org/10.1103/PhysRevLett.125.100602} {\bibfield  {journal} {\bibinfo  {journal} {Phys. Rev. Lett.}\ }\textbf {\bibinfo {volume} {125}},\ \bibinfo {pages} {100602} (\bibinfo {year} {2020})}\BibitemShut {NoStop}%
\bibitem [{\citenamefont {Zhen}\ \emph {et~al.}(2021)\citenamefont {Zhen}, \citenamefont {Egloff}, \citenamefont {Modi},\ and\ \citenamefont {Dahlsten}}]{Zhen2021boundbitreset}%
  \BibitemOpen
  \bibfield  {author} {\bibinfo {author} {\bibfnamefont {Y.-Z.}\ \bibnamefont {Zhen}}, \bibinfo {author} {\bibfnamefont {D.}~\bibnamefont {Egloff}}, \bibinfo {author} {\bibfnamefont {K.}~\bibnamefont {Modi}},\ and\ \bibinfo {author} {\bibfnamefont {O.}~\bibnamefont {Dahlsten}},\ }\bibfield  {title} {\bibinfo {title} {Universal bound on energy cost of bit reset in finite time},\ }\href {https://doi.org/10.1103/PhysRevLett.127.190602} {\bibfield  {journal} {\bibinfo  {journal} {Phys. Rev. Lett.}\ }\textbf {\bibinfo {volume} {127}},\ \bibinfo {pages} {190602} (\bibinfo {year} {2021})}\BibitemShut {NoStop}%
\bibitem [{\citenamefont {Lee}\ \emph {et~al.}(2022)\citenamefont {Lee}, \citenamefont {Lee}, \citenamefont {Kwon},\ and\ \citenamefont {Park}}]{Lee-Park_highly2022}%
  \BibitemOpen
  \bibfield  {author} {\bibinfo {author} {\bibfnamefont {J.~S.}\ \bibnamefont {Lee}}, \bibinfo {author} {\bibfnamefont {S.}~\bibnamefont {Lee}}, \bibinfo {author} {\bibfnamefont {H.}~\bibnamefont {Kwon}},\ and\ \bibinfo {author} {\bibfnamefont {H.}~\bibnamefont {Park}},\ }\bibfield  {title} {\bibinfo {title} {Speed limit for a highly irreversible process and tight finite-time landauer's bound},\ }\href {https://doi.org/10.1103/PhysRevLett.129.120603} {\bibfield  {journal} {\bibinfo  {journal} {Phys. Rev. Lett.}\ }\textbf {\bibinfo {volume} {129}},\ \bibinfo {pages} {120603} (\bibinfo {year} {2022})}\BibitemShut {NoStop}%
\bibitem [{\citenamefont {Van~Vu}\ and\ \citenamefont {Saito}(2022)}]{Van-Saito2022quantumLandauer}%
  \BibitemOpen
  \bibfield  {author} {\bibinfo {author} {\bibfnamefont {T.}~\bibnamefont {Van~Vu}}\ and\ \bibinfo {author} {\bibfnamefont {K.}~\bibnamefont {Saito}},\ }\bibfield  {title} {\bibinfo {title} {Finite-time quantum landauer principle and quantum coherence},\ }\href {https://doi.org/10.1103/PhysRevLett.128.010602} {\bibfield  {journal} {\bibinfo  {journal} {Phys. Rev. Lett.}\ }\textbf {\bibinfo {volume} {128}},\ \bibinfo {pages} {010602} (\bibinfo {year} {2022})}\BibitemShut {NoStop}%
\bibitem [{\citenamefont {Scandi}\ \emph {et~al.}(2022)\citenamefont {Scandi}, \citenamefont {Barker}, \citenamefont {Lehmann}, \citenamefont {Dick}, \citenamefont {Maisi},\ and\ \citenamefont {Perarnau-Llobet}}]{Sacdi-2022erasurelength}%
  \BibitemOpen
  \bibfield  {author} {\bibinfo {author} {\bibfnamefont {M.}~\bibnamefont {Scandi}}, \bibinfo {author} {\bibfnamefont {D.}~\bibnamefont {Barker}}, \bibinfo {author} {\bibfnamefont {S.}~\bibnamefont {Lehmann}}, \bibinfo {author} {\bibfnamefont {K.~A.}\ \bibnamefont {Dick}}, \bibinfo {author} {\bibfnamefont {V.~F.}\ \bibnamefont {Maisi}},\ and\ \bibinfo {author} {\bibfnamefont {M.}~\bibnamefont {Perarnau-Llobet}},\ }\bibfield  {title} {\bibinfo {title} {Minimally dissipative information erasure in a quantum dot via thermodynamic length},\ }\href {https://doi.org/10.1103/PhysRevLett.129.270601} {\bibfield  {journal} {\bibinfo  {journal} {Phys. Rev. Lett.}\ }\textbf {\bibinfo {volume} {129}},\ \bibinfo {pages} {270601} (\bibinfo {year} {2022})}\BibitemShut {NoStop}%
\bibitem [{\citenamefont {Parrondo}\ \emph {et~al.}(2015)\citenamefont {Parrondo}, \citenamefont {Horowitz},\ and\ \citenamefont {Sagawa}}]{Parrondo-Horowitz-Sagawa2015}%
  \BibitemOpen
  \bibfield  {author} {\bibinfo {author} {\bibfnamefont {J.~M.}\ \bibnamefont {Parrondo}}, \bibinfo {author} {\bibfnamefont {J.~M.}\ \bibnamefont {Horowitz}},\ and\ \bibinfo {author} {\bibfnamefont {T.}~\bibnamefont {Sagawa}},\ }\bibfield  {title} {\bibinfo {title} {Thermodynamics of information},\ }\href@noop {} {\bibfield  {journal} {\bibinfo  {journal} {Nature physics}\ }\textbf {\bibinfo {volume} {11}},\ \bibinfo {pages} {131} (\bibinfo {year} {2015})}\BibitemShut {NoStop}%
\bibitem [{\citenamefont {Leff}\ and\ \citenamefont {Rex}(2002)}]{leff2002maxwell}%
  \BibitemOpen
  \bibfield  {author} {\bibinfo {author} {\bibfnamefont {H.}~\bibnamefont {Leff}}\ and\ \bibinfo {author} {\bibfnamefont {A.~F.}\ \bibnamefont {Rex}},\ }\href@noop {} {\emph {\bibinfo {title} {Maxwell's Demon 2 Entropy, Classical and Quantum Information, Computing}}}\ (\bibinfo  {publisher} {CRC Press},\ \bibinfo {year} {2002})\BibitemShut {NoStop}%
\bibitem [{\citenamefont {Sagawa}\ and\ \citenamefont {Ueda}(2008)}]{Sagawa-Ueda2008quantumfeedback}%
  \BibitemOpen
  \bibfield  {author} {\bibinfo {author} {\bibfnamefont {T.}~\bibnamefont {Sagawa}}\ and\ \bibinfo {author} {\bibfnamefont {M.}~\bibnamefont {Ueda}},\ }\bibfield  {title} {\bibinfo {title} {Second law of thermodynamics with discrete quantum feedback control},\ }\href {https://doi.org/10.1103/PhysRevLett.100.080403} {\bibfield  {journal} {\bibinfo  {journal} {Phys. Rev. Lett.}\ }\textbf {\bibinfo {volume} {100}},\ \bibinfo {pages} {080403} (\bibinfo {year} {2008})}\BibitemShut {NoStop}%
\bibitem [{\citenamefont {Sagawa}\ and\ \citenamefont {Ueda}(2009)}]{Sagawa-Ueda2009erasure}%
  \BibitemOpen
  \bibfield  {author} {\bibinfo {author} {\bibfnamefont {T.}~\bibnamefont {Sagawa}}\ and\ \bibinfo {author} {\bibfnamefont {M.}~\bibnamefont {Ueda}},\ }\bibfield  {title} {\bibinfo {title} {Minimal energy cost for thermodynamic information processing: Measurement and information erasure},\ }\href {https://doi.org/10.1103/PhysRevLett.102.250602} {\bibfield  {journal} {\bibinfo  {journal} {Phys. Rev. Lett.}\ }\textbf {\bibinfo {volume} {102}},\ \bibinfo {pages} {250602} (\bibinfo {year} {2009})}\BibitemShut {NoStop}%
\bibitem [{\citenamefont {Sagawa}\ and\ \citenamefont {Ueda}(2010)}]{Sagawa-Ueda2010feedback}%
  \BibitemOpen
  \bibfield  {author} {\bibinfo {author} {\bibfnamefont {T.}~\bibnamefont {Sagawa}}\ and\ \bibinfo {author} {\bibfnamefont {M.}~\bibnamefont {Ueda}},\ }\bibfield  {title} {\bibinfo {title} {Generalized jarzynski equality under nonequilibrium feedback control},\ }\href {https://doi.org/10.1103/PhysRevLett.104.090602} {\bibfield  {journal} {\bibinfo  {journal} {Phys. Rev. Lett.}\ }\textbf {\bibinfo {volume} {104}},\ \bibinfo {pages} {090602} (\bibinfo {year} {2010})}\BibitemShut {NoStop}%
\bibitem [{\citenamefont {Ito}\ and\ \citenamefont {Sagawa}(2013)}]{ItoSagawa2013causal}%
  \BibitemOpen
  \bibfield  {author} {\bibinfo {author} {\bibfnamefont {S.}~\bibnamefont {Ito}}\ and\ \bibinfo {author} {\bibfnamefont {T.}~\bibnamefont {Sagawa}},\ }\bibfield  {title} {\bibinfo {title} {Information thermodynamics on causal networks},\ }\href {https://doi.org/10.1103/PhysRevLett.111.180603} {\bibfield  {journal} {\bibinfo  {journal} {Phys. Rev. Lett.}\ }\textbf {\bibinfo {volume} {111}},\ \bibinfo {pages} {180603} (\bibinfo {year} {2013})}\BibitemShut {NoStop}%
\bibitem [{\citenamefont {Horowitz}\ and\ \citenamefont {Esposito}(2014)}]{infoflow-Holowitz-2014}%
  \BibitemOpen
  \bibfield  {author} {\bibinfo {author} {\bibfnamefont {J.~M.}\ \bibnamefont {Horowitz}}\ and\ \bibinfo {author} {\bibfnamefont {M.}~\bibnamefont {Esposito}},\ }\bibfield  {title} {\bibinfo {title} {Thermodynamics with continuous information flow},\ }\href {https://doi.org/10.1103/PhysRevX.4.031015} {\bibfield  {journal} {\bibinfo  {journal} {Phys. Rev. X}\ }\textbf {\bibinfo {volume} {4}},\ \bibinfo {pages} {031015} (\bibinfo {year} {2014})}\BibitemShut {NoStop}%
\bibitem [{\citenamefont {Toyabe}\ \emph {et~al.}(2010)\citenamefont {Toyabe}, \citenamefont {Sagawa}, \citenamefont {Ueda}, \citenamefont {Muneyuki},\ and\ \citenamefont {Sano}}]{toyabe-2010exp-Jarzynski}%
  \BibitemOpen
  \bibfield  {author} {\bibinfo {author} {\bibfnamefont {S.}~\bibnamefont {Toyabe}}, \bibinfo {author} {\bibfnamefont {T.}~\bibnamefont {Sagawa}}, \bibinfo {author} {\bibfnamefont {M.}~\bibnamefont {Ueda}}, \bibinfo {author} {\bibfnamefont {E.}~\bibnamefont {Muneyuki}},\ and\ \bibinfo {author} {\bibfnamefont {M.}~\bibnamefont {Sano}},\ }\bibfield  {title} {\bibinfo {title} {Experimental demonstration of information-to-energy conversion and validation of the generalized jarzynski equality},\ }\href@noop {} {\bibfield  {journal} {\bibinfo  {journal} {Nature physics}\ }\textbf {\bibinfo {volume} {6}},\ \bibinfo {pages} {988} (\bibinfo {year} {2010})}\BibitemShut {NoStop}%
\bibitem [{\citenamefont {Koski}\ \emph {et~al.}(2014)\citenamefont {Koski}, \citenamefont {Maisi}, \citenamefont {Sagawa},\ and\ \citenamefont {Pekola}}]{Koski-Sagawa2014experiment}%
  \BibitemOpen
  \bibfield  {author} {\bibinfo {author} {\bibfnamefont {J.~V.}\ \bibnamefont {Koski}}, \bibinfo {author} {\bibfnamefont {V.~F.}\ \bibnamefont {Maisi}}, \bibinfo {author} {\bibfnamefont {T.}~\bibnamefont {Sagawa}},\ and\ \bibinfo {author} {\bibfnamefont {J.~P.}\ \bibnamefont {Pekola}},\ }\bibfield  {title} {\bibinfo {title} {Experimental observation of the role of mutual information in the nonequilibrium dynamics of a maxwell demon},\ }\href {https://doi.org/10.1103/PhysRevLett.113.030601} {\bibfield  {journal} {\bibinfo  {journal} {Phys. Rev. Lett.}\ }\textbf {\bibinfo {volume} {113}},\ \bibinfo {pages} {030601} (\bibinfo {year} {2014})}\BibitemShut {NoStop}%
\bibitem [{\citenamefont {Ribezzi-Crivellari}\ and\ \citenamefont {Ritort}(2019)}]{Ribezzi-Marco2019exp-Maxwell}%
  \BibitemOpen
  \bibfield  {author} {\bibinfo {author} {\bibfnamefont {M.}~\bibnamefont {Ribezzi-Crivellari}}\ and\ \bibinfo {author} {\bibfnamefont {F.}~\bibnamefont {Ritort}},\ }\bibfield  {title} {\bibinfo {title} {Large work extraction and the landauer limit in a continuous maxwell demon},\ }\href@noop {} {\bibfield  {journal} {\bibinfo  {journal} {Nature Physics}\ }\textbf {\bibinfo {volume} {15}},\ \bibinfo {pages} {660} (\bibinfo {year} {2019})}\BibitemShut {NoStop}%
\bibitem [{\citenamefont {Abreu}\ and\ \citenamefont {Seifert}(2011)}]{abreu-Seifert2011feedback}%
  \BibitemOpen
  \bibfield  {author} {\bibinfo {author} {\bibfnamefont {D.}~\bibnamefont {Abreu}}\ and\ \bibinfo {author} {\bibfnamefont {U.}~\bibnamefont {Seifert}},\ }\bibfield  {title} {\bibinfo {title} {Extracting work from a single heat bath through feedback},\ }\href@noop {} {\bibfield  {journal} {\bibinfo  {journal} {Europhysics Letters}\ }\textbf {\bibinfo {volume} {94}},\ \bibinfo {pages} {10001} (\bibinfo {year} {2011})}\BibitemShut {NoStop}%
\bibitem [{\citenamefont {Taghvaei}\ \emph {et~al.}(2022)\citenamefont {Taghvaei}, \citenamefont {Miangolarra}, \citenamefont {Fu}, \citenamefont {Chen},\ and\ \citenamefont {Georgiou}}]{Taghvaei-2022-demon}%
  \BibitemOpen
  \bibfield  {author} {\bibinfo {author} {\bibfnamefont {A.}~\bibnamefont {Taghvaei}}, \bibinfo {author} {\bibfnamefont {O.~M.}\ \bibnamefont {Miangolarra}}, \bibinfo {author} {\bibfnamefont {R.}~\bibnamefont {Fu}}, \bibinfo {author} {\bibfnamefont {Y.}~\bibnamefont {Chen}},\ and\ \bibinfo {author} {\bibfnamefont {T.~T.}\ \bibnamefont {Georgiou}},\ }\bibfield  {title} {\bibinfo {title} {On the relation between information and power in stochastic thermodynamic engines},\ }\href {https://doi.org/10.1109/LCSYS.2021.3078716} {\bibfield  {journal} {\bibinfo  {journal} {IEEE Control Systems Letters}\ }\textbf {\bibinfo {volume} {6}},\ \bibinfo {pages} {434} (\bibinfo {year} {2022})}\BibitemShut {NoStop}%
\bibitem [{\citenamefont {Fujimoto}\ and\ \citenamefont {Ito}(2024)}]{FujimotoIto2023}%
  \BibitemOpen
  \bibfield  {author} {\bibinfo {author} {\bibfnamefont {Y.}~\bibnamefont {Fujimoto}}\ and\ \bibinfo {author} {\bibfnamefont {S.}~\bibnamefont {Ito}},\ }\bibfield  {title} {\bibinfo {title} {Game-theoretical approach to minimum entropy productions in information thermodynamics},\ }\href {https://doi.org/10.1103/PhysRevResearch.6.013023} {\bibfield  {journal} {\bibinfo  {journal} {Phys. Rev. Res.}\ }\textbf {\bibinfo {volume} {6}},\ \bibinfo {pages} {013023} (\bibinfo {year} {2024})}\BibitemShut {NoStop}%
\bibitem [{\citenamefont {Nagase}\ and\ \citenamefont {Sagawa}(2024)}]{Nagase-Sagawa2023infogain}%
  \BibitemOpen
  \bibfield  {author} {\bibinfo {author} {\bibfnamefont {R.}~\bibnamefont {Nagase}}\ and\ \bibinfo {author} {\bibfnamefont {T.}~\bibnamefont {Sagawa}},\ }\bibfield  {title} {\bibinfo {title} {Thermodynamically optimal information gain in finite-time measurement},\ }\href {https://doi.org/10.1103/PhysRevResearch.6.033239} {\bibfield  {journal} {\bibinfo  {journal} {Phys. Rev. Res.}\ }\textbf {\bibinfo {volume} {6}},\ \bibinfo {pages} {033239} (\bibinfo {year} {2024})}\BibitemShut {NoStop}%
\bibitem [{\citenamefont {Van~Kampen}(1992)}]{vanKampen}%
  \BibitemOpen
  \bibfield  {author} {\bibinfo {author} {\bibfnamefont {N.~G.}\ \bibnamefont {Van~Kampen}},\ }\href@noop {} {\emph {\bibinfo {title} {Stochastic processes in physics and chemistry}}},\ Vol.~\bibinfo {volume} {1}\ (\bibinfo  {publisher} {Elsevier},\ \bibinfo {year} {1992})\BibitemShut {NoStop}%
\bibitem [{\citenamefont {Gardiner}(2009)}]{gardiner2009stochastic}%
  \BibitemOpen
  \bibfield  {author} {\bibinfo {author} {\bibfnamefont {C.}~\bibnamefont {Gardiner}},\ }\href@noop {} {\emph {\bibinfo {title} {Stochastic methods}}},\ Vol.~\bibinfo {volume} {4}\ (\bibinfo  {publisher} {Springer Berlin},\ \bibinfo {year} {2009})\BibitemShut {NoStop}%
\bibitem [{\citenamefont {Hartich}\ \emph {et~al.}(2014)\citenamefont {Hartich}, \citenamefont {Barato},\ and\ \citenamefont {Seifert}}]{hartich2014transfer}%
  \BibitemOpen
  \bibfield  {author} {\bibinfo {author} {\bibfnamefont {D.}~\bibnamefont {Hartich}}, \bibinfo {author} {\bibfnamefont {A.~C.}\ \bibnamefont {Barato}},\ and\ \bibinfo {author} {\bibfnamefont {U.}~\bibnamefont {Seifert}},\ }\bibfield  {title} {\bibinfo {title} {Stochastic thermodynamics of bipartite systems: transfer entropy inequalities and a maxwell's demon interpretation},\ }\href@noop {} {\bibfield  {journal} {\bibinfo  {journal} {Journal of Statistical Mechanics: Theory and Experiment}\ }\textbf {\bibinfo {volume} {2014}},\ \bibinfo {pages} {P02016} (\bibinfo {year} {2014})}\BibitemShut {NoStop}%
\bibitem [{\citenamefont {Maes}(2020)}]{Maes2020}%
  \BibitemOpen
  \bibfield  {author} {\bibinfo {author} {\bibfnamefont {C.}~\bibnamefont {Maes}},\ }\bibfield  {title} {\bibinfo {title} {Frenesy: Time-symmetric dynamical activity in nonequilibria},\ }\href {https://doi.org/https://doi.org/10.1016/j.physrep.2020.01.002} {\bibfield  {journal} {\bibinfo  {journal} {Physics Reports}\ }\textbf {\bibinfo {volume} {850}},\ \bibinfo {pages} {1} (\bibinfo {year} {2020})}\BibitemShut {NoStop}%
\bibitem [{\citenamefont {Shiraishi}\ and\ \citenamefont {Sagawa}(2015)}]{Shiraishi-Sagawa2015masked}%
  \BibitemOpen
  \bibfield  {author} {\bibinfo {author} {\bibfnamefont {N.}~\bibnamefont {Shiraishi}}\ and\ \bibinfo {author} {\bibfnamefont {T.}~\bibnamefont {Sagawa}},\ }\bibfield  {title} {\bibinfo {title} {Fluctuation theorem for partially masked nonequilibrium dynamics},\ }\href {https://doi.org/10.1103/PhysRevE.91.012130} {\bibfield  {journal} {\bibinfo  {journal} {Phys. Rev. E}\ }\textbf {\bibinfo {volume} {91}},\ \bibinfo {pages} {012130} (\bibinfo {year} {2015})}\BibitemShut {NoStop}%
\bibitem [{\citenamefont {Van~Vu}\ and\ \citenamefont {Hasegawa}(2022)}]{Van-Hasegawa2022unifiedTKUR}%
  \BibitemOpen
  \bibfield  {author} {\bibinfo {author} {\bibfnamefont {T.}~\bibnamefont {Van~Vu}}\ and\ \bibinfo {author} {\bibfnamefont {Y.}~\bibnamefont {Hasegawa}},\ }\bibfield  {title} {\bibinfo {title} {Unified thermodynamic--kinetic uncertainty relation},\ }\href@noop {} {\bibfield  {journal} {\bibinfo  {journal} {Journal of Physics A: Mathematical and Theoretical}\ }\textbf {\bibinfo {volume} {55}},\ \bibinfo {pages} {405004} (\bibinfo {year} {2022})}\BibitemShut {NoStop}%
\bibitem [{\citenamefont {Sudakov}(1979)}]{sudakov1979geometric}%
  \BibitemOpen
  \bibfield  {author} {\bibinfo {author} {\bibfnamefont {V.~N.}\ \bibnamefont {Sudakov}},\ }\href@noop {} {\emph {\bibinfo {title} {Geometric problems in the theory of infinite-dimensional probability distributions}}},\ \bibinfo {number} {141}\ (\bibinfo  {publisher} {American Mathematical Soc.},\ \bibinfo {year} {1979})\BibitemShut {NoStop}%
\bibitem [{\citenamefont {Ngatchou}\ \emph {et~al.}(2005)\citenamefont {Ngatchou}, \citenamefont {Zarei},\ and\ \citenamefont {El-Sharkawi}}]{ngatchou2005pareto}%
  \BibitemOpen
  \bibfield  {author} {\bibinfo {author} {\bibfnamefont {P.}~\bibnamefont {Ngatchou}}, \bibinfo {author} {\bibfnamefont {A.}~\bibnamefont {Zarei}},\ and\ \bibinfo {author} {\bibfnamefont {A.}~\bibnamefont {El-Sharkawi}},\ }\bibfield  {title} {\bibinfo {title} {Pareto multi objective optimization},\ }in\ \href@noop {} {\emph {\bibinfo {booktitle} {Proceedings of the 13th International Conference on, Intelligent Systems Application to Power Systems}}}\ (\bibinfo {organization} {IEEE},\ \bibinfo {year} {2005})\ pp.\ \bibinfo {pages} {84--91}\BibitemShut {NoStop}%
\bibitem [{\citenamefont {Coello}(2007)}]{coello2007evolutionary}%
  \BibitemOpen
  \bibfield  {author} {\bibinfo {author} {\bibfnamefont {C.}~\bibnamefont {Coello}},\ }\href@noop {} {\emph {\bibinfo {title} {Evolutionary Algorithms for solving Multi-Objective Problems}}}\ (\bibinfo  {publisher} {Springer},\ \bibinfo {year} {2007})\BibitemShut {NoStop}%
\bibitem [{\citenamefont {Benenti}\ \emph {et~al.}(2017)\citenamefont {Benenti}, \citenamefont {Casati}, \citenamefont {Saito},\ and\ \citenamefont {Whitney}}]{Benenti-2017Fundamental}%
  \BibitemOpen
  \bibfield  {author} {\bibinfo {author} {\bibfnamefont {G.}~\bibnamefont {Benenti}}, \bibinfo {author} {\bibfnamefont {G.}~\bibnamefont {Casati}}, \bibinfo {author} {\bibfnamefont {K.}~\bibnamefont {Saito}},\ and\ \bibinfo {author} {\bibfnamefont {R.}~\bibnamefont {Whitney}},\ }\bibfield  {title} {\bibinfo {title} {Fundamental aspects of steady-state conversion of heat to work at the nanoscale},\ }\href {https://doi.org/https://doi.org/10.1016/j.physrep.2017.05.008} {\bibfield  {journal} {\bibinfo  {journal} {Phys. Rep.}\ }\textbf {\bibinfo {volume} {694}},\ \bibinfo {pages} {1 } (\bibinfo {year} {2017})}\BibitemShut {NoStop}%
\bibitem [{\citenamefont {Solon}\ and\ \citenamefont {Horowitz}(2018)}]{solon-holowitz2018}%
  \BibitemOpen
  \bibfield  {author} {\bibinfo {author} {\bibfnamefont {A.~P.}\ \bibnamefont {Solon}}\ and\ \bibinfo {author} {\bibfnamefont {J.~M.}\ \bibnamefont {Horowitz}},\ }\bibfield  {title} {\bibinfo {title} {Phase transition in protocols minimizing work fluctuations},\ }\href {https://doi.org/10.1103/PhysRevLett.120.180605} {\bibfield  {journal} {\bibinfo  {journal} {Phys. Rev. Lett.}\ }\textbf {\bibinfo {volume} {120}},\ \bibinfo {pages} {180605} (\bibinfo {year} {2018})}\BibitemShut {NoStop}%
\bibitem [{\citenamefont {Ashida}\ and\ \citenamefont {Sagawa}(2021)}]{Ashida-Sagawa2021}%
  \BibitemOpen
  \bibfield  {author} {\bibinfo {author} {\bibfnamefont {Y.}~\bibnamefont {Ashida}}\ and\ \bibinfo {author} {\bibfnamefont {T.}~\bibnamefont {Sagawa}},\ }\bibfield  {title} {\bibinfo {title} {Learning the best nanoscale heat engines through evolving network topology},\ }\href {https://doi.org/10.1038/s42005-021-00553-z} {\bibfield  {journal} {\bibinfo  {journal} {Commun. Phys.}\ }\textbf {\bibinfo {volume} {4}},\ \bibinfo {pages} {45} (\bibinfo {year} {2021})}\BibitemShut {NoStop}%
\bibitem [{\citenamefont {Marler}\ and\ \citenamefont {Arora}(2010)}]{marler2010weightedsum}%
  \BibitemOpen
  \bibfield  {author} {\bibinfo {author} {\bibfnamefont {R.~T.}\ \bibnamefont {Marler}}\ and\ \bibinfo {author} {\bibfnamefont {J.~S.}\ \bibnamefont {Arora}},\ }\bibfield  {title} {\bibinfo {title} {The weighted sum method for multi-objective optimization: new insights},\ }\href@noop {} {\bibfield  {journal} {\bibinfo  {journal} {Structural and multidisciplinary optimization}\ }\textbf {\bibinfo {volume} {41}},\ \bibinfo {pages} {853} (\bibinfo {year} {2010})}\BibitemShut {NoStop}%
\bibitem [{\citenamefont {Cover}(1999)}]{cover1999elements}%
  \BibitemOpen
  \bibfield  {author} {\bibinfo {author} {\bibfnamefont {T.~M.}\ \bibnamefont {Cover}},\ }\href@noop {} {\emph {\bibinfo {title} {Elements of information theory}}}\ (\bibinfo  {publisher} {John Wiley \& Sons},\ \bibinfo {year} {1999})\BibitemShut {NoStop}%
\bibitem [{\citenamefont {Sagawa}\ and\ \citenamefont {Ueda}(2013)}]{sagawaNJP2013}%
  \BibitemOpen
  \bibfield  {author} {\bibinfo {author} {\bibfnamefont {T.}~\bibnamefont {Sagawa}}\ and\ \bibinfo {author} {\bibfnamefont {M.}~\bibnamefont {Ueda}},\ }\bibfield  {title} {\bibinfo {title} {Role of mutual information in entropy production under information exchanges},\ }\href@noop {} {\bibfield  {journal} {\bibinfo  {journal} {New Journal of Physics}\ }\textbf {\bibinfo {volume} {15}},\ \bibinfo {pages} {125012} (\bibinfo {year} {2013})}\BibitemShut {NoStop}%
\bibitem [{\citenamefont {Strasberg}\ \emph {et~al.}(2013)\citenamefont {Strasberg}, \citenamefont {Schaller}, \citenamefont {Brandes},\ and\ \citenamefont {Esposito}}]{Strasberg2013demon}%
  \BibitemOpen
  \bibfield  {author} {\bibinfo {author} {\bibfnamefont {P.}~\bibnamefont {Strasberg}}, \bibinfo {author} {\bibfnamefont {G.}~\bibnamefont {Schaller}}, \bibinfo {author} {\bibfnamefont {T.}~\bibnamefont {Brandes}},\ and\ \bibinfo {author} {\bibfnamefont {M.}~\bibnamefont {Esposito}},\ }\bibfield  {title} {\bibinfo {title} {Thermodynamics of a physical model implementing a maxwell demon},\ }\href {https://doi.org/10.1103/PhysRevLett.110.040601} {\bibfield  {journal} {\bibinfo  {journal} {Phys. Rev. Lett.}\ }\textbf {\bibinfo {volume} {110}},\ \bibinfo {pages} {040601} (\bibinfo {year} {2013})}\BibitemShut {NoStop}%
\bibitem [{\citenamefont {Diana}\ and\ \citenamefont {Esposito}(2014)}]{Diana-Esposito2014bipartite}%
  \BibitemOpen
  \bibfield  {author} {\bibinfo {author} {\bibfnamefont {G.}~\bibnamefont {Diana}}\ and\ \bibinfo {author} {\bibfnamefont {M.}~\bibnamefont {Esposito}},\ }\bibfield  {title} {\bibinfo {title} {Mutual entropy production in bipartite systems},\ }\href@noop {} {\bibfield  {journal} {\bibinfo  {journal} {Journal of Statistical Mechanics: Theory and Experiment}\ }\textbf {\bibinfo {volume} {2014}},\ \bibinfo {pages} {P04010} (\bibinfo {year} {2014})}\BibitemShut {NoStop}%
\bibitem [{\citenamefont {Salazar}(2022)}]{Salazar2022EPbound}%
  \BibitemOpen
  \bibfield  {author} {\bibinfo {author} {\bibfnamefont {D.~S.~P.}\ \bibnamefont {Salazar}},\ }\bibfield  {title} {\bibinfo {title} {Lower bound for entropy production rate in stochastic systems far from equilibrium},\ }\href {https://doi.org/10.1103/PhysRevE.106.L032101} {\bibfield  {journal} {\bibinfo  {journal} {Phys. Rev. E}\ }\textbf {\bibinfo {volume} {106}},\ \bibinfo {pages} {L032101} (\bibinfo {year} {2022})}\BibitemShut {NoStop}%
\bibitem [{\citenamefont {Kutvonen}\ \emph {et~al.}(2016)\citenamefont {Kutvonen}, \citenamefont {Sagawa},\ and\ \citenamefont {Ala-Nissila}}]{Kutvonen-Sagawa2016}%
  \BibitemOpen
  \bibfield  {author} {\bibinfo {author} {\bibfnamefont {A.}~\bibnamefont {Kutvonen}}, \bibinfo {author} {\bibfnamefont {T.}~\bibnamefont {Sagawa}},\ and\ \bibinfo {author} {\bibfnamefont {T.}~\bibnamefont {Ala-Nissila}},\ }\bibfield  {title} {\bibinfo {title} {Thermodynamics of information exchange between two coupled quantum dots},\ }\href {https://doi.org/10.1103/PhysRevE.93.032147} {\bibfield  {journal} {\bibinfo  {journal} {Phys. Rev. E}\ }\textbf {\bibinfo {volume} {93}},\ \bibinfo {pages} {032147} (\bibinfo {year} {2016})}\BibitemShut {NoStop}%
\bibitem [{\citenamefont {Remlein}\ and\ \citenamefont {Seifert}(2021)}]{Remlein-Seifert2021}%
  \BibitemOpen
  \bibfield  {author} {\bibinfo {author} {\bibfnamefont {B.}~\bibnamefont {Remlein}}\ and\ \bibinfo {author} {\bibfnamefont {U.}~\bibnamefont {Seifert}},\ }\bibfield  {title} {\bibinfo {title} {Optimality of nonconservative driving for finite-time processes with discrete states},\ }\href {https://doi.org/10.1103/PhysRevE.103.L050105} {\bibfield  {journal} {\bibinfo  {journal} {Phys. Rev. E}\ }\textbf {\bibinfo {volume} {103}},\ \bibinfo {pages} {L050105} (\bibinfo {year} {2021})}\BibitemShut {NoStop}%
\bibitem [{\citenamefont {Lan}\ \emph {et~al.}(2012{\natexlab{a}})\citenamefont {Lan}, \citenamefont {Sartori}, \citenamefont {Neumann}, \citenamefont {Sourjik},\ and\ \citenamefont {Tu}}]{lan-2012ESA}%
  \BibitemOpen
  \bibfield  {author} {\bibinfo {author} {\bibfnamefont {G.}~\bibnamefont {Lan}}, \bibinfo {author} {\bibfnamefont {P.}~\bibnamefont {Sartori}}, \bibinfo {author} {\bibfnamefont {S.}~\bibnamefont {Neumann}}, \bibinfo {author} {\bibfnamefont {V.}~\bibnamefont {Sourjik}},\ and\ \bibinfo {author} {\bibfnamefont {Y.}~\bibnamefont {Tu}},\ }\bibfield  {title} {\bibinfo {title} {The energy--speed--accuracy trade-off in sensory adaptation},\ }\href@noop {} {\bibfield  {journal} {\bibinfo  {journal} {Nature physics}\ }\textbf {\bibinfo {volume} {8}},\ \bibinfo {pages} {422} (\bibinfo {year} {2012}{\natexlab{a}})}\BibitemShut {NoStop}%
\bibitem [{\citenamefont {Sartori}\ \emph {et~al.}(2014)\citenamefont {Sartori}, \citenamefont {Granger}, \citenamefont {Lee},\ and\ \citenamefont {Horowitz}}]{sartori-2014sensory}%
  \BibitemOpen
  \bibfield  {author} {\bibinfo {author} {\bibfnamefont {P.}~\bibnamefont {Sartori}}, \bibinfo {author} {\bibfnamefont {L.}~\bibnamefont {Granger}}, \bibinfo {author} {\bibfnamefont {C.~F.}\ \bibnamefont {Lee}},\ and\ \bibinfo {author} {\bibfnamefont {J.~M.}\ \bibnamefont {Horowitz}},\ }\bibfield  {title} {\bibinfo {title} {Thermodynamic costs of information processing in sensory adaptation},\ }\href@noop {} {\bibfield  {journal} {\bibinfo  {journal} {PLoS computational biology}\ }\textbf {\bibinfo {volume} {10}},\ \bibinfo {pages} {e1003974} (\bibinfo {year} {2014})}\BibitemShut {NoStop}%
\bibitem [{\citenamefont {Flamary}\ \emph {et~al.}(2021)\citenamefont {Flamary}, \citenamefont {Courty}, \citenamefont {Gramfort}, \citenamefont {Alaya}, \citenamefont {Boisbunon}, \citenamefont {Chambon}, \citenamefont {Chapel}, \citenamefont {Corenflos}, \citenamefont {Fatras}, \citenamefont {Fournier}, \citenamefont {Gautheron}, \citenamefont {Gayraud}, \citenamefont {Janati}, \citenamefont {Rakotomamonjy}, \citenamefont {Redko}, \citenamefont {Rolet}, \citenamefont {Schutz}, \citenamefont {Seguy}, \citenamefont {Sutherland}, \citenamefont {Tavenard}, \citenamefont {Tong},\ and\ \citenamefont {Vayer}}]{flamary2021pot}%
  \BibitemOpen
  \bibfield  {author} {\bibinfo {author} {\bibfnamefont {R.}~\bibnamefont {Flamary}}, \bibinfo {author} {\bibfnamefont {N.}~\bibnamefont {Courty}}, \bibinfo {author} {\bibfnamefont {A.}~\bibnamefont {Gramfort}}, \bibinfo {author} {\bibfnamefont {M.~Z.}\ \bibnamefont {Alaya}}, \bibinfo {author} {\bibfnamefont {A.}~\bibnamefont {Boisbunon}}, \bibinfo {author} {\bibfnamefont {S.}~\bibnamefont {Chambon}}, \bibinfo {author} {\bibfnamefont {L.}~\bibnamefont {Chapel}}, \bibinfo {author} {\bibfnamefont {A.}~\bibnamefont {Corenflos}}, \bibinfo {author} {\bibfnamefont {K.}~\bibnamefont {Fatras}}, \bibinfo {author} {\bibfnamefont {N.}~\bibnamefont {Fournier}}, \bibinfo {author} {\bibfnamefont {L.}~\bibnamefont {Gautheron}}, \bibinfo {author} {\bibfnamefont {N.~T.}\ \bibnamefont {Gayraud}}, \bibinfo {author} {\bibfnamefont {H.}~\bibnamefont {Janati}}, \bibinfo {author} {\bibfnamefont {A.}~\bibnamefont {Rakotomamonjy}}, \bibinfo {author} {\bibfnamefont {I.}~\bibnamefont {Redko}}, \bibinfo {author} {\bibfnamefont
  {A.}~\bibnamefont {Rolet}}, \bibinfo {author} {\bibfnamefont {A.}~\bibnamefont {Schutz}}, \bibinfo {author} {\bibfnamefont {V.}~\bibnamefont {Seguy}}, \bibinfo {author} {\bibfnamefont {D.~J.}\ \bibnamefont {Sutherland}}, \bibinfo {author} {\bibfnamefont {R.}~\bibnamefont {Tavenard}}, \bibinfo {author} {\bibfnamefont {A.}~\bibnamefont {Tong}},\ and\ \bibinfo {author} {\bibfnamefont {T.}~\bibnamefont {Vayer}},\ }\bibfield  {title} {\bibinfo {title} {Pot: Python optimal transport},\ }\href@noop {} {\bibfield  {journal} {\bibinfo  {journal} {Journal of Machine Learning Research}\ }\textbf {\bibinfo {volume} {22}},\ \bibinfo {pages} {1} (\bibinfo {year} {2021})}\BibitemShut {NoStop}%
\bibitem [{\citenamefont {Pele}\ and\ \citenamefont {Werman}(2009)}]{Pele-Werman2009EMD}%
  \BibitemOpen
  \bibfield  {author} {\bibinfo {author} {\bibfnamefont {O.}~\bibnamefont {Pele}}\ and\ \bibinfo {author} {\bibfnamefont {M.}~\bibnamefont {Werman}},\ }\bibfield  {title} {\bibinfo {title} {Fast and robust earth mover's distances},\ }in\ \href@noop {} {\emph {\bibinfo {booktitle} {2009 IEEE 12th international conference on computer vision}}}\ (\bibinfo {organization} {IEEE},\ \bibinfo {year} {2009})\ pp.\ \bibinfo {pages} {460--467}\BibitemShut {NoStop}%
\bibitem [{\citenamefont {Cuturi}(2013)}]{cuturi2013sinkhorn}%
  \BibitemOpen
  \bibfield  {author} {\bibinfo {author} {\bibfnamefont {M.}~\bibnamefont {Cuturi}},\ }\bibfield  {title} {\bibinfo {title} {{S}inkhorn distances: Lightspeed computation of optimal transport},\ }\href@noop {} {\bibfield  {journal} {\bibinfo  {journal} {Advances in neural information processing systems}\ }\textbf {\bibinfo {volume} {26}} (\bibinfo {year} {2013})}\BibitemShut {NoStop}%
\bibitem [{\citenamefont {Pham}\ \emph {et~al.}(2020)\citenamefont {Pham}, \citenamefont {Le}, \citenamefont {Ho}, \citenamefont {Pham},\ and\ \citenamefont {Bui}}]{Pham-2020sinkhorn}%
  \BibitemOpen
  \bibfield  {author} {\bibinfo {author} {\bibfnamefont {K.}~\bibnamefont {Pham}}, \bibinfo {author} {\bibfnamefont {K.}~\bibnamefont {Le}}, \bibinfo {author} {\bibfnamefont {N.}~\bibnamefont {Ho}}, \bibinfo {author} {\bibfnamefont {T.}~\bibnamefont {Pham}},\ and\ \bibinfo {author} {\bibfnamefont {H.}~\bibnamefont {Bui}},\ }\bibfield  {title} {\bibinfo {title} {On unbalanced optimal transport: An analysis of {S}inkhorn algorithm},\ }in\ \href@noop {} {\emph {\bibinfo {booktitle} {International Conference on Machine Learning}}}\ (\bibinfo {organization} {PMLR},\ \bibinfo {year} {2020})\ pp.\ \bibinfo {pages} {7673--7682}\BibitemShut {NoStop}%
\bibitem [{\citenamefont {Kamijima}\ \emph {et~al.}(2024)\citenamefont {Kamijima}, \citenamefont {Takatsu}, \citenamefont {Funo},\ and\ \citenamefont {Sagawa}}]{kamijima-2024continuous}%
  \BibitemOpen
  \bibfield  {author} {\bibinfo {author} {\bibfnamefont {T.}~\bibnamefont {Kamijima}}, \bibinfo {author} {\bibfnamefont {A.}~\bibnamefont {Takatsu}}, \bibinfo {author} {\bibfnamefont {K.}~\bibnamefont {Funo}},\ and\ \bibinfo {author} {\bibfnamefont {T.}~\bibnamefont {Sagawa}},\ }\bibfield  {title} {\bibinfo {title} {Optimal finite-time maxwell's demons in langevin systems},\ }\href@noop {} {\bibfield  {journal} {\bibinfo  {journal} {arXiv preprint arXiv:2410.11603}\ } (\bibinfo {year} {2024})}\BibitemShut {NoStop}%
\bibitem [{\citenamefont {Wolpert}(2019)}]{Wolpert2019computation}%
  \BibitemOpen
  \bibfield  {author} {\bibinfo {author} {\bibfnamefont {D.~H.}\ \bibnamefont {Wolpert}},\ }\bibfield  {title} {\bibinfo {title} {The stochastic thermodynamics of computation},\ }\href@noop {} {\bibfield  {journal} {\bibinfo  {journal} {Journal of Physics A: Mathematical and Theoretical}\ }\textbf {\bibinfo {volume} {52}},\ \bibinfo {pages} {193001} (\bibinfo {year} {2019})}\BibitemShut {NoStop}%
\bibitem [{\citenamefont {Freitas}\ \emph {et~al.}(2021)\citenamefont {Freitas}, \citenamefont {Delvenne},\ and\ \citenamefont {Esposito}}]{Freitas-2021Circuits}%
  \BibitemOpen
  \bibfield  {author} {\bibinfo {author} {\bibfnamefont {N.}~\bibnamefont {Freitas}}, \bibinfo {author} {\bibfnamefont {J.-C.}\ \bibnamefont {Delvenne}},\ and\ \bibinfo {author} {\bibfnamefont {M.}~\bibnamefont {Esposito}},\ }\bibfield  {title} {\bibinfo {title} {Stochastic thermodynamics of nonlinear electronic circuits: A realistic framework for computing around $kt$},\ }\href {https://doi.org/10.1103/PhysRevX.11.031064} {\bibfield  {journal} {\bibinfo  {journal} {Phys. Rev. X}\ }\textbf {\bibinfo {volume} {11}},\ \bibinfo {pages} {031064} (\bibinfo {year} {2021})}\BibitemShut {NoStop}%
\bibitem [{\citenamefont {Rao}\ and\ \citenamefont {Esposito}(2016)}]{Rao-Esposito2016CRN}%
  \BibitemOpen
  \bibfield  {author} {\bibinfo {author} {\bibfnamefont {R.}~\bibnamefont {Rao}}\ and\ \bibinfo {author} {\bibfnamefont {M.}~\bibnamefont {Esposito}},\ }\bibfield  {title} {\bibinfo {title} {Nonequilibrium thermodynamics of chemical reaction networks: Wisdom from stochastic thermodynamics},\ }\href {https://doi.org/10.1103/PhysRevX.6.041064} {\bibfield  {journal} {\bibinfo  {journal} {Phys. Rev. X}\ }\textbf {\bibinfo {volume} {6}},\ \bibinfo {pages} {041064} (\bibinfo {year} {2016})}\BibitemShut {NoStop}%
\bibitem [{\citenamefont {Lan}\ \emph {et~al.}(2012{\natexlab{b}})\citenamefont {Lan}, \citenamefont {Sartori}, \citenamefont {Neumann}, \citenamefont {Sourjik},\ and\ \citenamefont {Tu}}]{Lan-2012energy-speed-accuracy}%
  \BibitemOpen
  \bibfield  {author} {\bibinfo {author} {\bibfnamefont {G.}~\bibnamefont {Lan}}, \bibinfo {author} {\bibfnamefont {P.}~\bibnamefont {Sartori}}, \bibinfo {author} {\bibfnamefont {S.}~\bibnamefont {Neumann}}, \bibinfo {author} {\bibfnamefont {V.}~\bibnamefont {Sourjik}},\ and\ \bibinfo {author} {\bibfnamefont {Y.}~\bibnamefont {Tu}},\ }\bibfield  {title} {\bibinfo {title} {The energy--speed--accuracy trade-off in sensory adaptation},\ }\href@noop {} {\bibfield  {journal} {\bibinfo  {journal} {Nature physics}\ }\textbf {\bibinfo {volume} {8}},\ \bibinfo {pages} {422} (\bibinfo {year} {2012}{\natexlab{b}})}\BibitemShut {NoStop}%
\bibitem [{\citenamefont {Ito}\ and\ \citenamefont {Sagawa}(2015)}]{Ito-Sgawa2015signaltransduction}%
  \BibitemOpen
  \bibfield  {author} {\bibinfo {author} {\bibfnamefont {S.}~\bibnamefont {Ito}}\ and\ \bibinfo {author} {\bibfnamefont {T.}~\bibnamefont {Sagawa}},\ }\bibfield  {title} {\bibinfo {title} {Maxwell's demon in biochemical signal transduction with feedback loop},\ }\href@noop {} {\bibfield  {journal} {\bibinfo  {journal} {Nature communications}\ }\textbf {\bibinfo {volume} {6}},\ \bibinfo {pages} {1} (\bibinfo {year} {2015})}\BibitemShut {NoStop}%
\bibitem [{\citenamefont {Murugan}\ \emph {et~al.}(2012)\citenamefont {Murugan}, \citenamefont {Huse},\ and\ \citenamefont {Leibler}}]{Arvind-2012speed-proofreading}%
  \BibitemOpen
  \bibfield  {author} {\bibinfo {author} {\bibfnamefont {A.}~\bibnamefont {Murugan}}, \bibinfo {author} {\bibfnamefont {D.~A.}\ \bibnamefont {Huse}},\ and\ \bibinfo {author} {\bibfnamefont {S.}~\bibnamefont {Leibler}},\ }\bibfield  {title} {\bibinfo {title} {Speed, dissipation, and error in kinetic proofreading},\ }\href@noop {} {\bibfield  {journal} {\bibinfo  {journal} {Proceedings of the National Academy of Sciences}\ }\textbf {\bibinfo {volume} {109}},\ \bibinfo {pages} {12034} (\bibinfo {year} {2012})}\BibitemShut {NoStop}%
\bibitem [{\citenamefont {Sartori}\ and\ \citenamefont {Pigolotti}(2015)}]{Sartori-Pigolotti2015Correction}%
  \BibitemOpen
  \bibfield  {author} {\bibinfo {author} {\bibfnamefont {P.}~\bibnamefont {Sartori}}\ and\ \bibinfo {author} {\bibfnamefont {S.}~\bibnamefont {Pigolotti}},\ }\bibfield  {title} {\bibinfo {title} {Thermodynamics of error correction},\ }\href {https://doi.org/10.1103/PhysRevX.5.041039} {\bibfield  {journal} {\bibinfo  {journal} {Phys. Rev. X}\ }\textbf {\bibinfo {volume} {5}},\ \bibinfo {pages} {041039} (\bibinfo {year} {2015})}\BibitemShut {NoStop}%
\bibitem [{\citenamefont {Ouldridge}\ \emph {et~al.}(2017)\citenamefont {Ouldridge}, \citenamefont {Govern},\ and\ \citenamefont {ten Wolde}}]{Ouldridge-2017Copying-Biochemical}%
  \BibitemOpen
  \bibfield  {author} {\bibinfo {author} {\bibfnamefont {T.~E.}\ \bibnamefont {Ouldridge}}, \bibinfo {author} {\bibfnamefont {C.~C.}\ \bibnamefont {Govern}},\ and\ \bibinfo {author} {\bibfnamefont {P.~R.}\ \bibnamefont {ten Wolde}},\ }\bibfield  {title} {\bibinfo {title} {Thermodynamics of computational copying in biochemical systems},\ }\href {https://doi.org/10.1103/PhysRevX.7.021004} {\bibfield  {journal} {\bibinfo  {journal} {Phys. Rev. X}\ }\textbf {\bibinfo {volume} {7}},\ \bibinfo {pages} {021004} (\bibinfo {year} {2017})}\BibitemShut {NoStop}%
\bibitem [{\citenamefont {Kolchinsky}\ and\ \citenamefont {Wolpert}(2021)}]{Kolchinsky-Wolpert2021Constraints}%
  \BibitemOpen
  \bibfield  {author} {\bibinfo {author} {\bibfnamefont {A.}~\bibnamefont {Kolchinsky}}\ and\ \bibinfo {author} {\bibfnamefont {D.~H.}\ \bibnamefont {Wolpert}},\ }\bibfield  {title} {\bibinfo {title} {Work, entropy production, and thermodynamics of information under protocol constraints},\ }\href {https://doi.org/10.1103/PhysRevX.11.041024} {\bibfield  {journal} {\bibinfo  {journal} {Phys. Rev. X}\ }\textbf {\bibinfo {volume} {11}},\ \bibinfo {pages} {041024} (\bibinfo {year} {2021})}\BibitemShut {NoStop}%
\bibitem [{\citenamefont {Zhong}\ and\ \citenamefont {DeWeese}(2022)}]{Zhong-DeWeese2022limited}%
  \BibitemOpen
  \bibfield  {author} {\bibinfo {author} {\bibfnamefont {A.}~\bibnamefont {Zhong}}\ and\ \bibinfo {author} {\bibfnamefont {M.~R.}\ \bibnamefont {DeWeese}},\ }\bibfield  {title} {\bibinfo {title} {Limited-control optimal protocols arbitrarily far from equilibrium},\ }\href {https://doi.org/10.1103/PhysRevE.106.044135} {\bibfield  {journal} {\bibinfo  {journal} {Phys. Rev. E}\ }\textbf {\bibinfo {volume} {106}},\ \bibinfo {pages} {044135} (\bibinfo {year} {2022})}\BibitemShut {NoStop}%
\end{thebibliography}
